\documentclass[preprint,amsmath,amssymb,aps,a4]{revtex4-2}
\usepackage{graphicx}
\usepackage{lineno}
\usepackage{bm}% bold math
\usepackage{txfonts}
\usepackage{hyperref}
\usepackage{caption}
\usepackage{amsmath}
\usepackage{amssymb}
\usepackage[section]{placeins}
\usepackage{tabularx}
\usepackage{braket}
\usepackage{multirow}
\usepackage{subcaption}
\usepackage{rotating}
\usepackage{xcolor}
\date{\today}
\newcommand{\be}{\begin{eqnarray}}
\newcommand{\ee}{\end{eqnarray}}

\newcommand{\bfk}{{\bf k}_{\perp}}

\newcommand{\Dp}{{\bf \Delta}_{\perp}}

\begin{document}
\title{Transverse and spatial structure of light to heavy pseudoscalar mesons in light-cone quark model}
\author{Satyajit Puhan$^{1}$}
\email{puhansatyajit@gmail.com}
\author{Navpreet Kaur$^{1}$}
\email{knavpreet.hep@gmail.com}
  \author{Harleen Dahiya$^{1}$}
\email{dahiyah@nitj.ac.in}
\affiliation{$^1$ Computational High Energy Physics Lab, Department of Physics, Dr. B.R. Ambedkar National
	Institute of Technology, Jalandhar, 144008, India}

\date{\today}% 
\begin{abstract}
In this work, we have investigated the transverse and spatial structure of light ($\pi^+ (u\bar d)$, $K^+(u\bar s)$) and heavy ($\eta_c(c\bar c)$, $\eta_b (b\bar b)$, $B$ and $D$) pseudoscalar mesons using the light-cone quark model. The transverse structure of these particles have been studied using the transverse momentum dependent parton distribution (TMDs). The leading twist unpolarized $f_1(x, \bfk^2)$ TMD has been solved using the quark-quark correlator. We have predicted the average momenta carried by the quark and antiquark of the considered mesons. For a complete description of mesons, we have also computed the leading twist unpolarized $H(x,0,-t)$ generalized parton distribution (GPD). Further, electromagnetic form factors (EMFFs), along with gravitational form factors (GFFs) have been derived by taking the zeroth and first moments of the unpolarized GPD. These EMFFs are found to be compatible with available lattice simulation results. Further, we have also calculated the parton distribution functions (PDFs) for both the quarks and the antiquarks of these mesons. The PDF sum rule has also been verified.

 \vspace{0.1cm}
    \noindent{\it Keywords}: Heavy mesons, transverse momentum dependent parton distributions (TMDs), generalized parton distributions (GPDs), parton distribution functions (PDFs), distribution amplitudes (DAs).
\end{abstract}
%====================================================
%
\maketitle
%
%\tableofcontents
%

\section{Introduction\label{secintro}}
Understanding the complexity of a hadron structure in terms of its constituent partons such as quarks, gluons and sea quarks with fundamental rules of interactions among them is still a difficult task. After the discovery of quarks, it has been very fascinating to understand the internal structure of hadrons and the properties of constituents within it. The strong interactions among partons within the hadron and the hadronic matrix elements of quark-gluon field operators can be understood by using Quantum Chromodynamics (QCD) \cite{Brodsky:1997de, Zhang:1997dd}. 
%The internal structure of a hadron can be understood using hadronic transition matrix elements of non-perturbative QCD. 
The one-dimensional parton distribution functions (PDFs) are the simplest distribution functions carrying information about the longitudinal momentum fraction ($x$) carried by the constituent quark from its parent hadron \cite{Collins:1981uw,Martin:1998sq,Gluck:1994uf}. PDFs are extracted from the deep inelastic scattering processes (DIS) in experiments \cite{Alekhin:2002fv,Polchinski:2002jw}. However, the transverse structure, spatial structure, quark densities, form factors, charge distributions etc of the quarks inside a hadron can not be accessed through the PDFs, the higher dimensional transverse momentum dependent parton distributions (TMDs) \cite{Diehl:2015uka,Angeles-Martinez:2015sea,Pasquini:2008ax,Kaur:2024wze}, generalized parton distributions (GPDs) \cite{Diehl:2003ny,Chavez:2021llq,Zhang:2021tnr,Broniowski:2022iip,Kaur:2023zhn,Guidal:2004nd}, generalized transverse momentum dependent parton distributions (GTMDs) \cite{Echevarria:2022ztg,Meissner:2009ww,Goeke:2005hb} are studied to achieve a complete description of a hadron. The three-dimensional ($3$D) TMDs are extended versions of PDFs with an addition of transverse structure. TMDs being  function of longitudinal momentum fraction $(x)$ and transverse momenta $(\bfk)$ of an active quark, carry information on azimuthal asymmetries, spin densities, spin-orbit correlations, transverse spin effects, QCD confinement and factorization. TMDs can be extracted from Drell-Yan experiments, semi-inclusive deep inelastic scattering (SIDIS) processes \cite{Bacchetta:2017gcc}, electron-positron annhilations \cite{Makris:2020ltr,Boer:1997mf}, $Z^0/W^{\pm}$ productions \cite{Catani:2015vma} and DIS at high energy. However, TMDs do not carry any information on the spatial structure of quarks inside a hadron. Therefore, $3$D GPDs play a pivotal role in describing the spatial structure of hadrons. GPDs are a function of longitudinal momentum fraction ($x$), skewness parameter ($\zeta$) and transverse momentum transfer $(\Delta_\perp)$ between initial and final hadron. Skewness parameter  $\zeta$ is the longitudinal momentum transferred between the initial and final parton. Elastic form factors (EFFs), mechanical properties, orbital angular momentum and physical properties of hadrons can be accessed through the GPDs \cite{Kaur:2024bgo,Freese:2020mcx,Luan:2023lmt} which can  be extracted further directly from deeply virtual Compton scattering (DVCS) \cite{Ji:1996nm,Xie:2023xkz} and deeply virtual meson production (DVMP) \cite{Favart:2015umi,Brooks:2018uqk}.

In this work, we have investigated the unpolarized quark TMD and GPD for spin-$0$ pseudoscalar mesons. We have introduced these distribution functions for the heavy quarkonia, $B$ and $D$-mesons along with the light pions and kaons. In case of spin-$0$ mesons, there are only two TMDs at the leading twist compared to $8$ quark TMDs for the case of spin-$1/2$ nucleons \cite{Meissner:2009ww,Meissner:2008ay}. However in this case, we have considered only the time-reversal $f^q(x,\bfk^2)$ quark unpolarized TMD \cite{Kaur:2019jfa}. There is no experimental data available for pion TMDs, but there have been pion TMD extractions from Drell-Yan data \cite{Cerutti:2022lmb} along with few lattice simulations observations \cite{LatticeParton:2023xdl,Engelhardt:2015xja}. Various models such as NJL model \cite{Pasquini:2014ppa}, DSE Model, light-front 
 holographic model \cite{Kaur:2019jfa}, maximum entropy method \cite{Han:2020vjp}, BSE model \cite{Kou:2023ady}, spectator model \cite{Pasquini:2014ppa} etc. have studied the pion TMDs successfully at the leading twist as well as for higher twists \cite{Pasquini:2014ppa,Puhan2023}. Similarly, only single unpolarized PDF exists in the case of spin-$0$ pseudoscalar mesons.
 However, in case of spin-$1/2$ and $1$ hadron \cite{Puhan:2023hio}, there are $3$ and $4$ PDFs respectively at the leading twist. For the case of pion, experimental data of  unpolarized PDFs along with lattice simulations is available \cite{Zhang:2018nsy}. Even though no experimental data of  the unpolarized PDFs for constituent quarks of kaon is available, some theoretical predictions along with model calculations are available \cite{Lan:2019rba,Cui:2020tdf}. Analogously for heavy mesons, only theoretical predictions are available for TMDs and PDFs \cite{Serna:2024vpn,Almeida-Zamora:2023bqb,Puhan:2024ckp,Acharyya:2024tql}. 
 
 Further, there are two GPDs at the leading twist for the spin-$0$ mesons, out of which $H_M(x,\zeta,-t)$ is the chiral-even unpolarized GPD. No experimental data and very few lattice simulation results \cite{Chen:2019lcm} are available for pion unpolarized GPD as far as our knowledge. There have been model calculations for pion and kaon GPDs in Refs. \cite{Kaur:2020vkq,Kaur:2018ewq}. This GPD is very special as it carries the information about electromagnetic form factors (EMFFs) and gravitational form factors (GFFs) of the mesons along with mechanical properties through $D$-term. In this current work, we have taken skewness parameter $\zeta$ as zero, therefore, the $D$-term vanishes. However, in the future we will try to calculate the mechanical properties like pressure, force and shear distribution through GPDs. The EMFF provide an insight of charge distribution inside a hadron. There have been some experimental data along with lattice simulations data for most of the mesons. Extraction of PDFs and EMFFs for  light mesons is one of the important goals of the upcoming electron-ion collider (EIC) project \cite{AbdulKhalek:2021gbh}. However, these can not be directly probed in the experiment but can be accessed through Sullivan process. Further, COMPASS and AMBER experiment will provide more insight on the PDFs of pion and kaon in the recent time \cite{Adams:2018pwt}. 

 \par  Valence quark distributions of spin-$0$ light and heavy pseudoscalar mesons in the form of TMDs, PDFs, GPDs and form factors has been calculated using light-cone quark model (LCQM) \cite{Arifi:2024mff,Weber:1992ww,Xiao:2002iv,Acharyya:2024enp,Xiao:2003wf,Puhan_2024}.   
 In case of heavy quarks, the higher Fock-state contributes very less compared to leading Fock-state \cite{Shi:2022erw}. So for this work, we have only considered the meson state as  $|M \rangle = \sum | q \bar{q} \rangle \psi_{q \bar{q}}$. LCQM is a non-perturbative framework for understanding the internal structure and properties of hadrons like mass spectra, radiative decay, decay constant etc. It is gauge invariant and relativistic by nature. The advantage of
LCQM is that it primarily focuses on the valence quarks of the hadrons and the valence quarks are the important constituents responsible for the overall structure and properties of hadrons. Even after proper QCD evolution of LCQM to perturbative limit, this model provides excellent results in describing the internal structure.

\par In this work, we have solved the quark-quark correlator for unpolarized valence quark TMDs and GPDs for light as well as heavy pseudoscalar mesons. As we are mainly focusing on the comparative analysis of light and heavy valence quarks, so we have not considered the gluon in the present work. We have obtained the overlap and explicit form of unpolarized $f^q_{M} (x,\bfk^2)$ quark TMD in the form of light front (LF) wave functions. We have considered both momentum space and spin wave functions for the LF wave functions. The collinear unpolarized TMD $f^q (x)$ quark PDF has been calculated by integrating the $f^q_{M} (x,\bfk^2)$ quark TMD over transverse momentum of valence quark. The pion and kaon valence quark PDFs have been evolved to $Q^2=16$ GeV$^2$ through DGLAP evolutions. These evolved PDFs have been compared with the available experimental data and other model predictions data. The average longitudinal momentum, transverse momentum and inverse momenta have also been calculated for all the mesons. To understand the spatial structure light as well as heavy pseudoscalar mesons, we have solved the GPDs correlator for valence quark at skewness $\zeta=0$. The $3$D structure of different mesons have been presented with respect to $x$ and $\Delta_{\perp}$. The $2$D EMFFs have been extracted from the GPDs and compared with the available experimental data as well as lattice simulation data. These EMFFs found to be sync with both experimental and lattice simulation data.

\par The paper is arranged as follows. In Sec. \ref{satya}, we have discussed the LCQM along with the spin and momentum wave function in LF formalism. The input parameters and wave form for different mesons have also been discussed in this section. The TMDs and PDFs have been solved using the quark-quark correlator in Sec. \ref{tmd}. The overlap form with explicit expressions for unpolarized quark TMD has been presented in this section. While in Sec. \ref{gpd}, we have presented the unpolarized GPDs and form factors derived from GPDs. We have finally summarized our work in Sec. \ref{con}.

 \section{Methodology}\label{satya}
 \subsection{Light-cone quark model}
 While describing the hadrons relativistically in terms of quark and gluon degrees of freedom, the light-cone (LC) formalism offers an useful framework. By using the LC Fock-state expansion, one can express the mesonic wave function as \cite{Lepage:1980fj,Brodsky:1997de,Pasquini:2023aaf}
 \begin{eqnarray}
|M\rangle &=& \sum |q\bar{q}\rangle \psi_{q\bar{q}}
        + \sum
        |q\bar{q}g\rangle \psi_{q\bar{q}g} + \sum
        |q\bar{q}gg\rangle \psi_{q\bar{q}g g} + \cdots  \, ,
\end{eqnarray}
where $|M\rangle$ denotes the meson eigenstate.
Since in this work we have not considered the higher Fock-state contribution \cite{Shi:2022erw}, the hadron wave function based on the  LC quantization of QCD  using multi-particle Fock-state expansion can be expressed as \cite{Puhan2023,Qian:2008px,Brodsky:2000xy}
\begin{eqnarray}\label{fockstate}
|M (P, S_z) \rangle
   &=&\sum_{n,\lambda_i}\int\prod_{i=1}^n \frac{\mathrm{d} x_i \mathrm{d}^2
        \mathbf{k}_{\perp i}}{\sqrt{x_i}~16\pi^3}
 16\pi^3 ~ \delta\Big(1-\sum_{i=1}^n x_i\Big)\nonumber\\
 &&\delta^{(2)}\Big(\sum_{i=1}^n \mathbf{k}_{\perp i}\Big) ~\psi_{n/M}(x_i,\mathbf{k}_{\perp i},\lambda_i)   | n ; \mathbf{k}^+_i, \mathbf{k}_{\perp i},
        \lambda_i \rangle.
\end{eqnarray}
Here $|M (P, S_z) \rangle$ is the hadron eigenstate with $P=(P^+,P^-,P_{\perp})$ as the meson's total momentum. $ \lambda_i$ and $S_z$ are the helicity of the $i$-th constituent and longitudinal spin projection of the target respectively. $\mathbf{k_i}=(\mathbf{k}^+_i,\mathbf{k}^-_i,\mathbf{k}_{i \perp})$ is the $i$-th constituent momentum of the meson. Longitudinal momentum fraction carried by an active quark is defined as $x=\frac{\mathbf{k}^+}{P^+}$. As we are dealing with lower Fock-state calculations, we have taken the minimal state description of Eq. (\ref{fockstate}) in the form of quark-antiquark and it can be expressed as
\begin{eqnarray}
|M(P, S_Z)\rangle &=& \sum_{\lambda_i,\lambda_j}\int
\frac{\mathrm{d} x \mathrm{d}^2
        \mathbf{k}_{\perp}}{\sqrt{x(1-x)}16\pi^3}
           \Psi_{S_Z}(x,\mathbf{k}_{\perp},\lambda_i,\lambda_j)|x,\mathbf{k}_{\perp},
        \lambda_i,\lambda_j \rangle
        .
        \label{meson}
\end{eqnarray}
The four-vector momenta of the meson ($P$), constituent quark ($k_1$) and anti-quark ($k_2$) in the LC frame are respectively defined as 
\begin{eqnarray}
P&\equiv&\bigg(P^+,\frac{\mathcal{M}^2}{P^+},\textbf{0}_\perp \bigg)\label{n1},\\
k_1&\equiv&\bigg(x P^+, \frac{\textbf{k}_\perp^2+m_q^2}{x P^+},\textbf{k}_\perp \bigg),\\
k_2&\equiv&\bigg((1-x) P^+, \frac{\textbf{k}_\perp^2+m_{\bar q}^2}{(1-x) P^+},-\textbf{k}_\perp \bigg),
\label{n3}
\end{eqnarray}
with $\mathcal{M}$ being the invariant mass of the composite meson system defined in terms of its quark mass $m_q$ and anti-quark mass $m_{\bar q}$ as
\begin{eqnarray}
    \mathcal{M}^2=\frac{\bfk^2+m^2_q}{x} +\frac{\bfk^2+m^2_{\bar q}}{1-x}\, .
\end{eqnarray}
 $\Psi_{S_Z}(x,\mathbf{k}_{\perp},\lambda_i,\lambda_j)$ in Eq. (\ref{meson}) is the LC meson wave function with different spin and helicity projections. It can be expressed as 
\begin{eqnarray}
\Psi_{S_z}(x,\textbf{k}_\perp, \lambda_i, \lambda_j)= J_{S_z}(x,\textbf{k}_\perp, \lambda_i, \lambda_j) \psi^{M}(x, \textbf{k}_\perp).\
\label{space}
\end{eqnarray}
Here $J_{S_z}(x,\textbf{k}_\perp, \lambda_i, \lambda_j)$ and $\psi^{M}(x, \textbf{k}_\perp)$ are the spin and momentum space wave functions of the mesons respectively.
The momentum space wave function in Eq. (\ref{space}) can be expressed using  Brodsky-Huang-Lepage
(BHL) \cite{Qian:2008px,Xiao:2002iv} as 
\begin{eqnarray}
\psi^M(x,\textbf{k}_\perp)= A \ {\rm exp} \Bigg[-\frac{\frac{\textbf{k}^2_\perp+m_q^2}{x}+\frac{\textbf{k}^2_\perp+m^2_{\bar q}}{1-x}}{8 \beta^2}
-\frac{(m_q^2-m_{\bar q}^2)^2}{8 \beta^2 \bigg(\frac{\textbf{k}^2_\perp+m_q^2}{x}+\frac{\textbf{k}^2_\perp+m_{\bar q}^2}{1-x}\bigg)}\Bigg]\, ,
\label{bhl-k}
\end{eqnarray}
%$m_{q (\bar q)}$ are the masses of quark and anti-quark of the meson. 
where $A$ and $\beta$ are the normalization constant and harmonic scale parameter of the mesons respectively.
$J_{S_z}$ in Eq. (\ref{bhl-k}) is front-form spin wave function derived from the instant form by Melosh-Wigner rotation \cite{Qian:2008px,Xiao:2002iv,Kaur:2020vkq}.
It is well known that to solve the ``proton spin puzzle", it is essential to comprehend the Melosh-Wigner rotation, which is fundamentally a relativistic phenomenon caused by the transverse motion of quarks inside the hadron \cite{Qian:2008px,Xiao:2002iv}.
This transformation of instant form state $\Phi(T)$ and front form state $\Phi(F)$ is expressed as 

\begin{eqnarray}
\Phi_i^\uparrow(T)&=&-\frac{[\textbf{k}_i^R \Phi_i^\downarrow(F)-(\textbf{k}_i^+ +m_{q(\bar q)})\Phi_i^\uparrow(F)]}{\omega_i},\label{instant-front1}\\
\Phi_i^\downarrow(T)&=&\frac{[\textbf{k}_i^L\Phi_i^\uparrow(F)+(\textbf{k}_i^+ +m_{q(\bar q)})\Phi_i^\downarrow(F)]}{\omega_i}.
\label{instant-front}
\end{eqnarray}
Here $\Phi(F)$ is a two-component Dirac spinor and $\textbf{k}_i^{R(L)}=\textbf{k}_i^1 \pm \iota \textbf{k}_i^2$. $\omega_i$ is defined as $\omega_i=1/ \sqrt{2 \textbf{k}^+_i (\textbf{k}^0+m_{q(\bar q)})}$. Now applying different momenta forms from Eqs. (\ref{n1})-(\ref{n3}) in  Melosh-Wigner rotation, the spin wave function is obtained in the form $\kappa_{0}^F(x,\textbf{k}_\perp, \lambda_1, \lambda_2)$ coefficient as 
\begin{eqnarray}
J_{S_z}(x,\textbf{k}_\perp, \lambda_i, \lambda_j)=\sum_{\lambda_1, \lambda_2}\kappa_{0}^F(x,\textbf{k}_\perp, \lambda_1, \lambda_2) \Phi_1^{\lambda_1}(F) \Phi_2^{\lambda_2}(F).
\end{eqnarray}
These spin-wave function coefficients satisfy the following normalization relation
\begin{eqnarray}
\sum_{\lambda_1,\lambda_2} \kappa_0^{F*}(x, \textbf{k}_\perp, \lambda_1, \lambda_2)\kappa_0^F(x, \textbf{k}_\perp, \lambda_1, \lambda_2)=1.
\end{eqnarray}
Similarly, the same spin-wave function can be calculated using the proper vertex chosen for the meson \cite{Choi:1996mq,Qian:2008px} as
\begin{eqnarray}
    J_{S_z}(x,\textbf{k}_\perp, \lambda_i, \lambda_j) = \bar u (k_1,\lambda_1) \frac{\gamma_5}{\sqrt{2}\sqrt{\mathcal{M}^2-(m^2_q-m^2_{\bar q}})} v(k_2,\lambda_2) \, .
\end{eqnarray}
Here $u$ and $v$ are the Dirac spinors.
Both the above methods give rise to same form of spin wave function. The spin wave function for pseudoscalar mesons ($S_z=0$) with different helicities is expressed as \cite{Qian:2008px}
\begin{equation}
\left\{
  \begin{array}{lll}
    J_{(S_z=0)}(x,\mathbf{k}_\perp,\uparrow,\uparrow)&=&\frac{1}{\sqrt{2}}\omega^{-1}(-\textbf{k}^L)(\mathcal{M}+m_q+m_{\bar q}),\\
    J_{(S_z=0)}(x,\mathbf{k}_\perp,\uparrow,\downarrow)&=&\frac{1}{\sqrt{2}}\omega^{-1}((1-x)m_q+x m_{\bar q})(\mathcal{M}+m_q+m_{\bar q}),\\
    J_{(S_z=0)}(x,\mathbf{k}_\perp,\downarrow,\uparrow)&=&\frac{1}{\sqrt{2}}\omega^{-1}(-(1-x)m_q-x m_{\bar q})(\mathcal{M}+m_q+m_{\bar q}),\\
    J_{(S_z=0)}(x,\mathbf{k}_\perp,\downarrow,\downarrow)&=&\frac{1}{\sqrt{2}}\omega^{-1}(-\textbf{k}^{R})(\mathcal{M}+m_q+m_{\bar q}),
  \end{array}
\right.
\end{equation}
with
$\omega=(M+m_q+m_{\bar q})\sqrt{x(1-x)[M^2-(m_q-m_{\bar q})^2]}$. The two-particle Fock-state in Eq. (\ref{meson}) can be written in the form of LC wave functions (LCWFs) with all possible helicities of its constituent quark and anti-quark as
\begin{eqnarray}
\ket{M (P^+,\textbf{P}_\perp,S_z=0)}&=&\int \frac{{ {\rm d}x  \rm d}^2\textbf{k}_\perp}{2 (2 \pi)^3 \sqrt{x(1-x)}}\big[\Psi_{S_z=0}(x,\textbf{k}_\perp, \uparrow, \uparrow)\ket{x P^+, \textbf{k}_\perp, \uparrow, \uparrow} \nonumber\\
&&+\Psi_{S_Z=0}(x,\textbf{k}_\perp, \downarrow, \downarrow)\ket{x P^+, \textbf{k}_\perp, \downarrow, \downarrow}+\Psi_{S_Z=0}(x,\textbf{k}_\perp, \downarrow, \uparrow)\nonumber\\
&&\ket{x P^+, \textbf{k}_\perp, \downarrow, \uparrow}+\Psi_{S_Z=0}(x,\textbf{k}_\perp, \uparrow, \downarrow)\ket{x P^+, \textbf{k}_\perp, \uparrow, \downarrow}\big].
\label{overlap}
\end{eqnarray}

\subsection{Input parameters and wave functions}
\begin{figure*}
\centering
\begin{minipage}[c]{0.98\textwidth}
(a)\includegraphics[width=7cm]{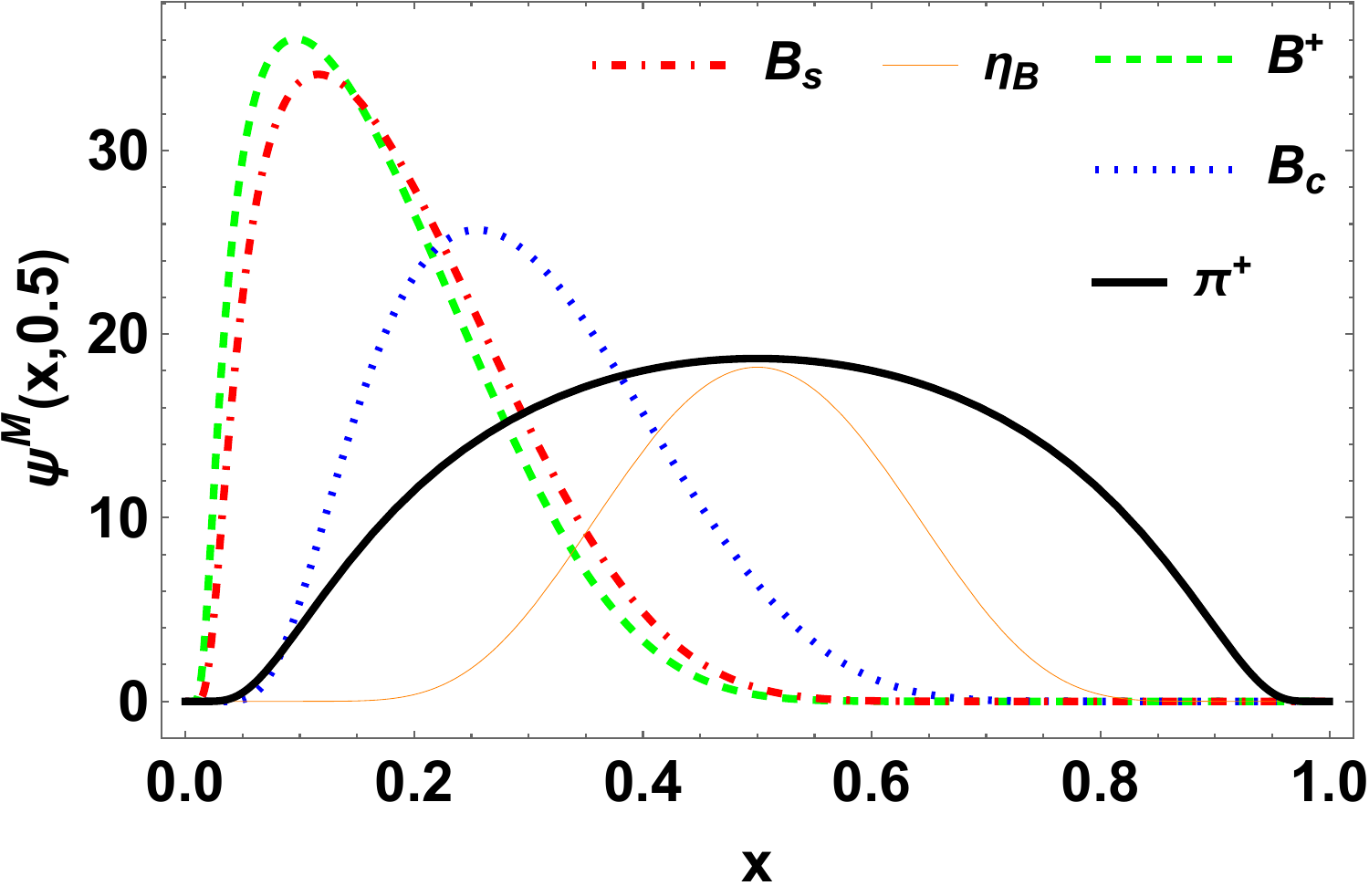}
\hspace{0.03cm}	
(b)\includegraphics[width=7cm]{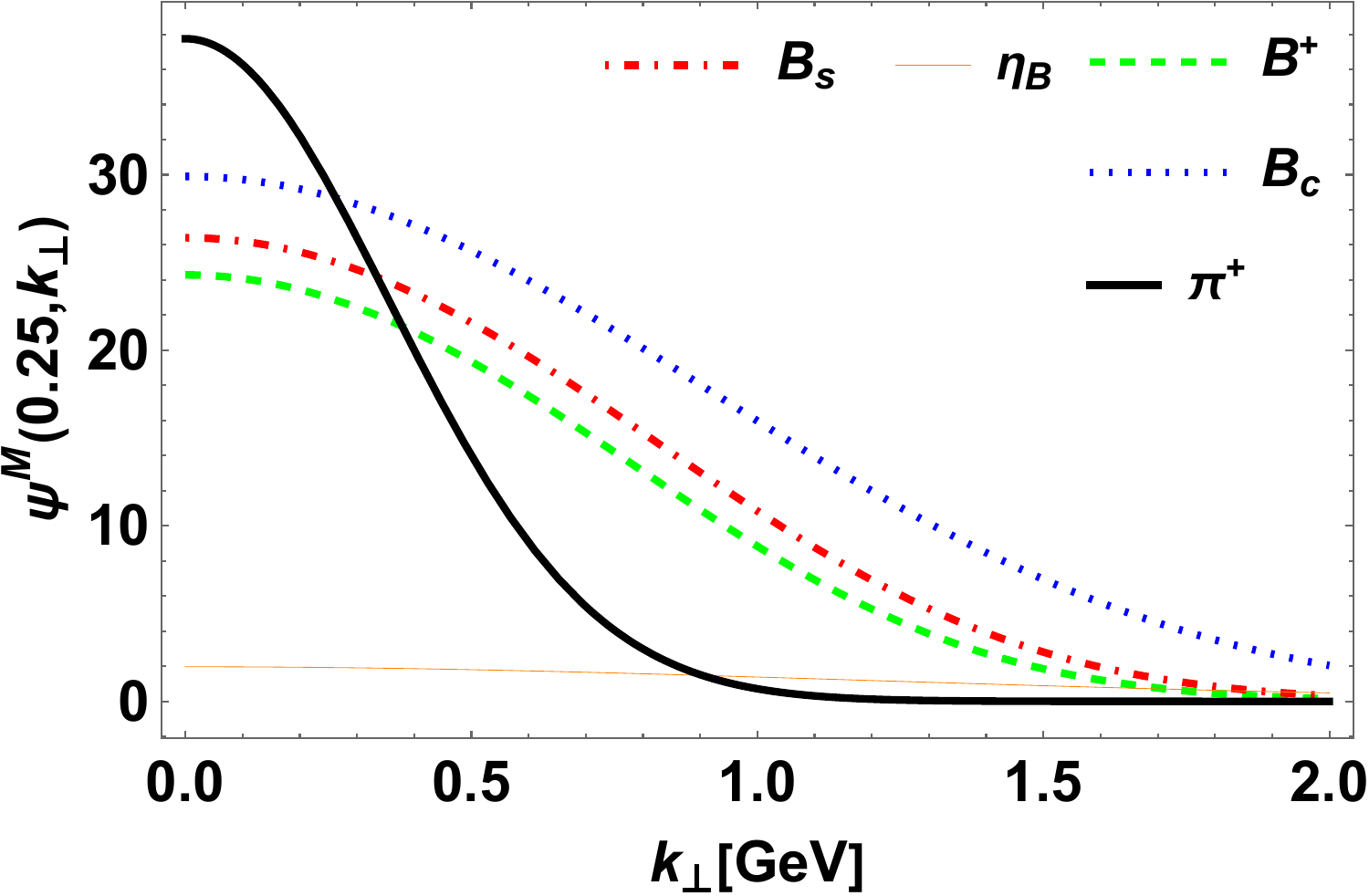} 
\hspace{0.03cm}
(c)\includegraphics[width=7cm]{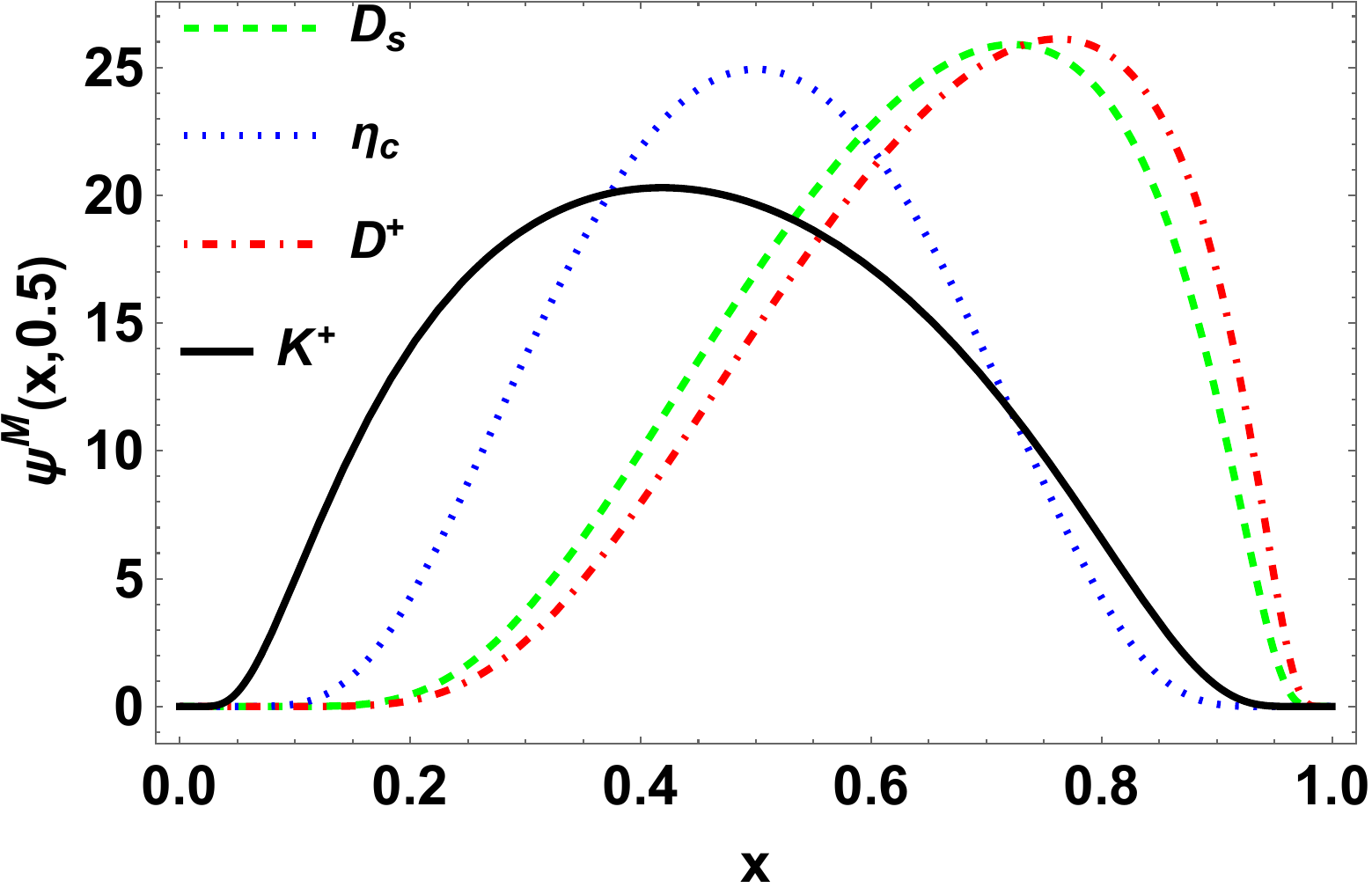}
\hspace{0.03cm}	
(d)\includegraphics[width=7cm]{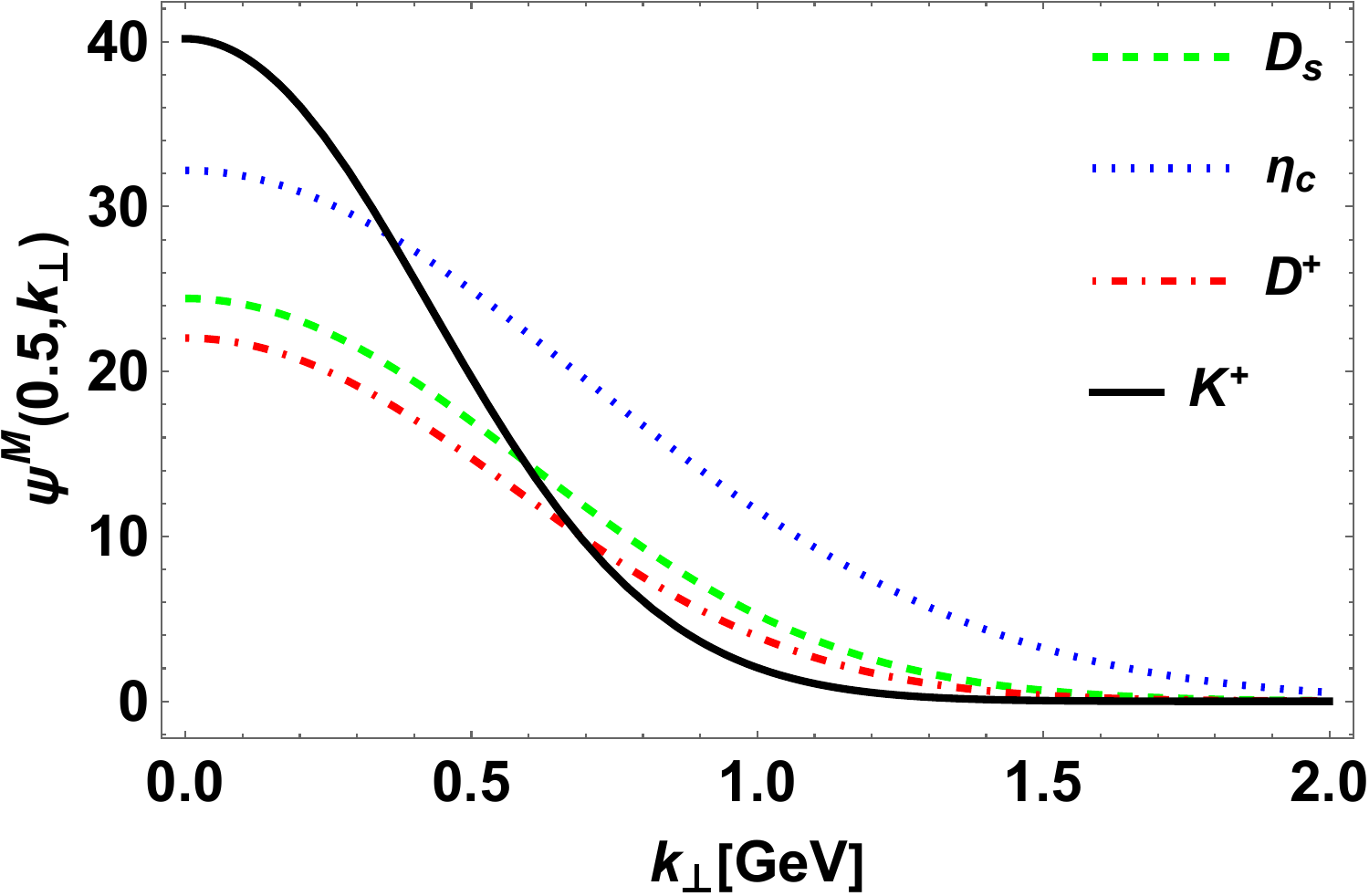} 
\hspace{0.03cm}
\end{minipage}
\caption{\label{wavefn} (Color online) Momentum space wave function of light and heavy mesons have been plotted with longitudinal momentum fraction $(x)$ at fixed value of transverse momentum ($\bfk=0.5$ GeV) in the left panel, while in the right panel these mesons have been plotted with $\bfk$ at fixed $x=0.25$ and $0.5$ respectively.}
\end{figure*}

For the numerical calculations in LCQM, we need only quark (antiquark) masses ($m_{q (\bar q)}$) and harmonic scale parameter ($\beta$) as the input parameters. These parameters have been adopted following Refs. \cite{Arifi:2022pal,Puhan2023} and have been presented in Table \ref{input}. These parameters have been calculated by fitting meson masses for pure state \cite{Puhan2023,Arifi:2022pal} and they provide excellent results for decay constant and DAs. For a complete descriptions of light and heavy quarks within a meson, we have taken $\pi^+ (u \bar d)$, $K^+(u \bar s)$, $\eta_c (c \bar c)$, $\eta_b (b \bar b)$, $D^+ (c \bar d)$, $D_s (c \bar s)$, $B^+ (u \bar b)$, $B_s (s \bar b)$ and $B_c (c \bar b)$ spin-$0$ mesons. 
In order to understand in detail the role of masses in the longitudinal momentum fraction carried by the parton, the momentum space wave functions $\psi^M(x,\textbf{k}_\perp)$ from Eq. (\ref{bhl-k}) have been plotted with respect to longitudinal momentum fraction $x$ at transverse quark momenta $\bfk^2=0.5$ GeV for all the mesons in the left panel of Fig. \ref{wavefn}. Momentum space wave functions of mesons with equal quark (antiquark) masses show symmetry about $x \longleftrightarrow (1-x)$. Mesons with a heavier antiquark show a shift of LCWF towards lower longitudinal momentum fraction $(x)$ and have a maximum distribution in between $0 \le x \le 5$, whereas opposite trend of distribution is observed for meson with heavier quark. Mesons like $B_c$, $B^+$, $D^+$ and $D_s$ with light-heavy quark-antiquark pair shows higher peak distributions than other mesons. $B$-meson shows no distribution after $x \le 0.6$ due to the presence of heavy $b$-anti-quark, while for the case of $D$-mesons it shows the trend below $x \le 0.2$. In case of $\eta_b$ meson, the distribution is higher than the $\eta_c$, however both have maximum distributions at $x=0.5$ along with pion.

\par In the right panel of Fig. \ref{wavefn}, we have plotted the momentum space wave function with respect to transverse momenta of quark $(\bfk)$ at a fixed value of longitudinal momentum fraction $(x=0.25)$ and $0.5$ respectively for different mesons. The wave function of pion vanishes after $\bfk=1$ GeV very sharply when compared to the drop in the case of other mesons but has a higher maximum distributions at $x=0.25$ as compared to the other mesons as seen in Fig. \ref{wavefn} (b). Similar observation can be made for the case of kaon in Fig. \ref{wavefn} (d). This clearly indicates that higher the mass of the meson, higher is the spread of the wave functions in the $\bfk$ direction. 
\par 

\begin{table}
    \centering
    \begin{tabular}{|c|c|c|c|c|c|c|c|c|c|c|c|c|}
    \hline
         $m_{u(d)}$& $m_{\textit{s}}$ & $m_{\textit{c}}$ & $m_{\textit{b}}$ & $\beta_{q \textit{q}}$ & $\beta_{q \textit{s}}$ & $\beta_{q \textit{c}}$ & $\beta_{c \textit{c}}$ &$\beta_{b \textit{b}}$ & $\beta_{\textit{s} \textit{c}}$ & $\beta_{ q \textit{b}}$ & $\beta_{\textit{s} \textit{b}}$ & $\beta_{\textit{b} \textit{c}}$ \\
         \hline
          0.22 & 0.45 & 1.68 & 5.10& 0.410 & 0.405& 0.500 & 0.699 & 1.376 & 0.537 & 0.585 & 0.636 & 0.906\\
          \hline
    \end{tabular}
    \caption{The constituent light quark masses $m_q$ ($q=u,d$ and $s$),  heavy quark masses $m_Q$ ($Q=c$ and $b$) and harmonic scale parameter $\beta$ are in unit of GeV. These parameters have been adopted from Ref. \cite{Arifi:2022pal,Puhan2023}}
    \label{input}
\end{table}

\section{Transverse momentum parton distribution functions}\label{tmd}
For spin-0 pseudoscalar mesons, the valence quark unpolarized TMDs can be expressed through the
quark-quark correlation function, which is defined as
\begin{eqnarray}
    f{^q}_{M}(x,\bfk^2)=\frac{1}{2} \int \frac{dz^- d^2z_\perp}{2 (2\pi)^3} e^{ix\bar{P} \cdot z} \bigg\langle M(P,\lambda^\prime)\bigg|~\bar{\Theta} \bigg(-\frac{z}{2}\bigg)~\mathcal{W}(-z/2,z/2) \gamma^+~ \Theta \bigg(\frac{z}{2}\bigg) ~\bigg| M(P,\lambda) \bigg\rangle \, , 
    \label{tmdeq}
\end{eqnarray}
where $z=(z^+,z^-,z^\perp)$ is the position four-vector. $\bar{P}$ is the average momentum of initial and final state momentum of the meson. $\Theta$ represents the quark field operator at two different positions $-z/2$ and $z/2$.  $\mathcal{W}(-z/2,z/2)$ is the Wilson line which preserves the gauge invariance of the bilocal quark field operators in the correlation functions \cite{Bacchetta:2020vty} which has been taken as $1$ in the present case. The function $f^q_{M}$ describes the momentum distribution of unpolarized valence quark within a meson. For the polarized quark momentum distributions, we have to study the Boer-Mulders $h^\perp_1 (x,\bfk^2)$ TMD \cite{Kaur:2019jfa}.
The overlap form of unpolarized $f^q_{M}(x,\bfk^2)$ TMD in the form of LCWFs, as expressed in Eq. (\ref{space}), are found to be 
\begin{eqnarray}
   f^q_{M}(x,\bfk^2)&=&\frac{1}{16 \pi^3} \big[\mid{ \Psi_{S_{z}=0}(x,\textbf{k}_\perp, \uparrow, \uparrow )}\mid^2 +\mid \Psi_{S_{z}=0}(x,\textbf{k}_\perp, \downarrow, \downarrow )\mid^2 + \mid \Psi_{S_{z}=0} (x,\textbf{k}_\perp, \downarrow, \uparrow )\mid^2 \nonumber\\
   &&+ \mid \Psi_{S_{z}=0}(x,\textbf{k}_\perp, \uparrow, \downarrow)\mid^2\big].
\end{eqnarray}
The explicit form of $f^q_{M}(x,\bfk^2)$ TMD by introducing space and spin wave functions can be expressed as

\begin{eqnarray}
    f^q_{M}(x,\bfk^2) &=& \frac{1}{16 \pi^3} \bigg[(\mathcal{M}+m_q+m_{\bar q})^2(\bfk^2+(1-x)m_q+x m_{\bar q})^2\bigg]
\frac{\mid \psi^M(x,\textbf{k}_\perp)\mid^2}{\omega^2}.\label{f1}
\end{eqnarray}
%-------------------------------------------------------
\begin{figure*}
\centering
\begin{minipage}[c]{0.98\textwidth}
(a)\includegraphics[width=7.5cm]{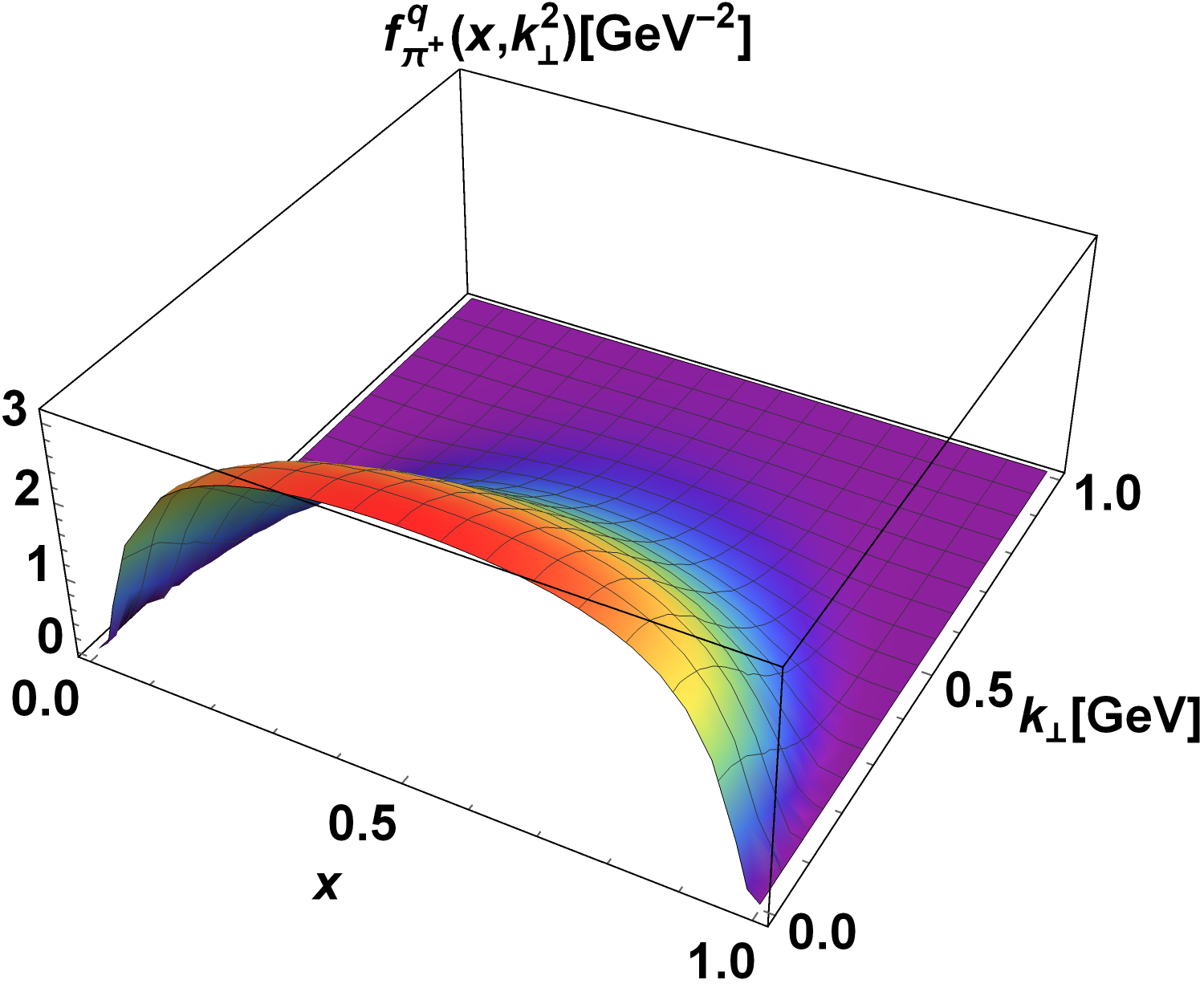}
\hspace{0.03cm}	
(b)\includegraphics[width=7.5cm]{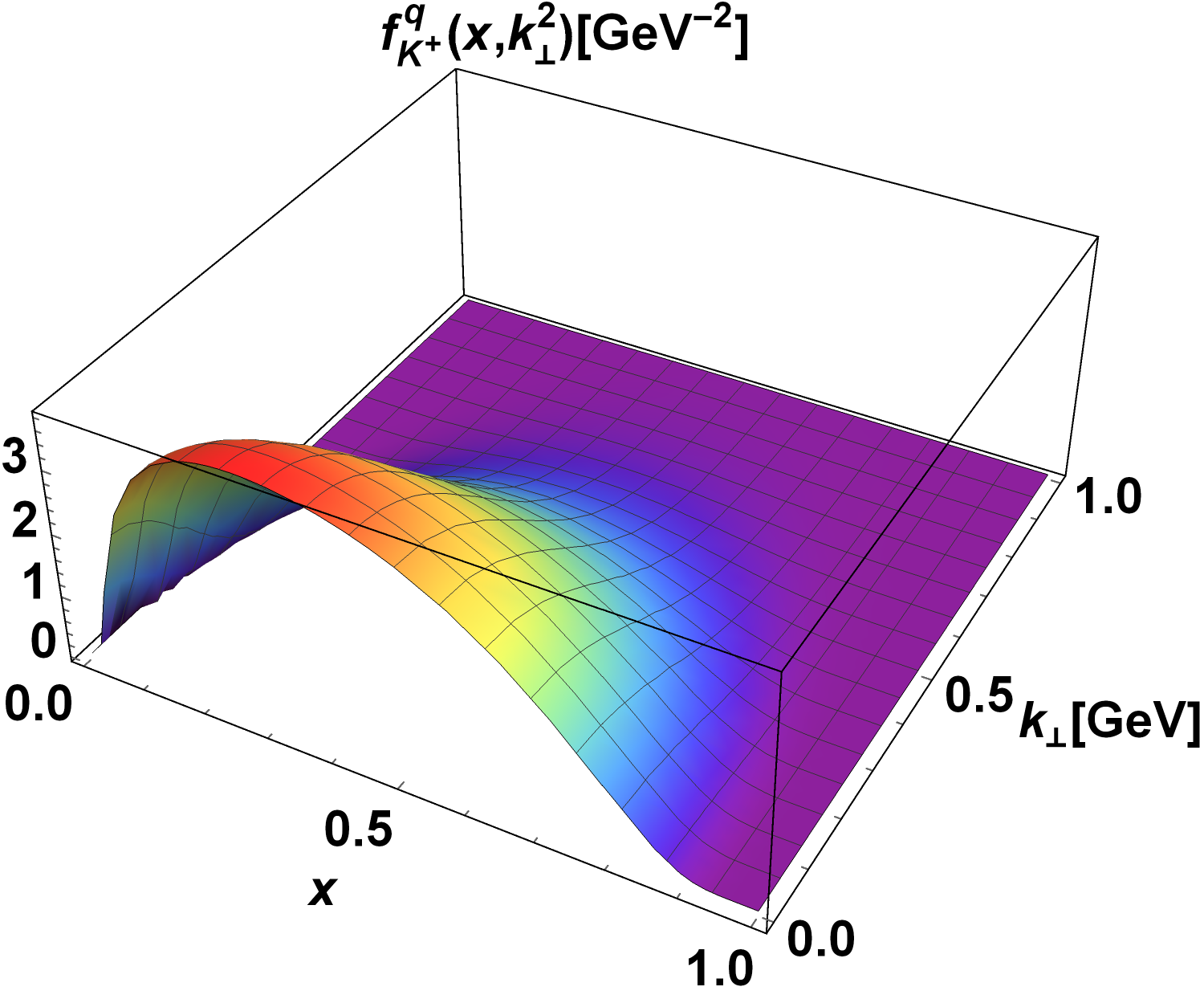} 
\hspace{0.03cm}
\end{minipage}
\caption{\label{tmd1} (Color online) Unpolarized quark transverse momentum parton distribution function as a function of longitudinal momentum fraction $x$ and transverse momentum $\bfk$ of (a) pion and (b) kaon.}
\end{figure*}
%---------------------------------------------------------------
\begin{figure*}
\centering
\begin{minipage}[c]{0.98\textwidth}
(a)\includegraphics[width=7.5cm]{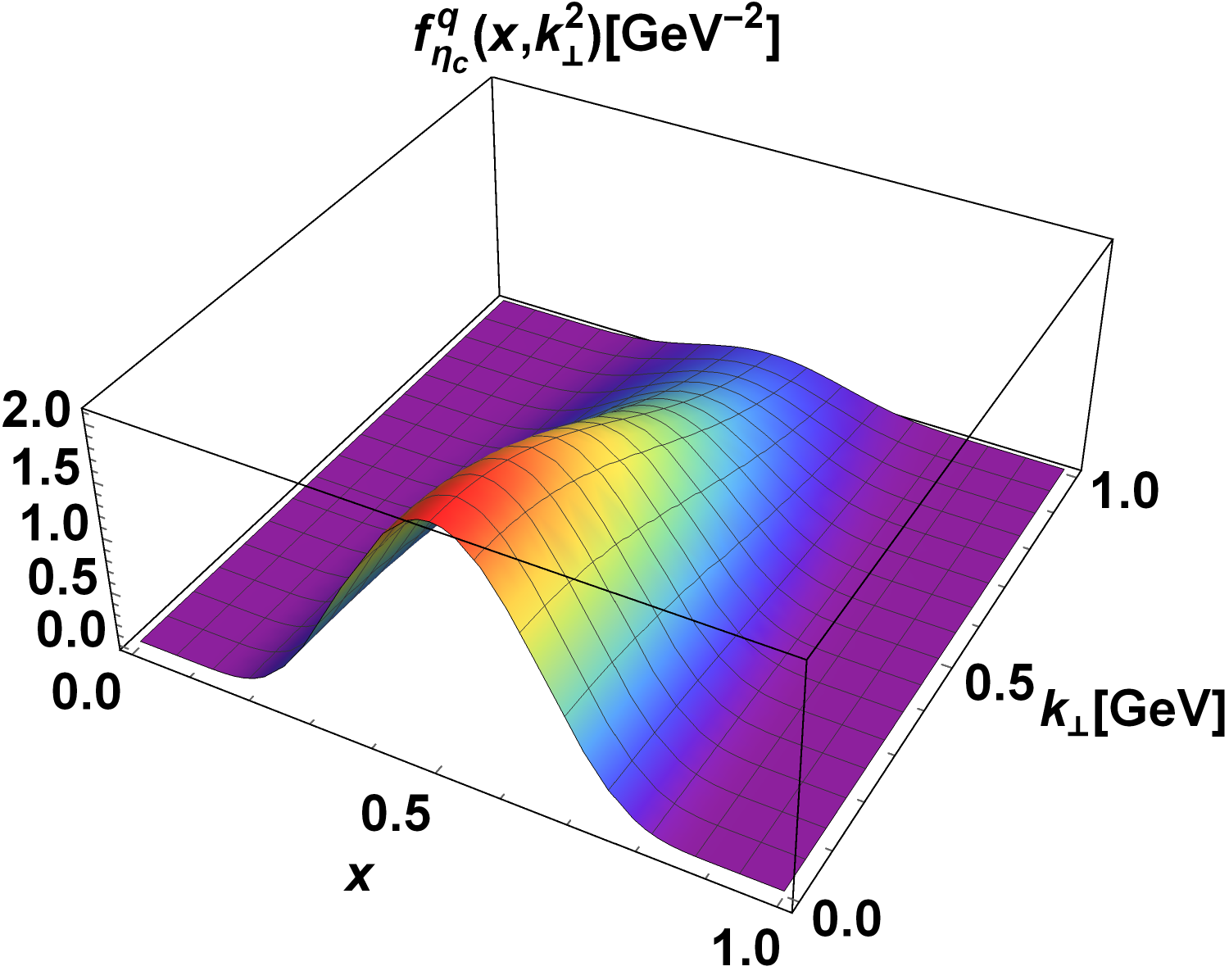}
\hspace{0.03cm}	
(b)\includegraphics[width=7.5cm]{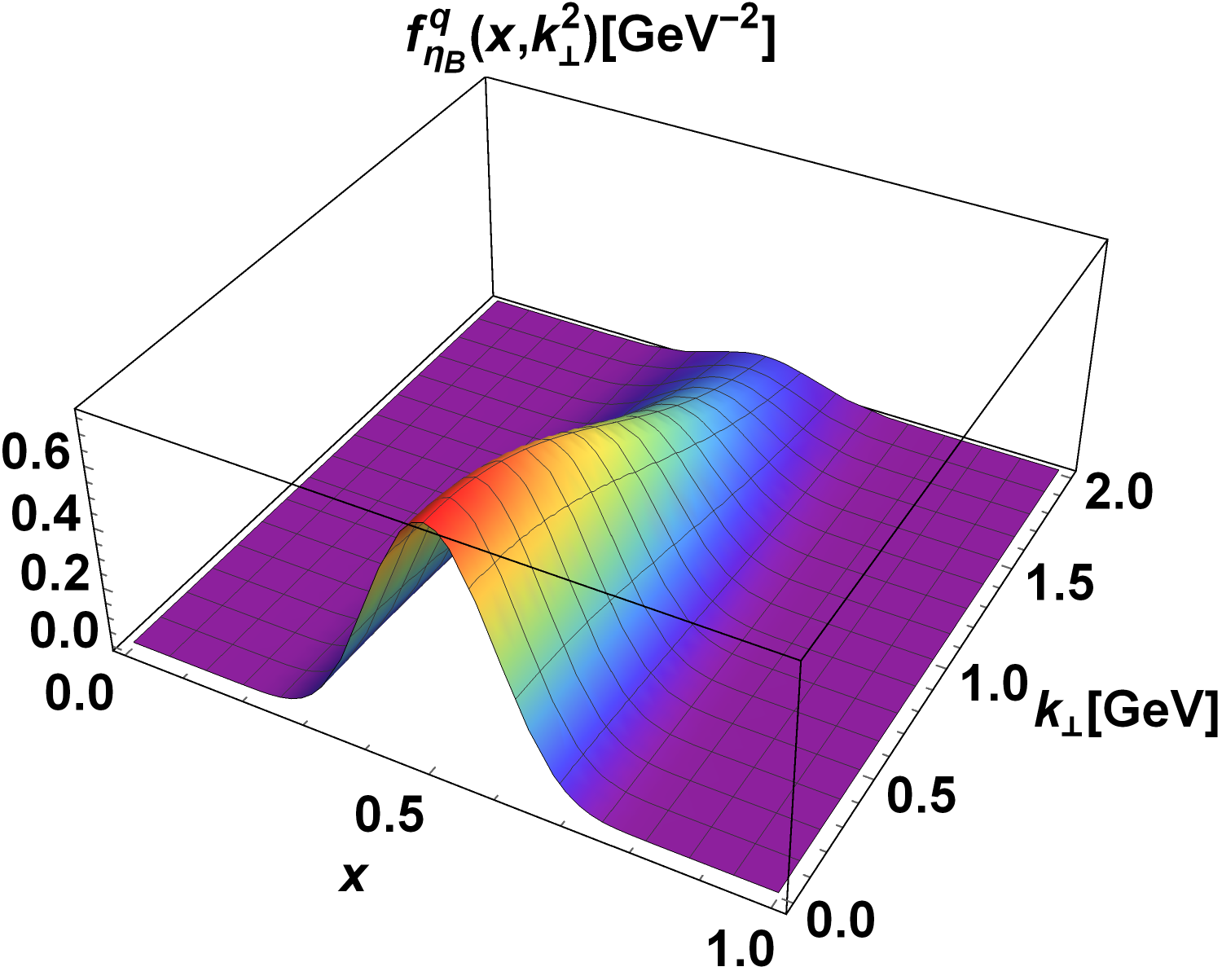} 
\hspace{0.03cm}
\end{minipage}
\caption{\label{tmd2} (Color online) Unpolarized quark transverse momentum parton distribution function as a function of longitudinal momentum fraction $x$ and transverse momentum $\bfk$ of (a) $\eta_c$ and (b) $\eta_b$ mesons.}
\end{figure*}
%--------------------------------------------------------------
\begin{figure*}
\centering
\begin{minipage}[c]{0.98\textwidth}
(a)\includegraphics[width=7.5cm]{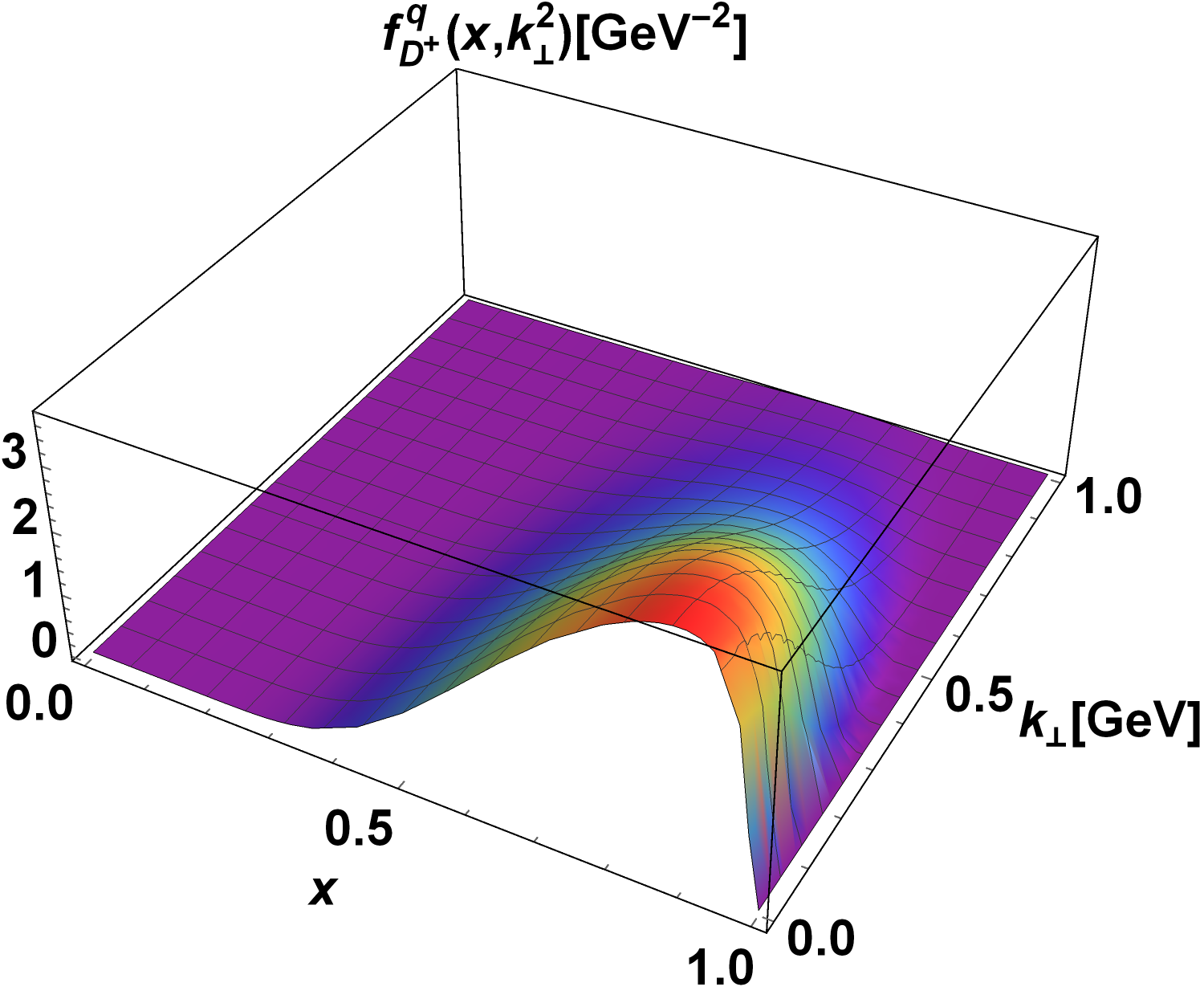}
\hspace{0.03cm}	
(b)\includegraphics[width=7.5cm]{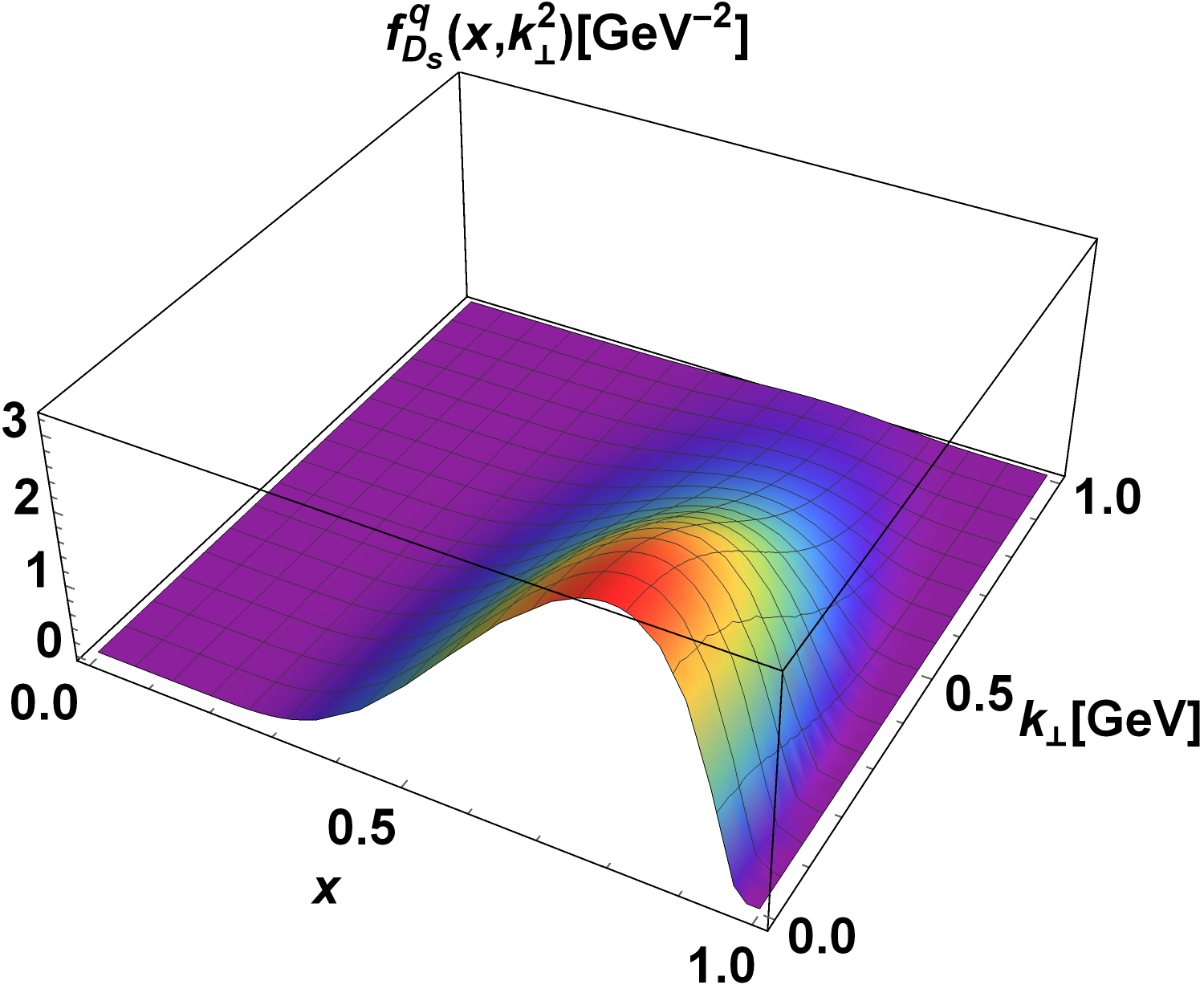} 
\hspace{0.03cm}
\end{minipage}
\caption{\label{tmd3} (Color online) Unpolarized quark transverse momentum parton distribution function as a function of longitudinal momentum fraction $x$ and transverse momentum $\bfk$ of (a) $D^+$ and (b) $D_s$ mesons.}
\end{figure*}
%--------------------------------------------------------------------
\begin{figure*}
\centering
\begin{minipage}[c]{0.98\textwidth}
(a)\includegraphics[width=7.5cm]{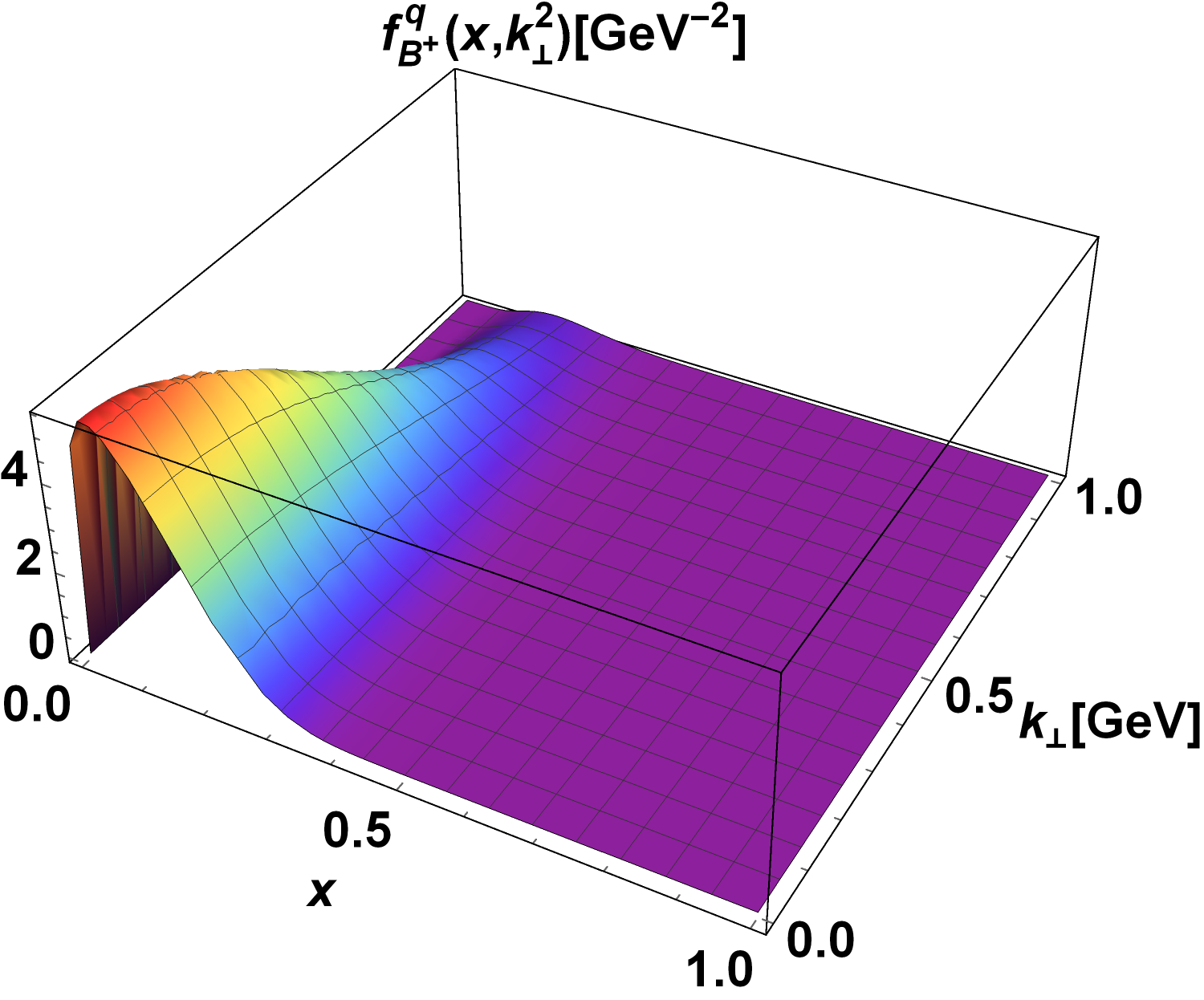}
\hspace{0.03cm}	
(b)\includegraphics[width=7.5cm]{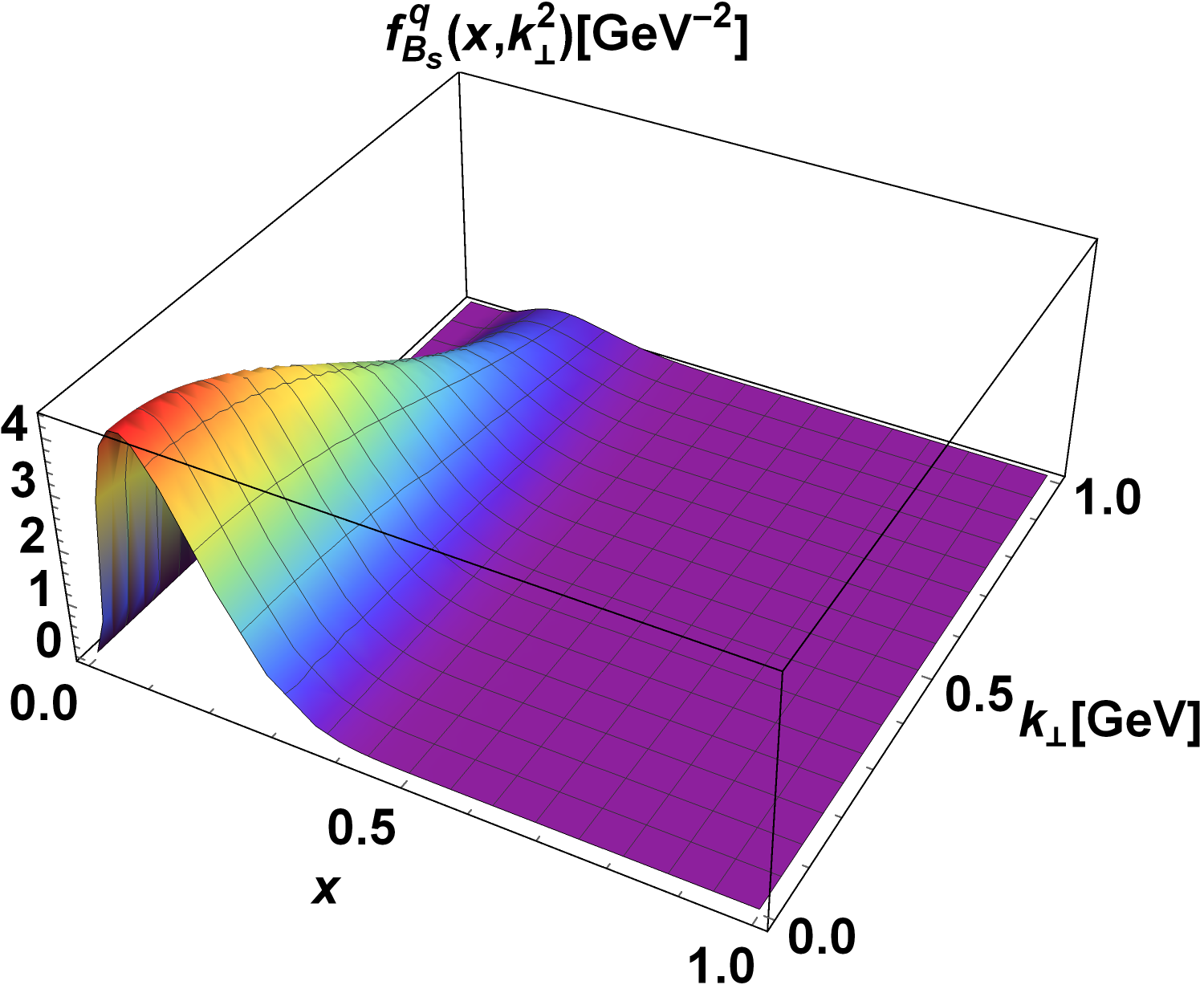} 
\hspace{0.03cm}
(c)\includegraphics[width=7.5cm]{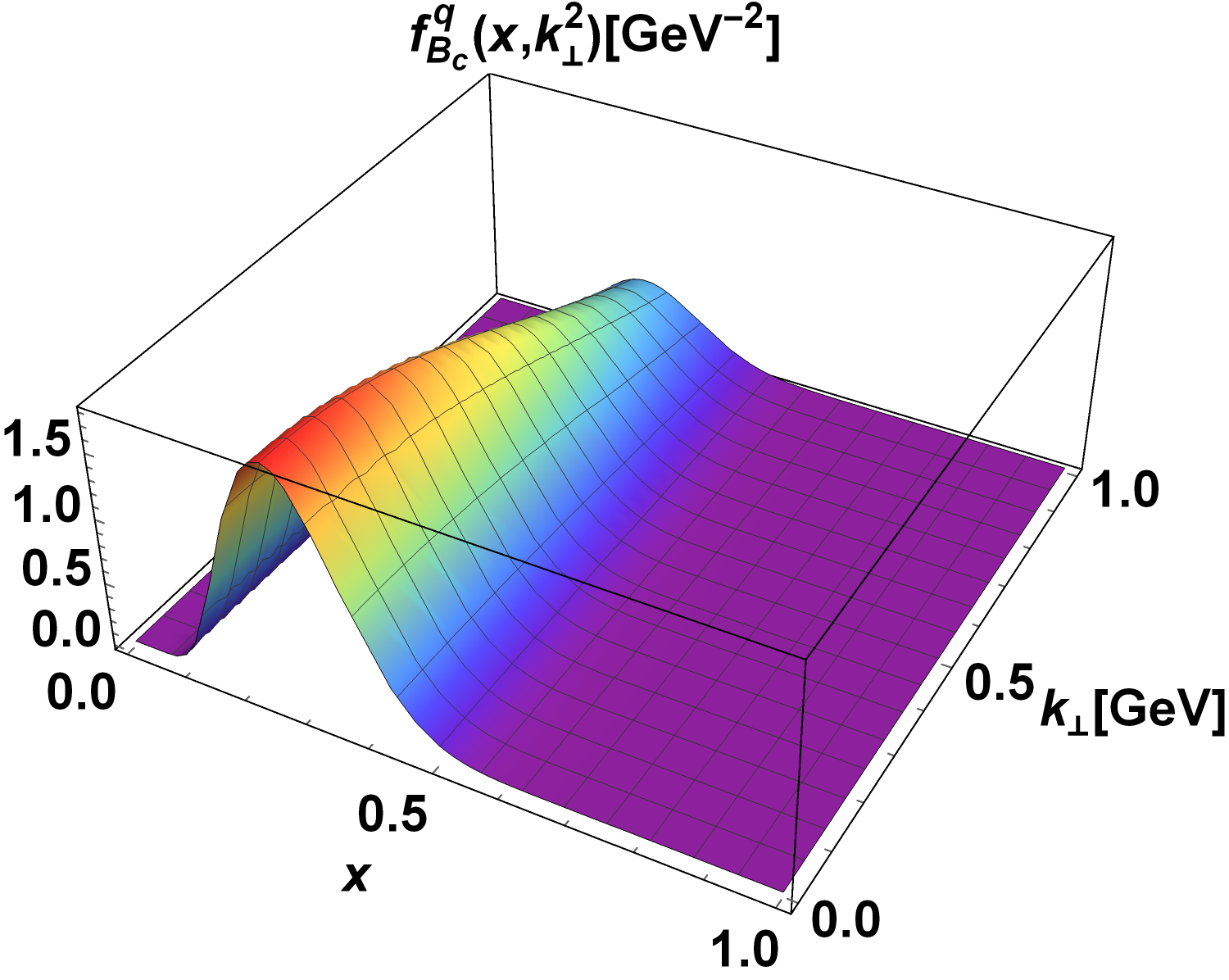} 
\hspace{0.03cm}
\end{minipage}
\caption{\label{tmd4} (Color online) Unpolarized quark transverse momentum parton distribution function as a function of longitudinal momentum fraction $x$ and transverse momentum $\bfk$ of (a) $B^+$  (b) $B_s$, and (c) $B_c$ mesons.}
\end{figure*}
%----------------------------
%--------------------------------------------------------------
%----------------------------------------------------------------
\par For the present work, we have considered only the unpolarized quark TMD. This TMD has been plotted with respect to longitudinal momentum fraction ($x$) and transverse momenta of quark ($\bfk^2$) for quark of light and heavy mesons in Figs. \ref{tmd1}, \ref{tmd2}, \ref{tmd3}, and \ref{tmd4}. For the case of pion (Fig. \ref{tmd1}), the quark TMD distribution is spread over the entire $x$ range and found to be a smooth decreasing function with increase in the transverse momentum $\bfk$ (GeV) of the quark. However, in case of other mesons (Figs. \ref{tmd2}, \ref{tmd3}, and \ref{tmd4}), the distributions are limited to certain range of $x$. It is observed that unpolarized quark TMD obeys the positivity constrain $f^q_{M} (x, \bfk^2) \ge 0$ for all the mesons \cite{Lorce:2016ugb, Lorce:2014hxa,Puhan2023}. Quark TMDs with lighter quark mass show a shift of distribution towards lower value of $x$, while the trend is opposite for the case of heavy quark TMDs. Further, the heavy meson distributions decrease slowly with increase in $\bfk$ as compared to that in  the light mesons. This kind of property has also been observed in the BS model \cite{Serna:2024vpn}, algebraic model \cite{Almeida-Zamora:2023bqb} and in DSE model \cite{Shi:2024laj}. Mesons with equal quark and antiquark mass show a symmetry about $x \leftrightarrow (1-x)$ with a peak distribution at $x=0.5$. Quark TMD of heavy mesons show narrow $x$ dependence and thin distributions compared to light mesons. Mesons with heavy quark and anti-quark masses show distributions spread in higher $\bfk$ values along with  narrower distributions in $x$. From Fig. \ref{tmd2}, it is observed that $\eta_b$ has higher distribution in $\bfk$ than $\eta_c$ due to the presence of heavy $\textit{b}$ quark in it. However both $D^+$ and $D_s$ mesons shows almost similar distributions. In case of $B$ mesons, $B_c$ meson shows quite different distribution than that of $B^+$ and $B_c$ mesons. However both $B^+$ and $B_c$ mesons show maximum distributions compared to other models. It is observed that dynamic chiral symmetry is less important for heavy mesons. For antiquark TMDs, distributions of these mesons can be obtained using the following relations \cite{Albino:2022gzs}
\begin{eqnarray}
    f^q_{M} (x, \bfk)= f^{\bar q}_{M} (1-x, -\bfk).
\end{eqnarray}
The above relations obey the conservation of momentum i.e, the total longitudinal momentum fraction and transverse momenta of carried by the constituent of a mesons are unity and zero respectively.

We have also calculated the average momenta $\langle \bfk \rangle$ and $\langle \bfk^2 \rangle$ carried by the quark inside a mesons using the following expression 
\begin{eqnarray}
    \langle \bfk^n \rangle = \frac{{\int} {\rm d}x ~{\rm d}^2{\bf k}_\perp | {\bf k}^n_\perp|  { f^q_{M}}(x,{\bf k}^2_\perp)}{\int {\rm d}x ~{\rm d}^2{\bf k}_\perp { f^q_{M}}(x,{\bf k}^2_\perp)},
\label{moments}
\end{eqnarray}
where $n=1$ and $2$. The calculated average $\langle \bfk \rangle$ and $\langle \bfk^2 \rangle$ values have been presented in Table \ref{table-moment}. It is observed that the mean transverse momenta $\langle \bfk \rangle$ of quark is minimum for the light mesons and it gradually increase with heavy quarks. It can be seen that $\eta_B$ valence quark carries the highest $\langle \bfk \rangle$ due to the presence of $\textit{b}$ quark and anti-quark in it. The $\langle \bfk^2 \rangle$ value is smaller than $\langle \bfk \rangle$ for the mesons except $\eta_b$ and $B_c$, clearly indicating a rare phenomena for heavy mesons. Similar kind of results have also been observed in the BSE model \cite{Shi:2024laj}. 

\subsection{Parton Distribution Functions} 

The probability of finding the constituent quark of a meson as a function of longitudinal momentum fraction $x$ is encoded in one-dimensional PDFs. In case of  spin-$0$ pseudoscalar mesons, there is only $f^q (x)$ unpolarized PDF at the leading twist \cite{Puhan2023, Kaur:2019jfa}. The unpolarized $f^q (x)$ can be obtained by integrating the $f^q_{M} (x, \bfk^2)$ TMD over transverse momentum of quark as
\begin{eqnarray}
    f^q (x)=\int d^2 \bfk f^q_{M} (x, \bfk^2).
\end{eqnarray}
PDFs can be calculated from quark GPDs by limiting our GPDs to $\Delta_\perp=0$, where $\Delta_\perp$ is difference between the transverse momentum of final and initial meson as 
\begin{eqnarray}
     f^q (x)= H_{M_q} (x, 0,0)
\end{eqnarray}

\par
\begin{figure*}
\centering
\begin{minipage}[c]{0.98\textwidth}
(a)\includegraphics[width=7.5cm]{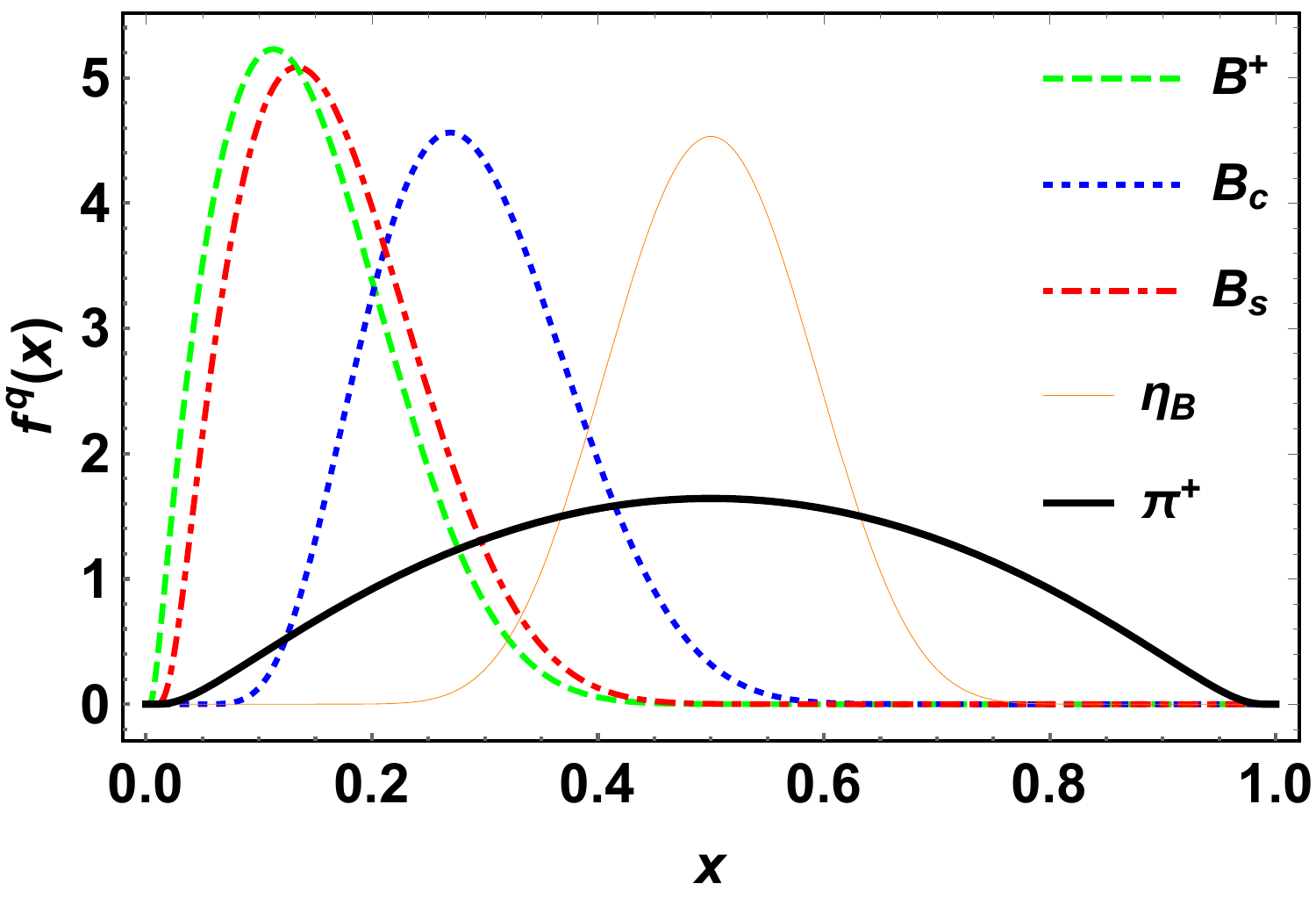}
\hspace{0.03cm}	
(b)\includegraphics[width=7.5cm]{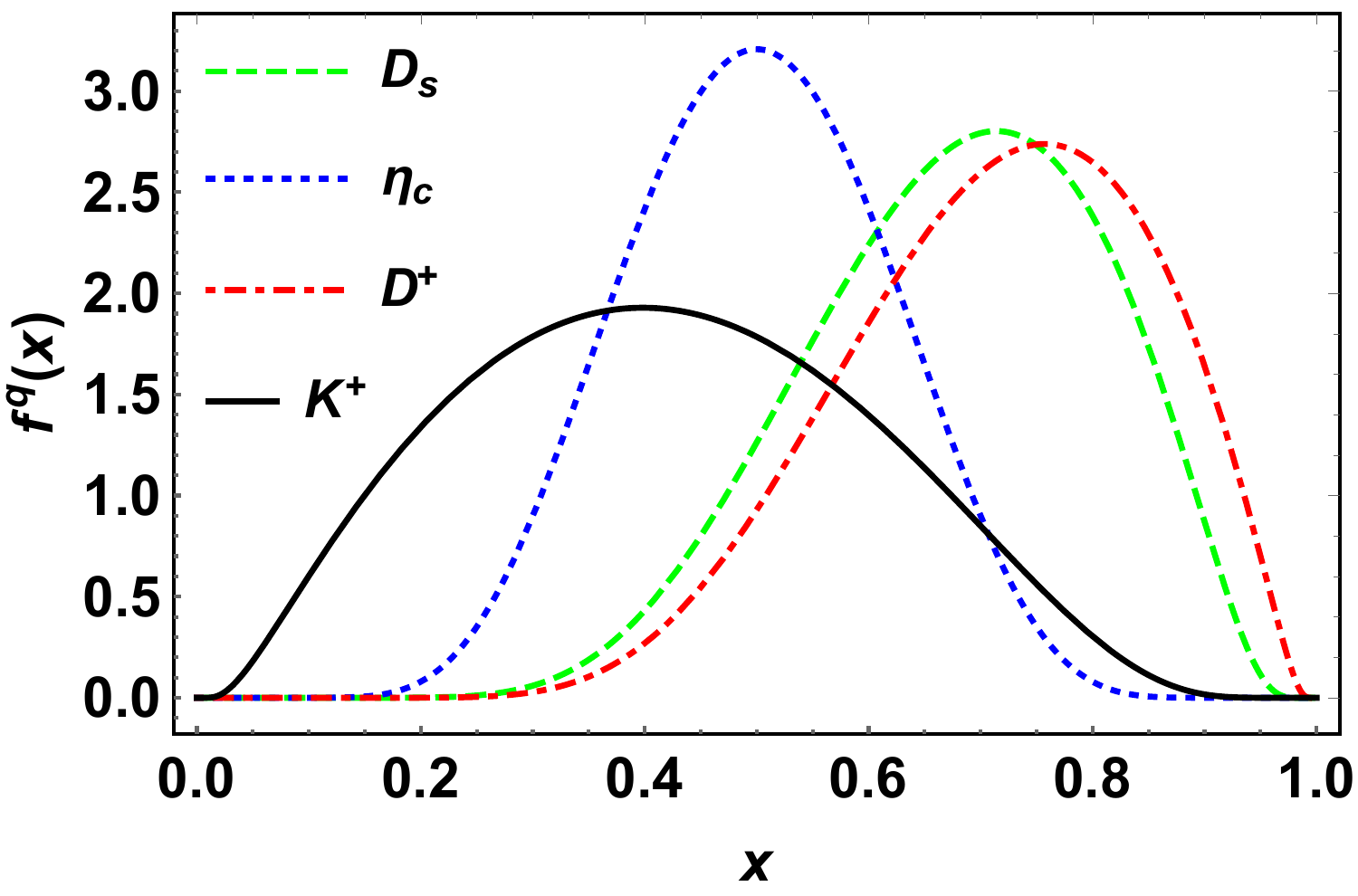} 
\hspace{0.03cm}
\end{minipage}
\caption{\label{pdf} (Color online) Unpolarized quark PDFs as a function of longitudinal momentum fraction $x$ for different mesons.}
\end{figure*}

\begin{figure*}
\centering
\begin{minipage}[c]{0.98\textwidth}
(a)\includegraphics[width=7.5cm]{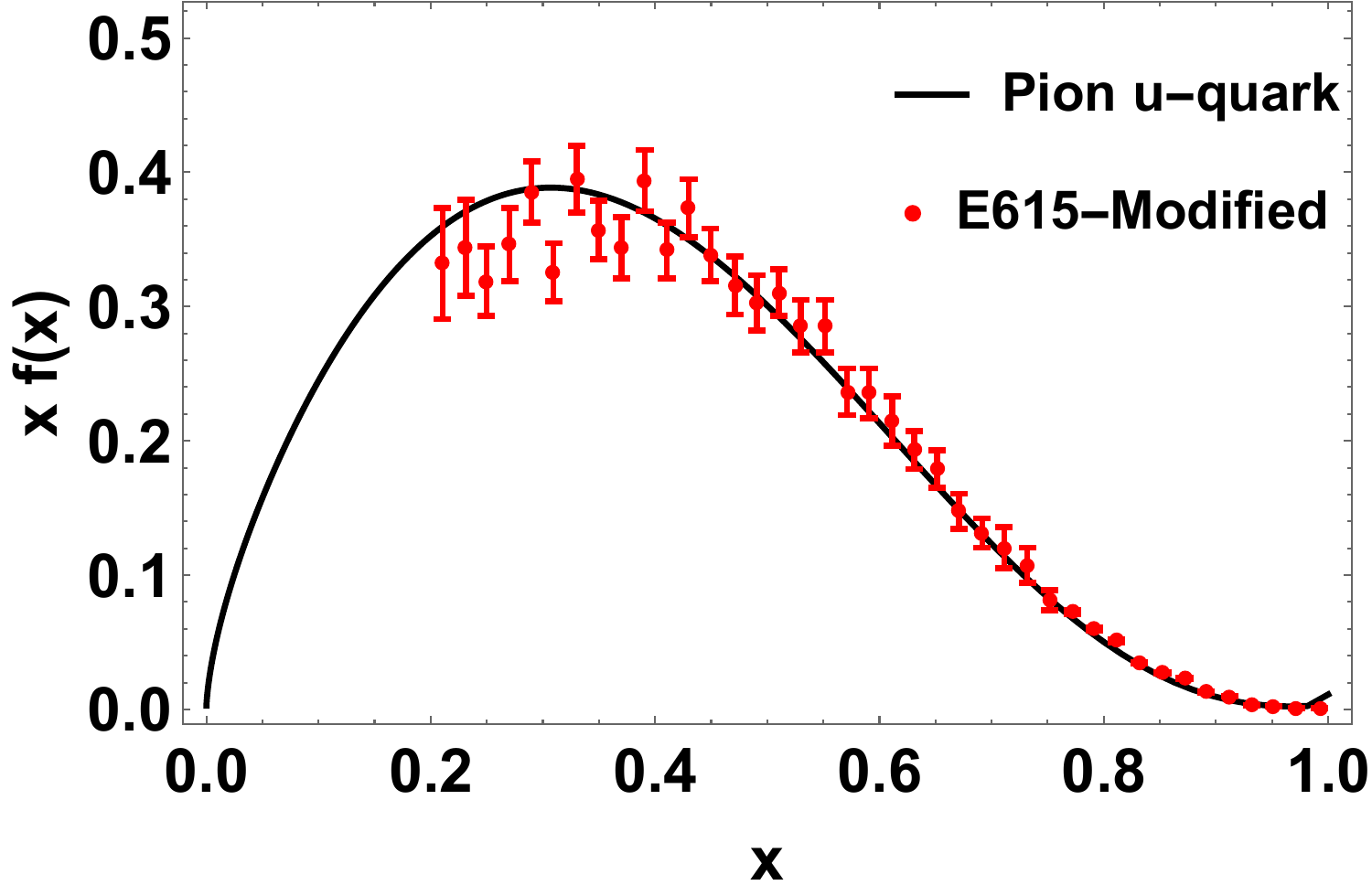}
\hspace{0.03cm}	
(b)\includegraphics[width=7.5cm]{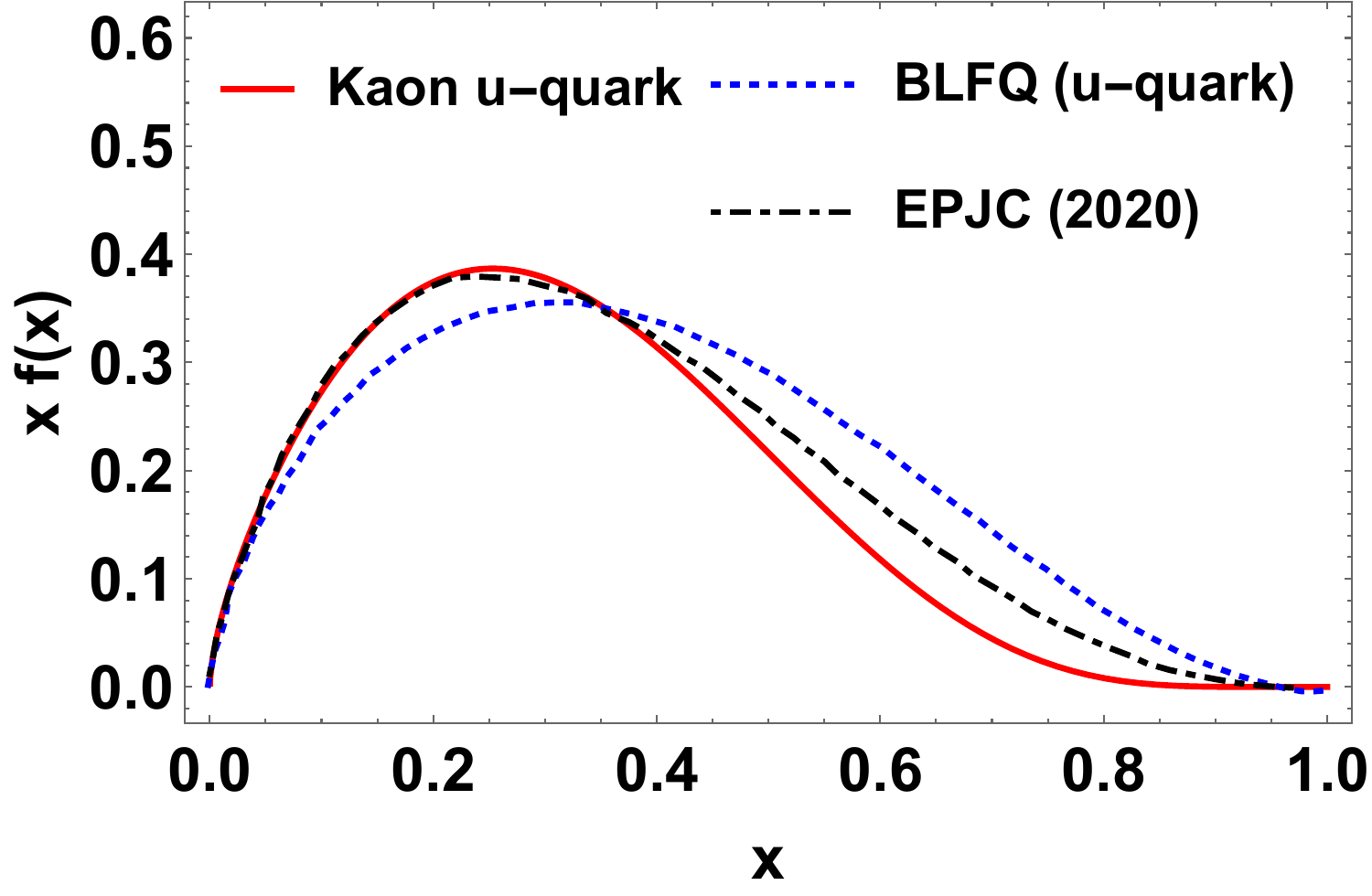} 
\hspace{0.03cm}
(c)\includegraphics[width=7.5cm]{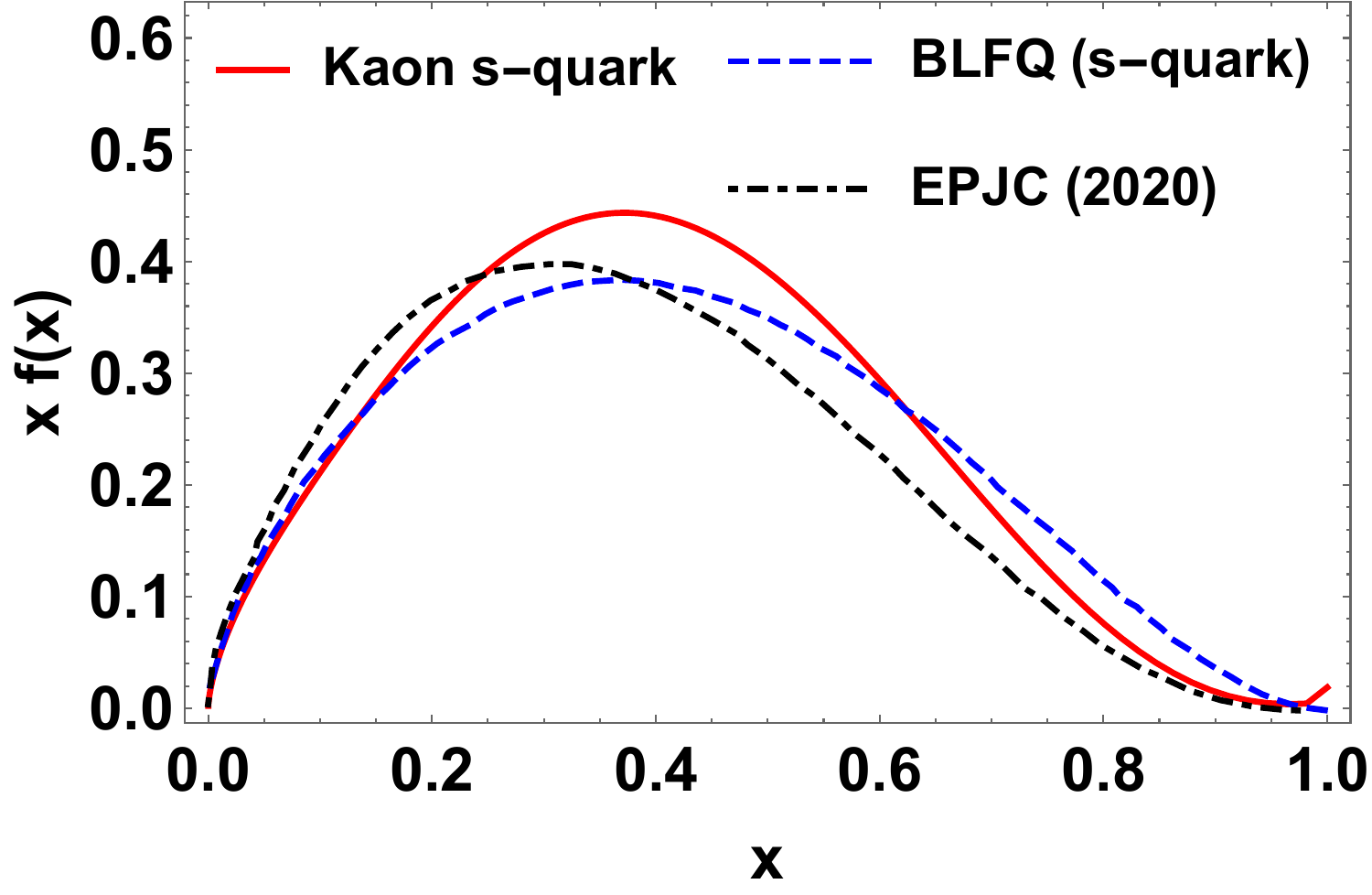} 
\hspace{0.03cm}
\end{minipage}
\caption{\label{evolve} (Color online) (a) pion $u$-quark PDF has been evolved to $16$ GeV$^2$ from an initial scale of $0.20$ GeV$^2$ and compared with modified FNAL-E$615$ data \cite{Aicher:2010cb}. In (b) and (c), we have plotted the kaon $u$-quark and $s$-quark PDFs at $16$ GeV$^2$ along with the BLFQ prediction data  \cite{Lan:2019rba} and data in Ref. \cite{Cui:2020tdf}.}
\end{figure*}

 In our calculations, the unpolarized PDF obeys all the PDF sum rules \cite{Lorce:2016ugb, Zhu:2023lst} which are given as follows

\begin{align}
  \int d x f^{q(\bar q)}(x) & = 1\,, 	\label{Eq:f1-sum-rule} \\
  \sum_q\int d x (x f^q(x)+(1-x) f^{\bar q}(x))& = 1. \label{Eq:mom-sum-rule} 
\end{align}
Here $f^{\bar q}$ is the anti-quark PDF. As in this work, we have not considered gluon contributions, therefore, the total momentum of a meson will be distributed between the quark and anti-quark. We have plotted the unpolarized $f^q (x)$ PDF for different mesons with respect to longitudinal momentum fraction $x$ in Fig. \ref{pdf}. It is observed that the pion $u$-quark PDF distributes all over $x$ and is symmetric about $x=0.5$ along with $\eta_c$ and $\eta_b$ mesons. The quark PDFs of $B$ mesons show  maximum distribution in the range of $0 \le x \le 0.6$, whereas for the case of $D$-mesons the distributions lie in the range $0.4 \le x \le 1$. This clearly indicates that light quark of heavy mesons shows PDF distribution towards lower $x$ and heavy quark of heavy mesons towards higher $x$. We have also evolved our pion and kaon constituent quark PDFs to ${Q}^2=16$ GeV$^2$ using next to leading order (NLO) Dokshitzer-Gribov-Lipatov-Altarelli-Parisi a (DGLAP) equations using Brute-Force method \cite{Miyama:1995bd,Hirai:1997gb,Hirai:1997mm,Hirai:2011si}. The initial scale of our model is ${Q}^2_0=0.20$ GeV$^2$. The evolved $u$-quark PDF of has been presented Fig. \ref{evolve} (a) with available modified FNAL-E$615$ experimental data \cite{Aicher:2010cb}. Our predictions are found to be consistent with the experimental data. As there is no experimental data available for kaon PDF, we have compared our $u$- and $s$-quark PDFs of kaon with BLFQ predictions \cite{Lan:2019rba} and with Ref. \cite{Cui:2020tdf} and presented the results of $u$- and $s$-quark PDFs of kaon  in Figs. \ref{evolve} (b) and (c) respectively. We observe that $u$-quark PDF of kaon vanishes after $x=0.8$ unlike in the case of BLFQ. Similar kind of observations were obtained in Ref. \cite{Cui:2020tdf}. The $s$-anti-quark PDF shows higher distributions as compared the $u$-quark, which has also been observed in Ref. \cite{Cui:2020tdf,Bourrely:2023yzi}. The kaon PDFs shows similar kind of distribution with other extractions \cite{Bourrely:2023yzi} and model predictions \cite{Han:2020vjp,Bednar:2018mtf}. 

\par Further, we have calculated the average longitudinal momentum $\langle x \rangle$ carried by the quark from its parent mesons as $\langle x \rangle = \int dx x f^q (x)$. The calculated $\langle x \rangle$ values at the model scale have been presented in Table. \ref{table-moment}. We observe that the quark PDF of mesons with maximum difference of quark and anti-quark masses carry very less momentum compared to other mesons where the difference of the masses is small. This indicates that $\langle x \rangle$ of quark PDF is inversely proportional to mass difference of quark and anti-quark. We have also calculated the inverse momenta as \cite{Brodsky:2007fr,Lorce:2016ugb}
\begin{eqnarray}
    \langle x^{-1} \rangle = \int dx x^{-1} f^q (x).
\end{eqnarray}
The inverse moment value of these mesons have also been presented in Table \ref{table-moment}.
These inverse moments play
an important role in describing the sum rules. It is observed that $\langle x^{-1} \rangle$ is maximum and minimum for $B^+$ and $B_s$ respectively. 
%------------------------------------------------------------
\begin{table}
\centering
\begin{tabular}{|c|c|c|c|c|c|}
\hline 
 & \multicolumn{4}{c|}{LCQM} & \multicolumn{1}{c|}{BSE model}\\
 Mesons & \multicolumn{4}{c|}{(This work)} & \multicolumn{1}{c|}{\cite{Shi:2024laj}} \\
\cline{2-6}
& ${\langle{{\bfk}}\rangle}$ (GeV) & ${\langle{{\bfk^2}}\rangle}$ (GeV$^2$)& $ ~~~\langle x \rangle$~~~ & ~$\langle x^{-1} \rangle$~ &  ${\langle{{\textbf{k}}_\perp}\rangle}$ (GeV) \\
\hline
$\pi^+ (u \bar d)$ & 0.328 & 0.1395 & 0.50 & 2.634 & 0.39\\
$K^+ (u \bar s)$ & 0.334 & 0.143 & 0.42 & 3.140 & - \\
$\eta_c (c \bar c)$ & 0.602 & 0.462 & 0.50 & 2.127 & 0.65\\
$\eta_B (b \bar b)$ & 1.201 & 1.839 & 0.50 & 2.062 & 1  \\
$D^+(c \bar d)$ & 0.426 & 0.232 & 0.71 & 1.459 & 0.43  \\
$D_s(c \bar s)$ & 0.677 & 0.458 & 0.27 & 1.543 & -  \\
$B^+(u \bar b)$ & 0.510 & 0.331 & 0.15 & 10.013 & 0.42 \\
$B_s (s \bar b)$ & 0.554 & 0.391 & 0.17 & 1.211 & - \\
$B_c (c \bar b)$ & 0.790 & 0.794 & 0.29 & 3.780 & 0.65 \\
\hline
\end{tabular}
\caption{The average quark momenta $\langle \bfk \rangle$ and $\langle \bfk^2 \rangle$, average longitudinal quark momenta $\langle x \rangle$, inverse momenta $\langle x^{-1} \rangle$ for all the mesons. The BSE results \cite{Shi:2024laj} for $\langle \bfk \rangle$ have also been presented for comparison.}
\label{table-moment}
\end{table}
%-------------------------------------------------------------

%---------------------------------------------------
%--------------------------
\section{Generalized parton distribution functions}\label{gpd}
The matrix elements of quark operators at a light-like separation are defined as GPDs \cite{Diehl:2003ny}. For spin-$0$ particles, we have only one chiral-even unpolarized GPD which can be defined in terms of the bilocal current as
\begin{eqnarray}
    H_{M}(x,\zeta,-t)=\frac{1}{2} \int \frac{dz^-}{2\pi} e^{ix\bar{P}^+z^-} \bigg\langle M(P^\prime,\lambda^\prime)\bigg|~\bar{\Theta} \bigg(-\frac{z}{2}\bigg)~ \gamma^+~ \Theta \bigg(\frac{z}{2}\bigg) ~\bigg| M(P,\lambda) \bigg\rangle \, . 
\end{eqnarray}
 Other kinematic variables which include the four-momentum transfer and skewness parameter are respectively expressed as $\Delta^\mu=P^{\prime \mu}-P^\mu$ with $t-\Delta^2=-\Delta_\perp$ and $\zeta=-\Delta^+/2P^+$. We have chosen light-front gauge as $A^+=0$ which in turn makes gauge link, appearing between the quark field operators, unity. The overlap form of GPD $H_{M_q}(x,0,-t)$ with zero skewness can be expressed as
%\begin{eqnarray}
   % H_{M}(x,0,-t)&=&\int \frac{d^2 \mathbf{k}}{16 \pi^3} \big[\Psi_{S_z=0}^\ast (x^{\prime\prime},\mathbf{k}^{\prime\prime}_\perp, \uparrow, \uparrow) \Psi_{S_z=0} (x^{\prime},\mathbf{k}^{\prime}_\perp, \uparrow, \uparrow) \nonumber \\ 
  %  &&+ \Psi_{S_z=0}^\ast (x^{\prime\prime},\mathbf{k}^{\prime\prime}_\perp, \uparrow, \downarrow) \Psi_{S_z=0} (x^{\prime},\mathbf{k}^{\prime}_\perp, \uparrow, \downarrow) \nonumber \\ 
  %  &&+ \Psi_{S_z=0}^\ast (x^{\prime\prime},\mathbf{k}^{\prime\prime}_\perp, \downarrow, \uparrow) \Psi_{S_z=0} (x^{\prime},\mathbf{k}^{\prime}_\perp, \downarrow, \uparrow) \nonumber \\
 %   &&+ \Psi_{S_z=0}^\ast (x^{\prime\prime},\mathbf{k}^{\prime\prime}_\perp, \downarrow, \downarrow) \Psi_{S_z=0} (x^{\prime},\mathbf{k}^{\prime}_\perp, \downarrow, \downarrow)\big] \, ,
%\label{GPDeq}
%\end{eqnarray}
\begin{eqnarray}
    H_{M}(x,0,-t)&=&\int \frac{d^2 \mathbf{k_\perp}}{16 \pi^3} \big[\Psi_{S_z=0}^\ast (x^{\prime\prime},\mathbf{k}^{\prime\prime}_\perp, \uparrow, \uparrow) \Psi_{S_z=0} (x^{\prime},\mathbf{k}^{\prime}_\perp, \uparrow, \uparrow) \nonumber \\ 
    &&+ \Psi_{S_z=0}^\ast (x^{\prime\prime},\mathbf{k}^{\prime\prime}_\perp, \uparrow, \downarrow) \Psi_{S_z=0} (x^{\prime},\mathbf{k}^{\prime}_\perp, \uparrow, \downarrow) + \Psi_{S_z=0}^\ast (x^{\prime\prime},\mathbf{k}^{\prime\prime}_\perp, \downarrow, \uparrow) \Psi_{S_z=0} (x^{\prime},\mathbf{k}^{\prime}_\perp, \downarrow, \uparrow) \nonumber \\
    &&+ \Psi_{S_z=0}^\ast (x^{\prime\prime},\mathbf{k}^{\prime\prime}_\perp, \downarrow, \downarrow) \Psi_{S_z=0} (x^{\prime},\mathbf{k}^{\prime}_\perp, \downarrow, \downarrow)\big] \, ,
\label{GPDeq}
\end{eqnarray}
where $\bfk^{\prime\prime}$ and $\bfk^{\prime}$ correspond to the final and initial state quark momentum respectively. In symmetric frame, they can be expressed as
\begin{eqnarray}
    \bfk^{\prime\prime}&=&\bfk-(1-x^{\prime\prime})~\frac{\Dp}{2} \, , \nonumber \\
    \bfk^{\prime}&=&\bfk+(1-x^{\prime})~\frac{\Dp}{2} \, .
\end{eqnarray}
Since we are dealing with zero skewness GPDs, the initial and final state longitudinal momentum fraction carried by an active quark of a meson remain the same. Hence, we can express initial and final state longitudinal momentum fraction by $x$ only. \par
%----------------------------
\begin{figure*}
\centering
\begin{minipage}[c]{0.98\textwidth}
(a)\includegraphics[width=7.5cm]{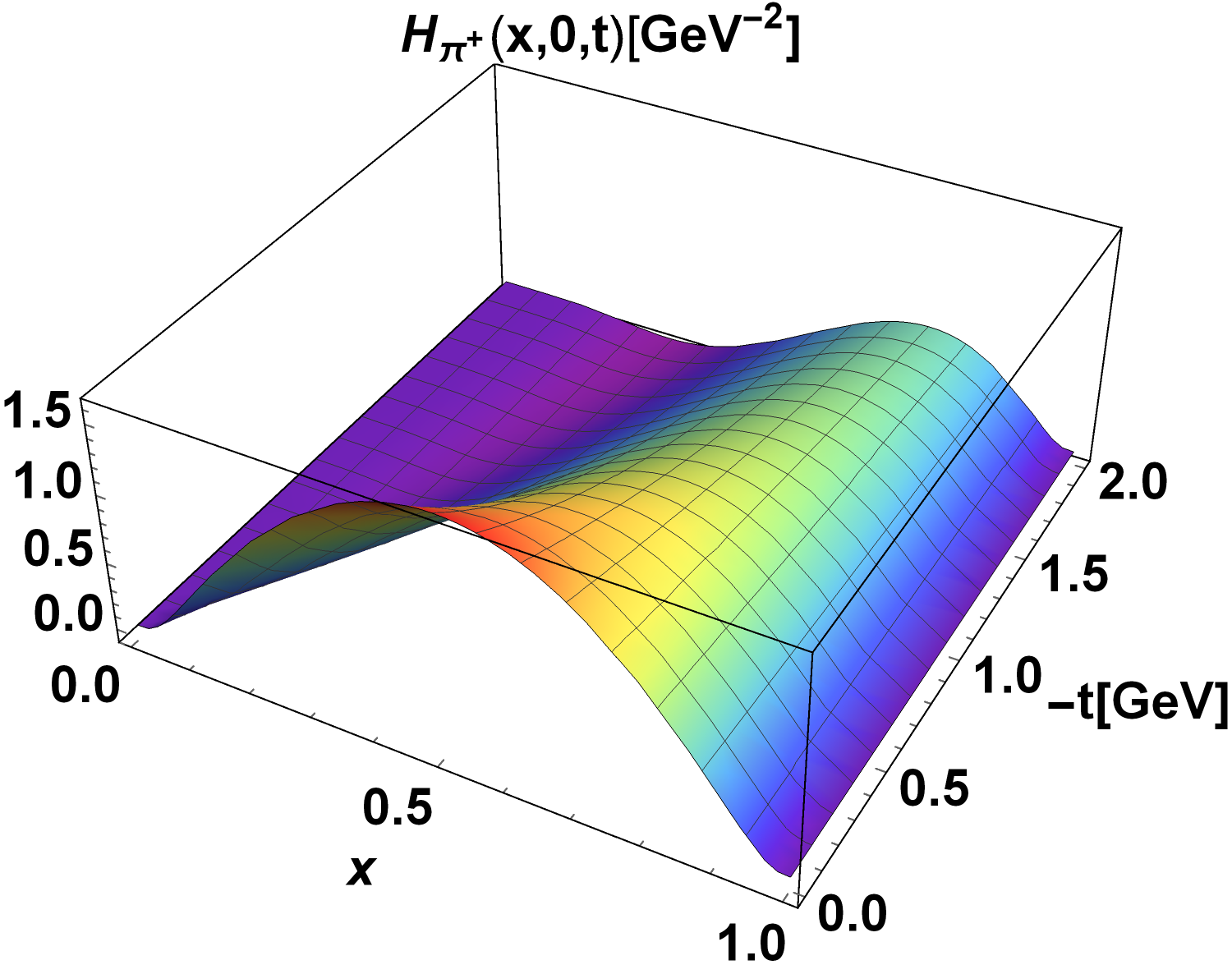}
\hspace{0.03cm}	
(b)\includegraphics[width=7.5cm]{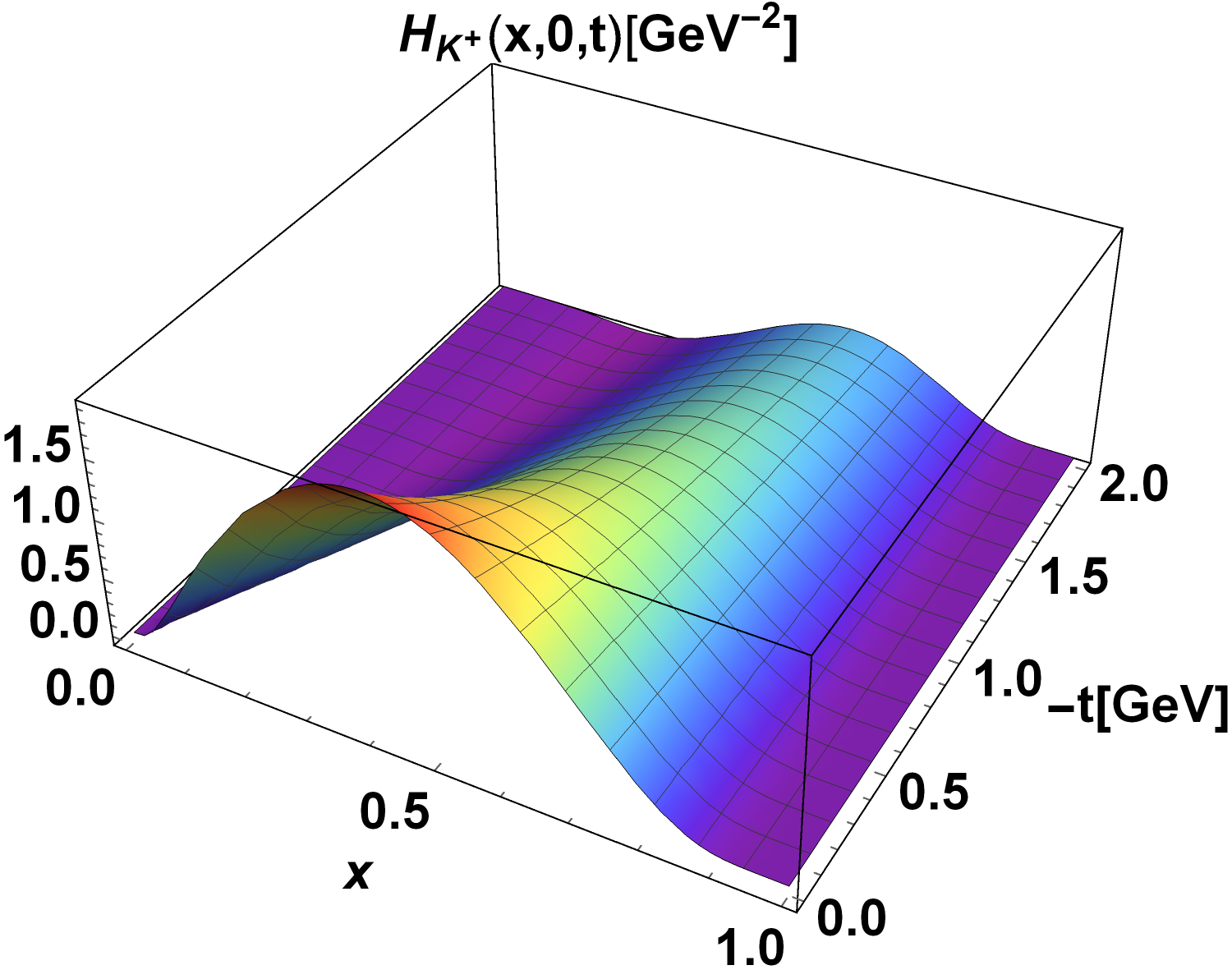} 
\hspace{0.03cm}
\end{minipage}
\caption{\label{GPDpk} (Color online) Unpolarized generalized parton distributions as a function of longitudinal momentum fraction $x$ and invariant momentum transfer $-t$ for an active $u$-quark of (a) $\pi^+$ and (b) $K^+$.}
\end{figure*}
%----------------------------
Fig. \ref{GPDpk} presents the unpolarized GPDs for an active $u$-quark of $\pi^+$ and $K^+$ with respect to longitudinal momentum fraction $x$ and invariant momentum transfer $-t$. It is observed that the distributions for both the mesons are intense at $-t=0$ and fall off smoothly with an increase in the value of $-t$. However, the difference in these mesons lies in the peak values of $x$. The peak of the distribution for the case of $K^+$ is found to be shifted towards smaller values of  $x$ with tapering down of its $x$-dependence as compared to that for the $\pi^+$ distribution. This difference is a consequence of their respective antiquark masses. The massive $\bar{s}$ antiquark of $K^+$ carries comparatively more longitudinal momentum fraction as compared to the $\bar{d}$ antiquark of $\pi^+$. \par 
%----------------------------
\begin{figure*}
\centering
\begin{minipage}[c]{0.98\textwidth}
(a)\includegraphics[width=7.5cm]{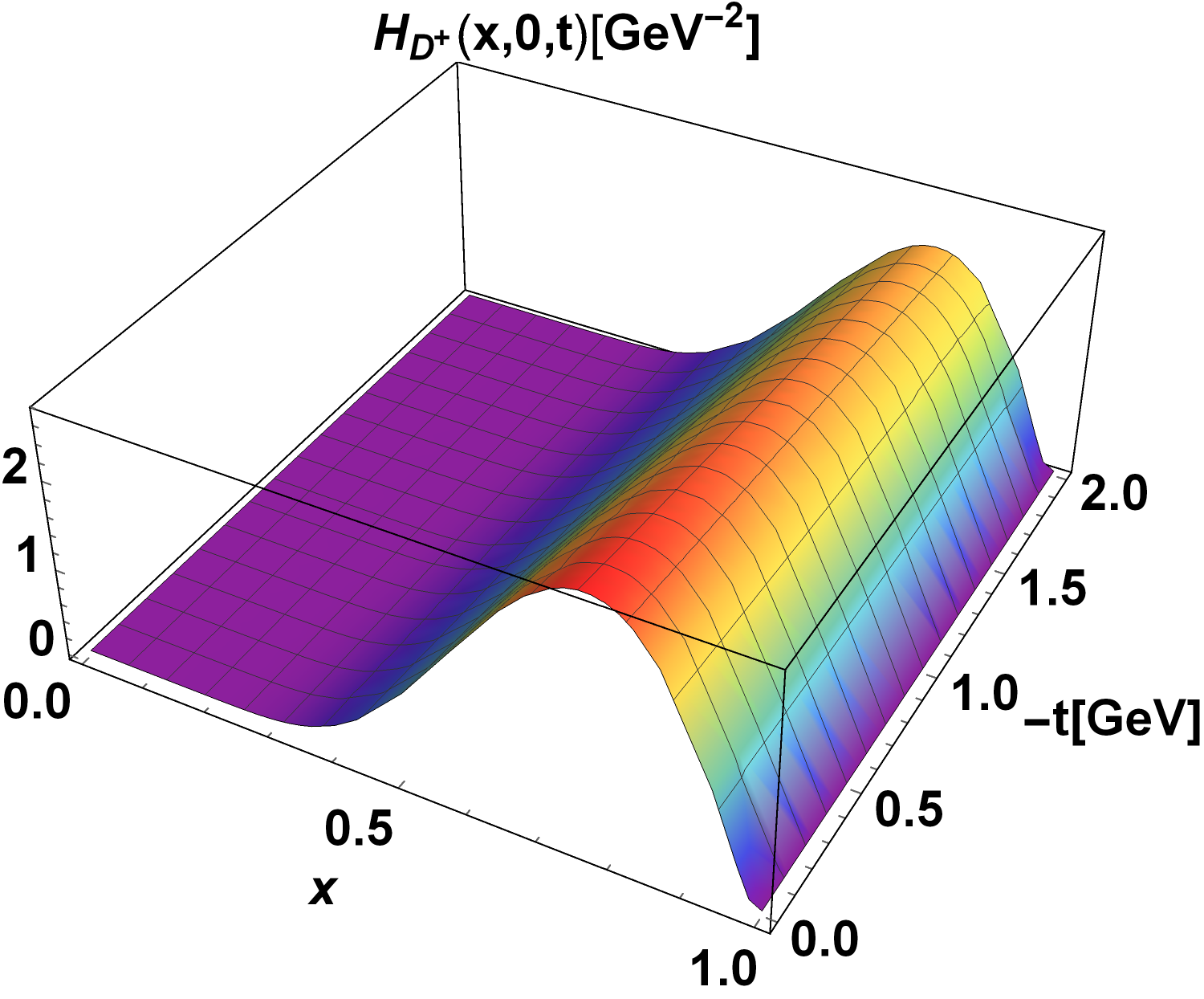}
\hspace{0.03cm}	
(b)\includegraphics[width=7.5cm]{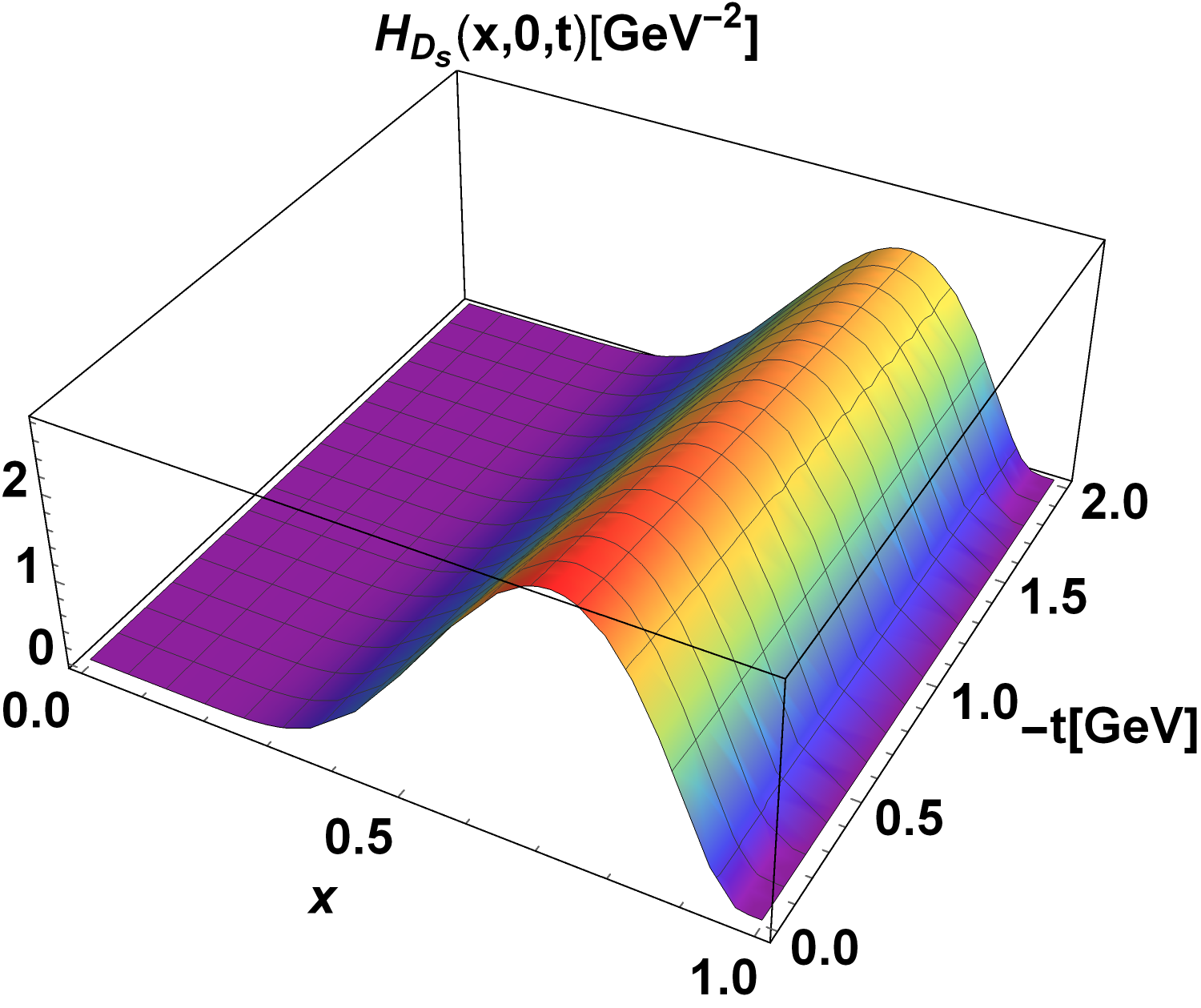} 
\hspace{0.03cm}
\end{minipage}
\caption{\label{GPDD} (Color online) Unpolarized generalized parton distributions as a function of longitudinal momentum fraction $x$ and invariant momentum transfer $-t$ for an active $c$-quark of (a) $D^+$ and (b) $D_s$ mesons.}
\end{figure*}
\begin{figure*}
\centering
\begin{minipage}[c]{0.98\textwidth}
(a)\includegraphics[width=6.5cm]{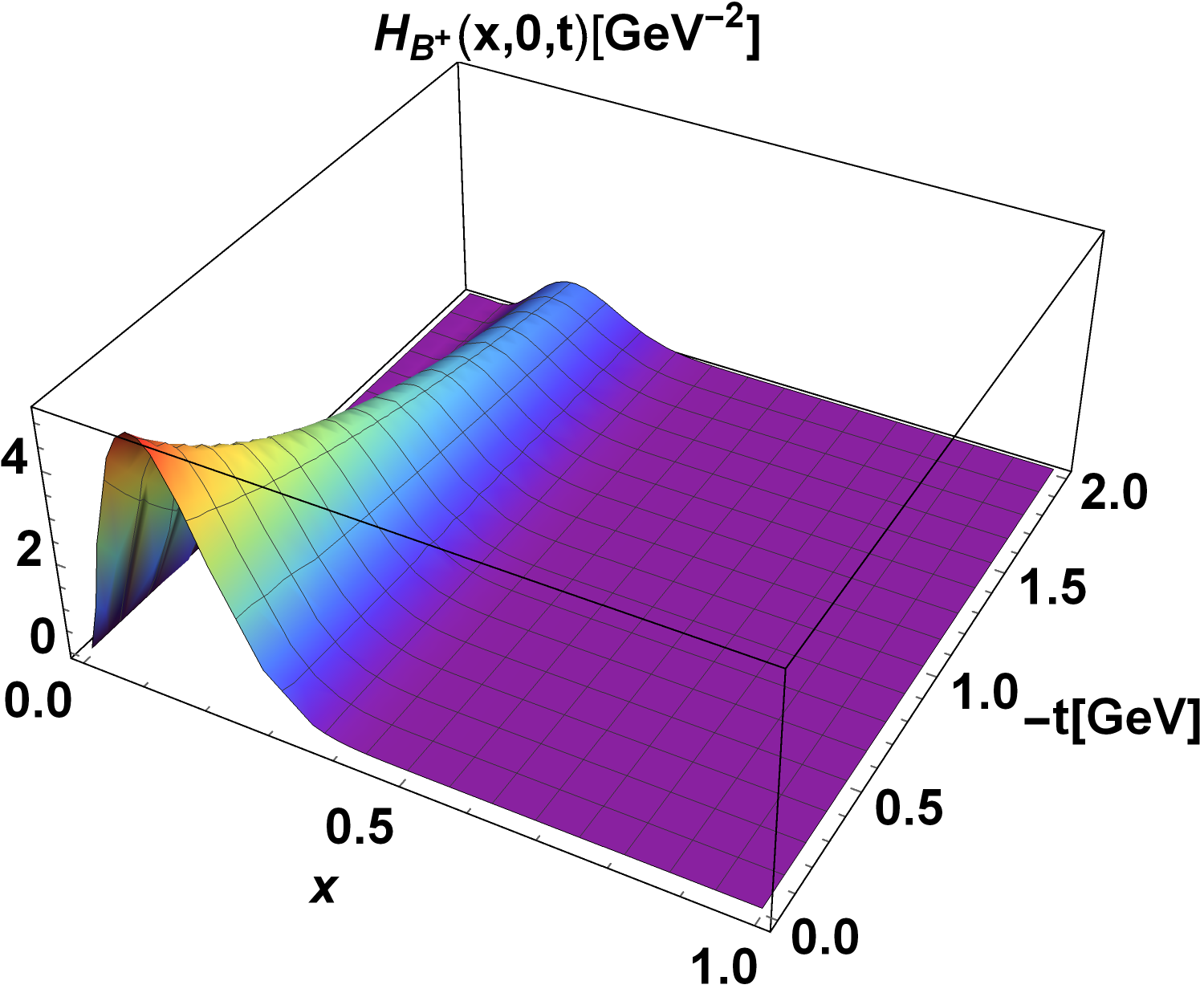}
\hspace{0.03cm}	
(b)\includegraphics[width=6.5cm]{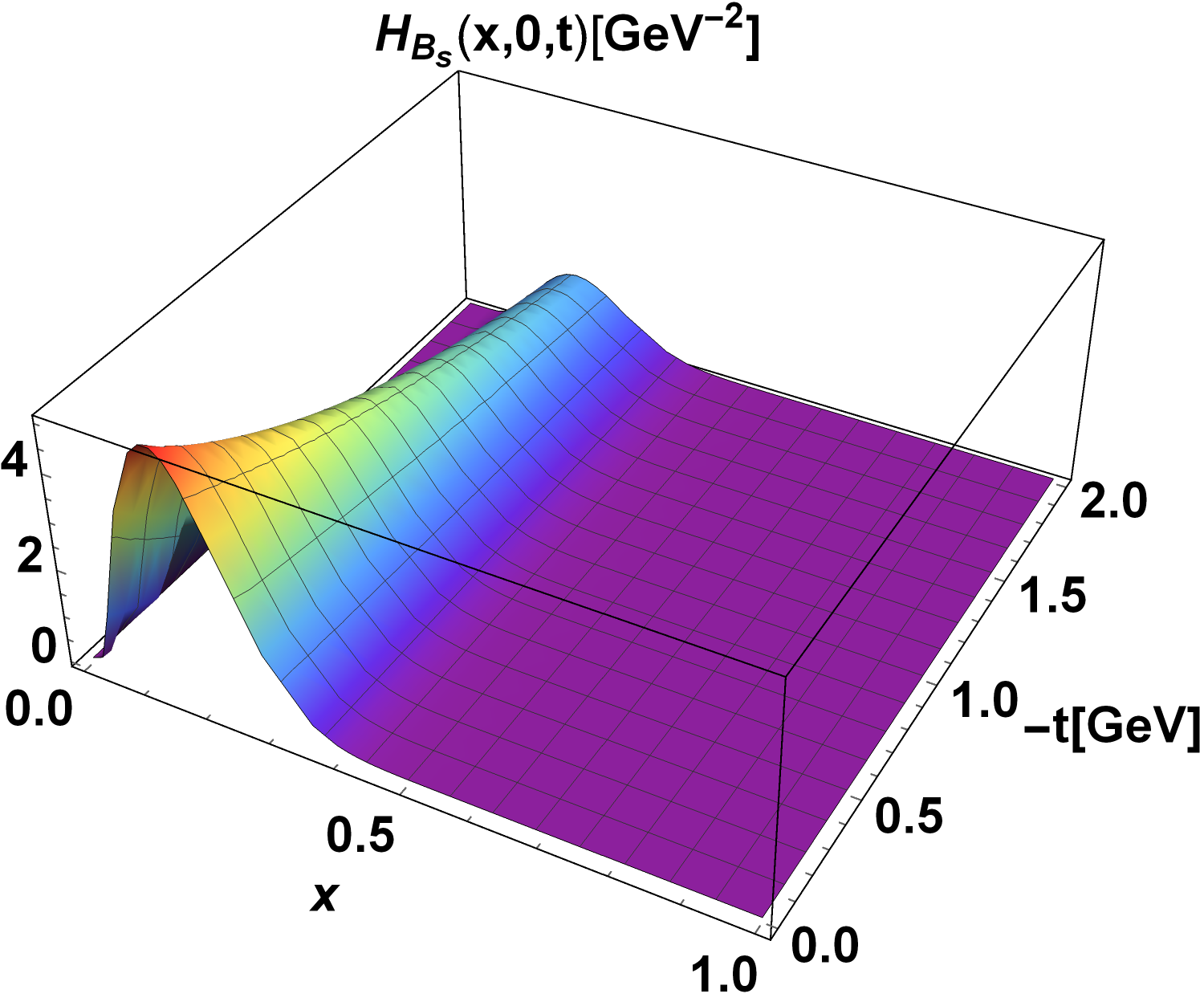} 
\hspace{0.03cm}
(c)\includegraphics[width=6.5cm]{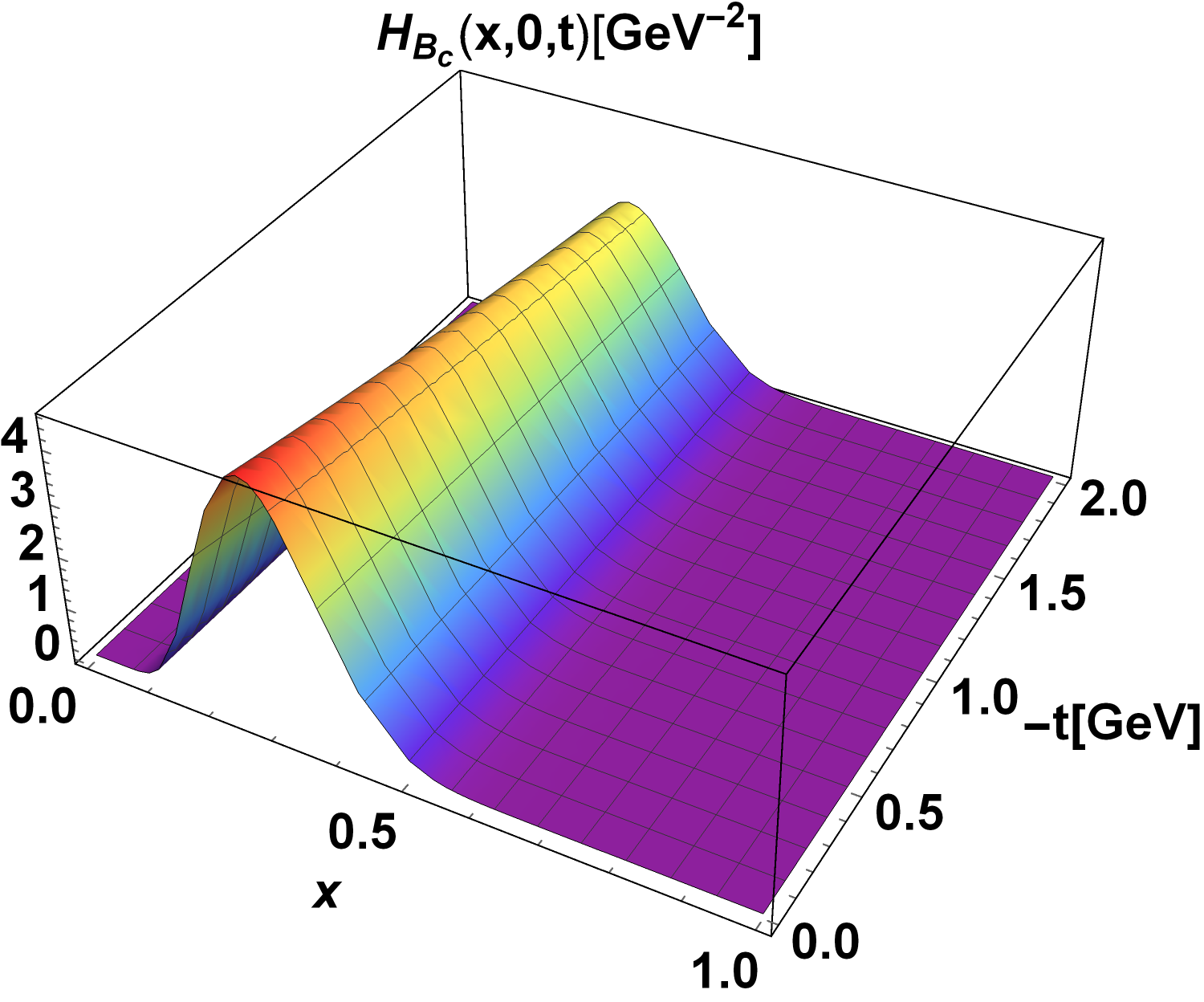} 
\hspace{0.03cm}
\end{minipage}
\caption{\label{GPDB} (Color online) Unpolarized generalized parton distributions as a function of longitudinal momentum fraction $x$ and invariant momentum transfer $-t$ for (a) an active $u$-quark of $B^+$ meson, (b) an active $s$-quark of $B_s$ meson and (c) an active $c$-quark of $B_c$ meson.}
\end{figure*}
\begin{figure*}
\centering
\begin{minipage}[c]{0.98\textwidth}
(a)\includegraphics[width=6.5cm]{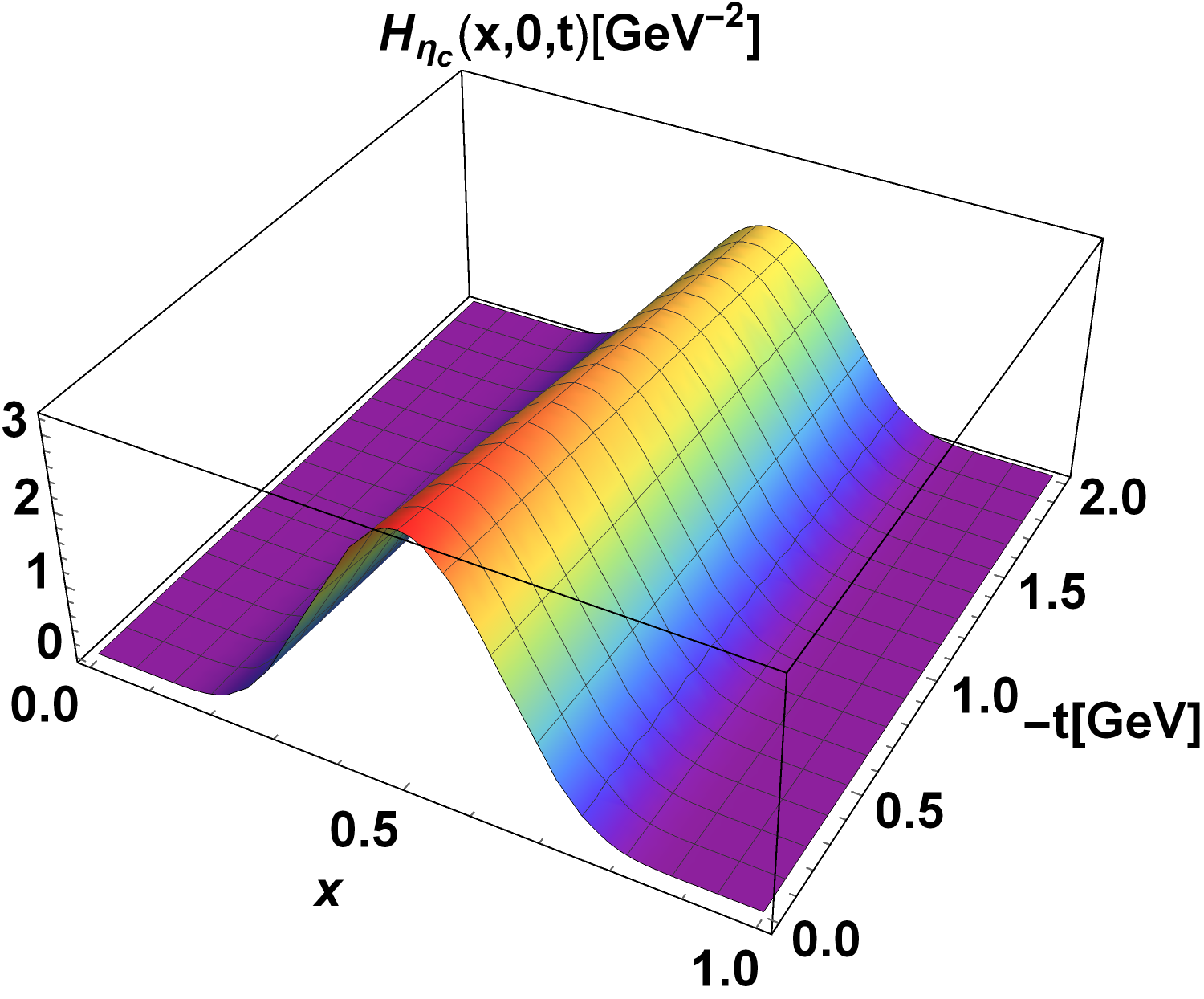}
\hspace{0.03cm}	
(b)\includegraphics[width=6.5cm]{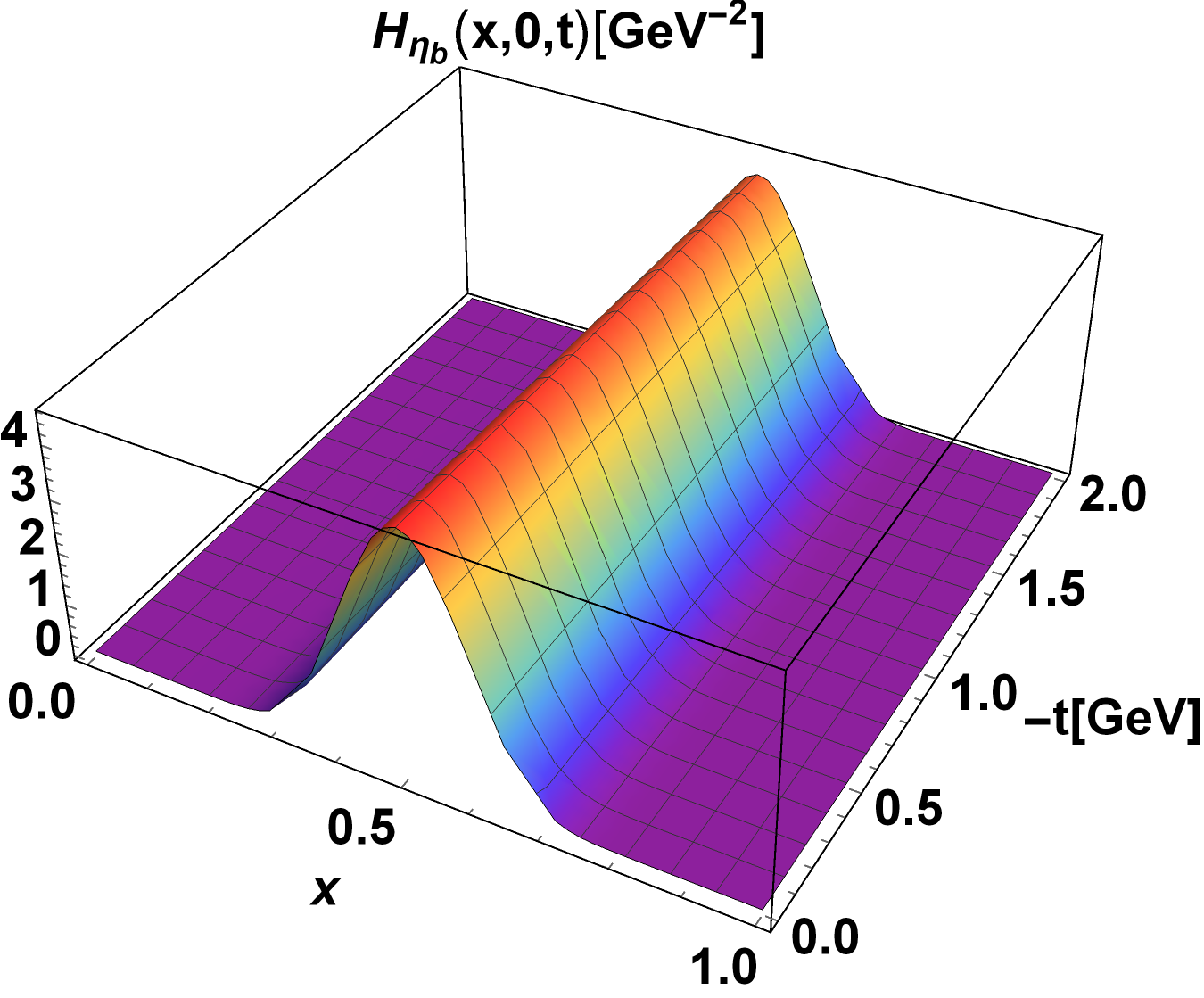} 
\hspace{0.03cm}
\end{minipage}
\caption{\label{GPDeta} (Color online) Unpolarized generalized parton distributions as a function of longitudinal momentum fraction $x$ and invariant momentum transfer $-t$ for an active $c$-quark of (a) $\eta_c$ and active $b$-quark of (b) $\eta_b$ mesons.}
\end{figure*}

%----------------------------
The unpolarized GPDs for an active $c$-quark of $D$ mesons ($D^+$ and $D_s$) with respect to longitudinal momentum fraction $x$ and invariant momentum transfer $-t$ are presented in Fig. \ref{GPDD}. As in the case of light mesons, the distributions are intense at $-t=0$. However, the distributions get tapered over a smaller region of $x$ with no significant fall off with respect to $-t$ for both the $D$ mesons. The presence of peak value in the higher region of $x$ as well as insignificant difference of peak values of $x$, for both $D^+$ and $D_s$ mesons, can be attributed to the heavy mass of an active $c$-quark. In the same context, the distributions of unpolarized GPDs for active quarks of $B$ mesons with respect to $x$ and $-t$ are demonstrated in Fig. \ref{GPDB}. Moving from a lighter $B^+$ meson to a comparatively heavier $B_s$ meson and the heaviest $B_c$ meson, the $\bar{b}$ antiquark remains the same. However, the quark content changes from $u$ to $s$ and then $c$, respectively. With the increase in the mass of an active quark, the distributions are shifted to a comparatively larger values of $x$ for zero momentum transfer with attenuation of fall off with respect to $-t$. Hence, the family of $B$ mesons also approves the dependency of these distributions on quark masses. Among $eta_c$ and $\eta_b$ mesons, the unpolarized GPDs for their active quarks are also investigated to analyze the impact of mass on the ability to carry longitudinal momentum fraction $x$ with and without momentum transfer $-t$. Their distributions show symmetric distribution around $x=0.5$ as both quark-antiquark flavors are identical, hence having an equal probability of carrying longitudinal momentum fraction. However, the tapering down of distribution over a smaller region of $x$ with almost negligible dependency on $-t$ for heavier $b$-quark flavor is also observed in Fig. \ref{GPDeta}. In short, the tapered $x$-dependence of distributions for heavy mesons implies less significance for dynamical chiral symmetry breaking for them. Distributions corresponding to unpolarized antiquark of a particular meson can be analyzed by replacing $x$ by ($1-x$) in Eq. (\ref{GPDeq}). \hspace{-2em}
%----------------------------
%----------------------------

%----------------------------
\section{Form factors}
\begin{figure*}
\centering
\begin{minipage}[c]{0.98\textwidth}
(a)\includegraphics[width=7.5cm]{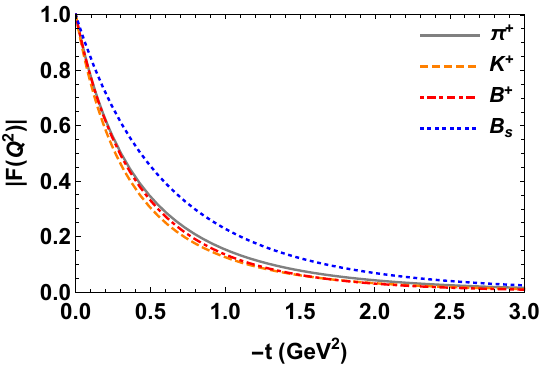}
\hspace{0.03cm}	
(b)\includegraphics[width=7.5cm]{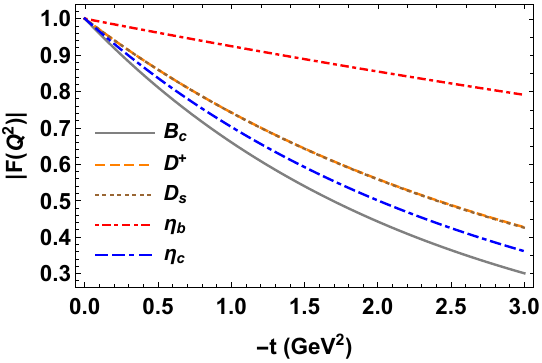} 
\hspace{0.03cm}
\end{minipage}
\caption{\label{EMFFs} (Color online) Electromagnetic form factors as a function of $-t$ for quarks of (a) light-light and light-heavy pair of mesons and (b) heavy-light and heavy-heavy pair of mesons.}
\end{figure*}

The zeroth moment of the unpolarized GPD $H_{M}(x,0,-t)$ provides an insight to the contribution of $q$-quark flavor to the total elastic EMFF of meson and can be described as
\be
F_{M_q}(-t)=\int dx ~ H_M(x,0,-t) \,, 
\ee
where $Q^2=-t$. An analogous expression can be written for the meson antiquark, whereas complete EMFF of a meson can be obtained by summing up the charge multiplied form factors of constituent quark and antiquark of a meson as 
\be
F_M(-t)=e_q F_{M_q}(-t)+e_{\bar{q}} F_{M_{\bar{q}}}(-t).
\label{FFs}
\ee
EMFFs for quark-antiquark pairs carrying the same or comparatively lighter quarks than their antiquark partners are presented in Fig. \ref{EMFFs} (a). A fast declination is found for EMFFs with an increase of $Q^2$ and a saturation near zero value of $|F_M(-t)|$. On the other hand, EMFFs for quark-antiquark pairs carrying the same or comparatively heavier quarks than their antiquark partners are presented in Fig. \ref{EMFFs} (b), representing a smooth fall off of EMFFs with $Q^2$. Between $\eta_c$ an $\eta_b$, the fall of EMFF for  $\eta_c$ comes out to be steeper than $\eta_b$, which is a result of the mass difference between them. Heavier the quark, more gradually it will fall. Among $D^+$, $D_s$ and $B_c$ mesons, $D^+=c\bar{d}$ and $D_s=c\bar{s}$ show a similar behavior whereas the fall in $B_c=c\bar{b}$ is steeper. This is a consequence of the mass difference between the quark-antiquark pair and the EMFFs fall off more smoothly with $Q^2$ when the mass difference of its quark-antiquark pair is smaller.  

\begin{figure*}
\centering
\begin{minipage}[c]{0.98\textwidth}
(a)\includegraphics[width=7.5cm]{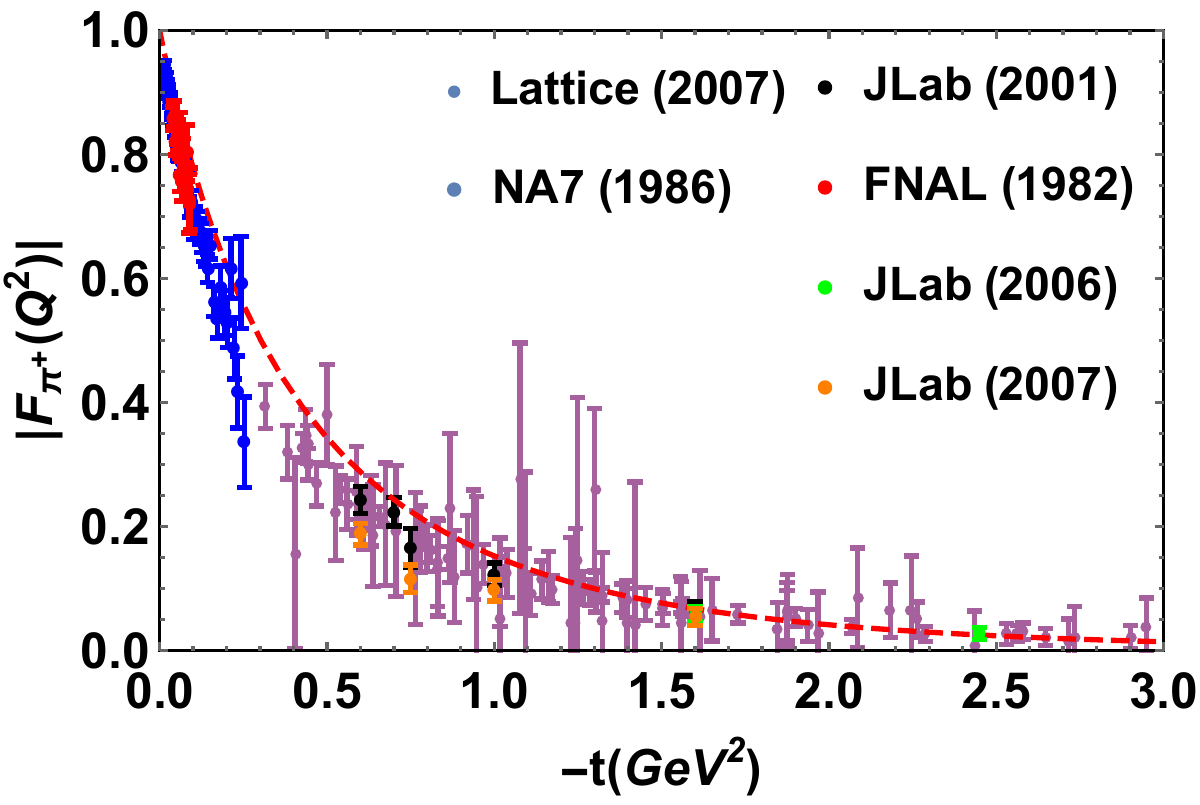}
\hspace{0.03cm}	
(b)\includegraphics[width=7.5cm]{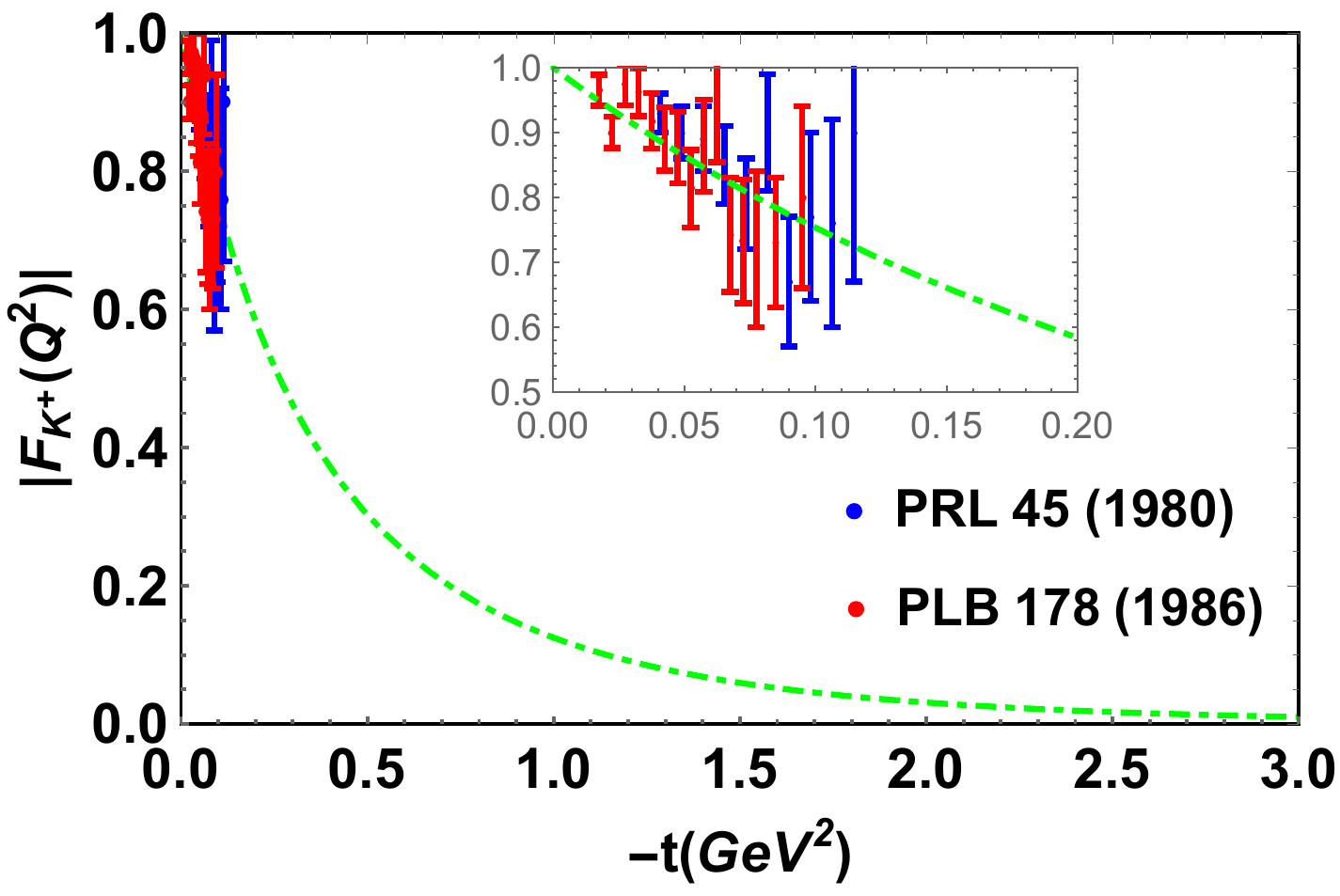} 
\hspace{0.03cm}
\end{minipage}
\caption{\label{EMFFpk} (Color online) Electromagnetic form factors of (a) pion compared with NA7 \cite{NA7:1986vav}, FNAL \cite{Dally:1982zk}, lattice simulation \cite{QCDSFUKQCD:2006gmg}, JLab (2001) \cite{JeffersonLabFpi:2000nlc}, JLab (2006) \cite{JeffersonLabFpi-2:2006ysh}, JLab (2007) \cite{JeffersonLabFpi:2007vir} and (b) kaon compared with the available experimental data \cite{Dally:1980dj,Amendolia:1986ui}.}
\end{figure*}

\begin{figure*}
\centering
(a)\includegraphics[width=7.5cm]{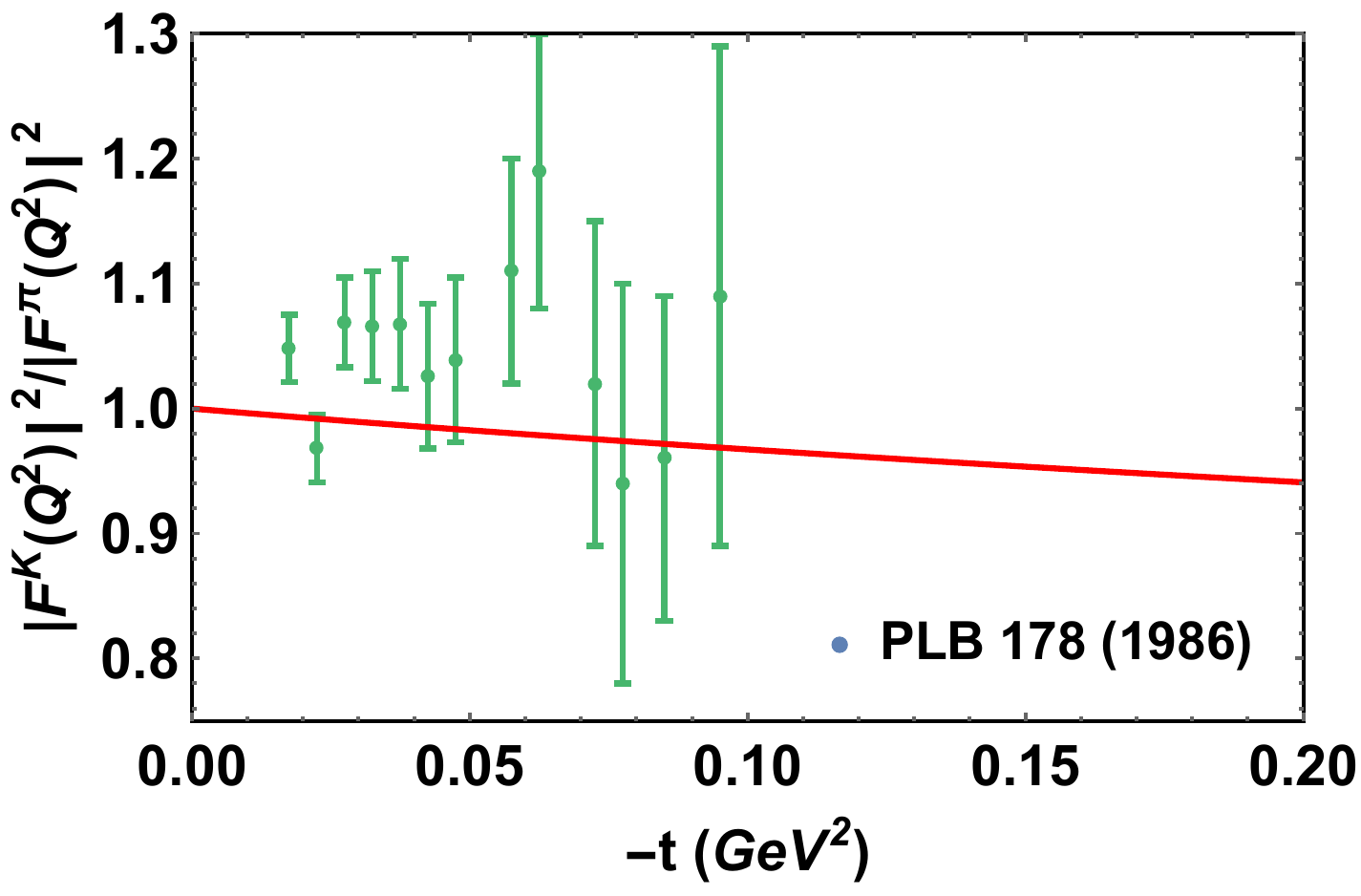}
\caption{\label{EMFF ratio} (Color online) Electromagnetic form factors ratio of kaon to pion compared with the available experimental data \cite{Amendolia:1986ui}.}
\end{figure*}

\begin{figure*}

\centering
\begin{minipage}[c]{0.98\textwidth}
(a)\includegraphics[width=7.5cm]{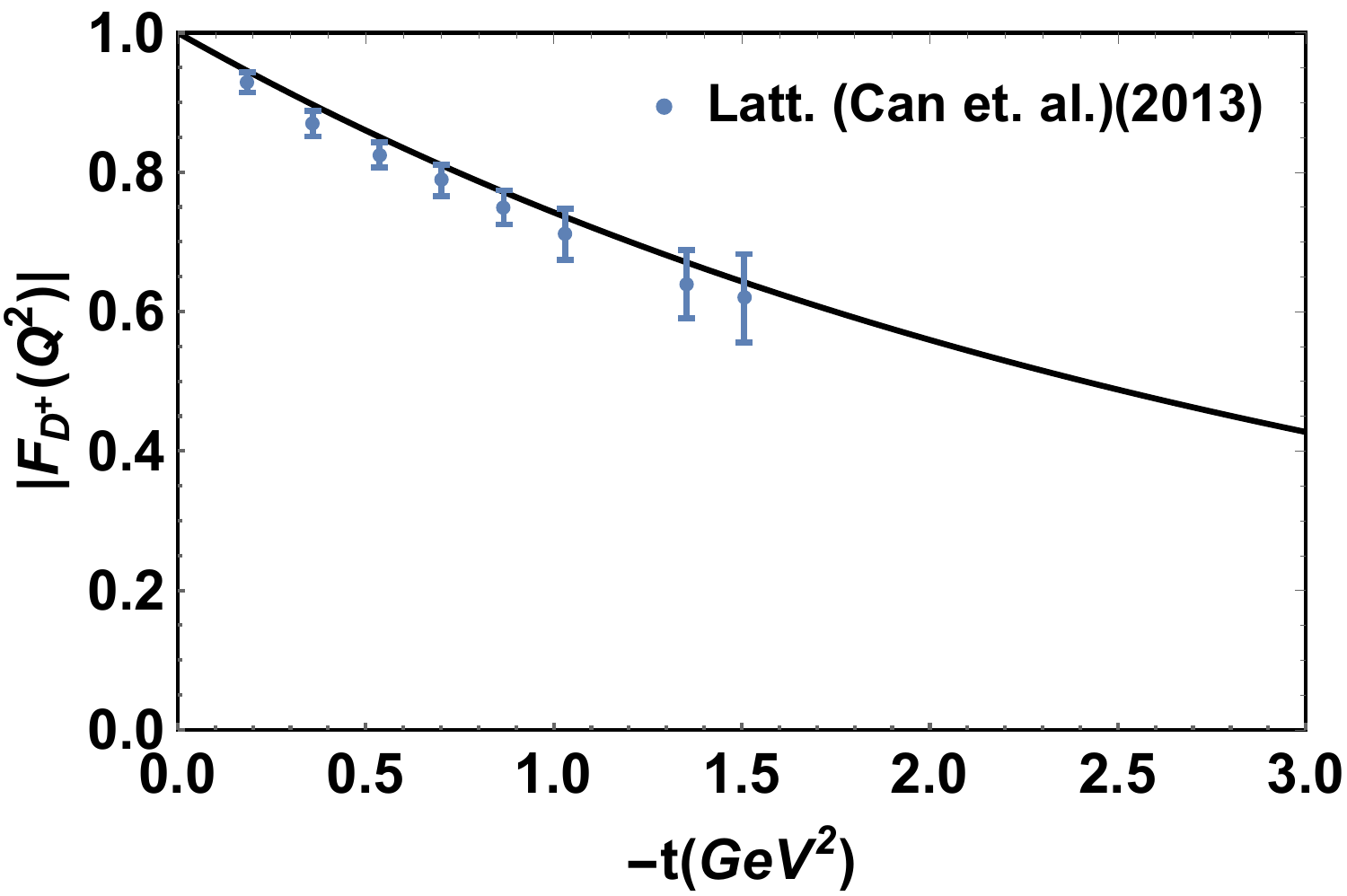}
\hspace{0.03cm}	
(b)\includegraphics[width=7.5cm]{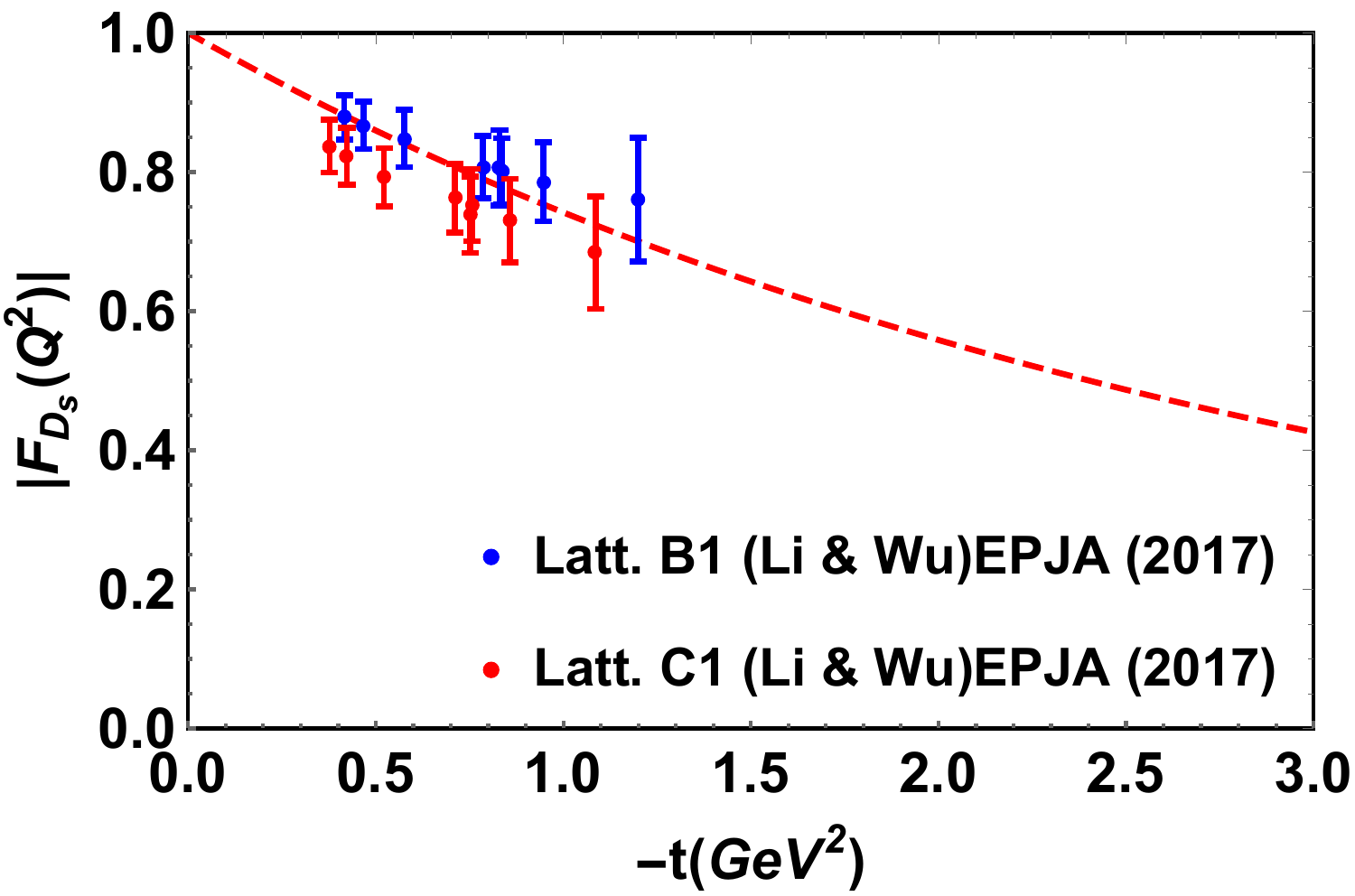} 
\hspace{0.03cm}
\end{minipage}
\caption{\label{EMFFd} (Color online) Electromagnetic form factors for an active $u$-quark of (a) $D^+$ compared with lattice simulation \cite{Can:2012tx} and (b) $D_s$ compared with available lattice data \cite{Li:2017eic}.}
\end{figure*}

\begin{figure*}
\centering
\begin{minipage}[c]{0.98\textwidth}
\includegraphics[width=7.5cm]{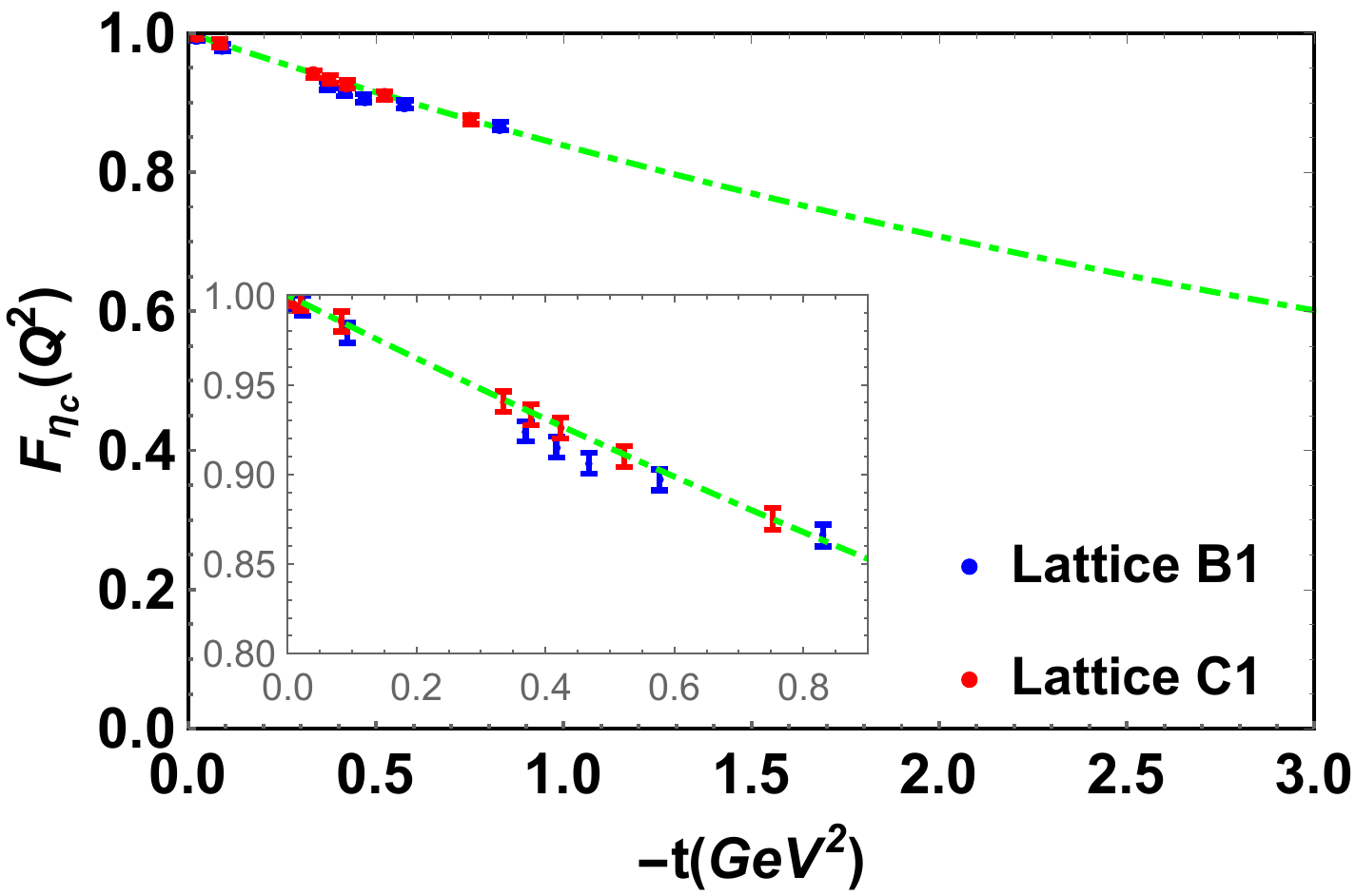}
\hspace{0.03cm}	
\end{minipage}

\caption{\label{EMFFc} (Color online) Electromagnetic form factors for an active $c$-quark of $\eta_c$ compared with available data \cite{Li:2020gau}.}
\end{figure*}

A comparison of EMFFs for $\pi^+$ and $K^+$ with available data is presented in Fig. \ref{EMFFpk}. Our model result for the case of pion overlaps with the available experimental data of JLab \cite{JeffersonLabFpi-2:2006ysh} for $1.5<Q^2<2.5$ region very well. In addition,  our results are also found to be within the error bars of other available data points of lattice and experimental data \cite{NA7:1986vav, Dally:1982zk, QCDSFUKQCD:2006gmg, JeffersonLabFpi:2000nlc, JeffersonLabFpi:2007vir}. For $K^+$ meson, data is available only for smaller region of $Q^2$ are our results are in good agreement with the available data \cite{Dally:1980dj, Amendolia:1986ui}. The ratio between the kaon and pion form factor has been portrayed in Fig. \ref{EMFF ratio}, which shows compatible results to the experimental data \cite{Amendolia:1986ui}. However, in Ref. \cite{Dias:2010sg}, the ratio between the kaon and pion form factor has a positive slope, which is in  contradiction with our result with a negative slope of this ratio. For few heavier mesons, lattice data \cite{Can:2012tx, Li:2017eic, Li:2020gau} is available for invariant momentum transfer $Q^2<2$ and the compatible comparison of our results with them are presented with them in Fig. \ref{EMFFd} and \Ref {EMFFc} for $D$ and $\eta_c$ mesons. 

%----------------------------
%----------------------------
%----------------------------
%------------------------------------
%--------------------------------
%-------------------------------

%----------------------------
\begin{figure*}
\centering
\begin{minipage}[c]{0.98\textwidth}
(a)\includegraphics[width=7.5cm]{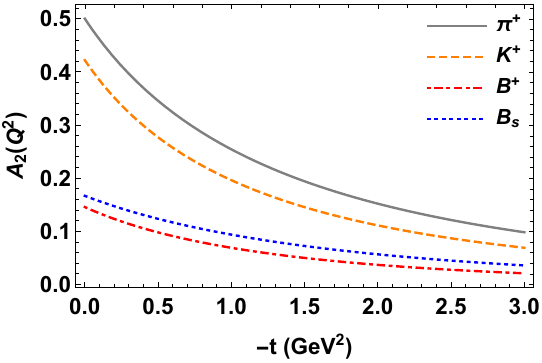}
\hspace{0.03cm}	
(b)\includegraphics[width=7.5cm]{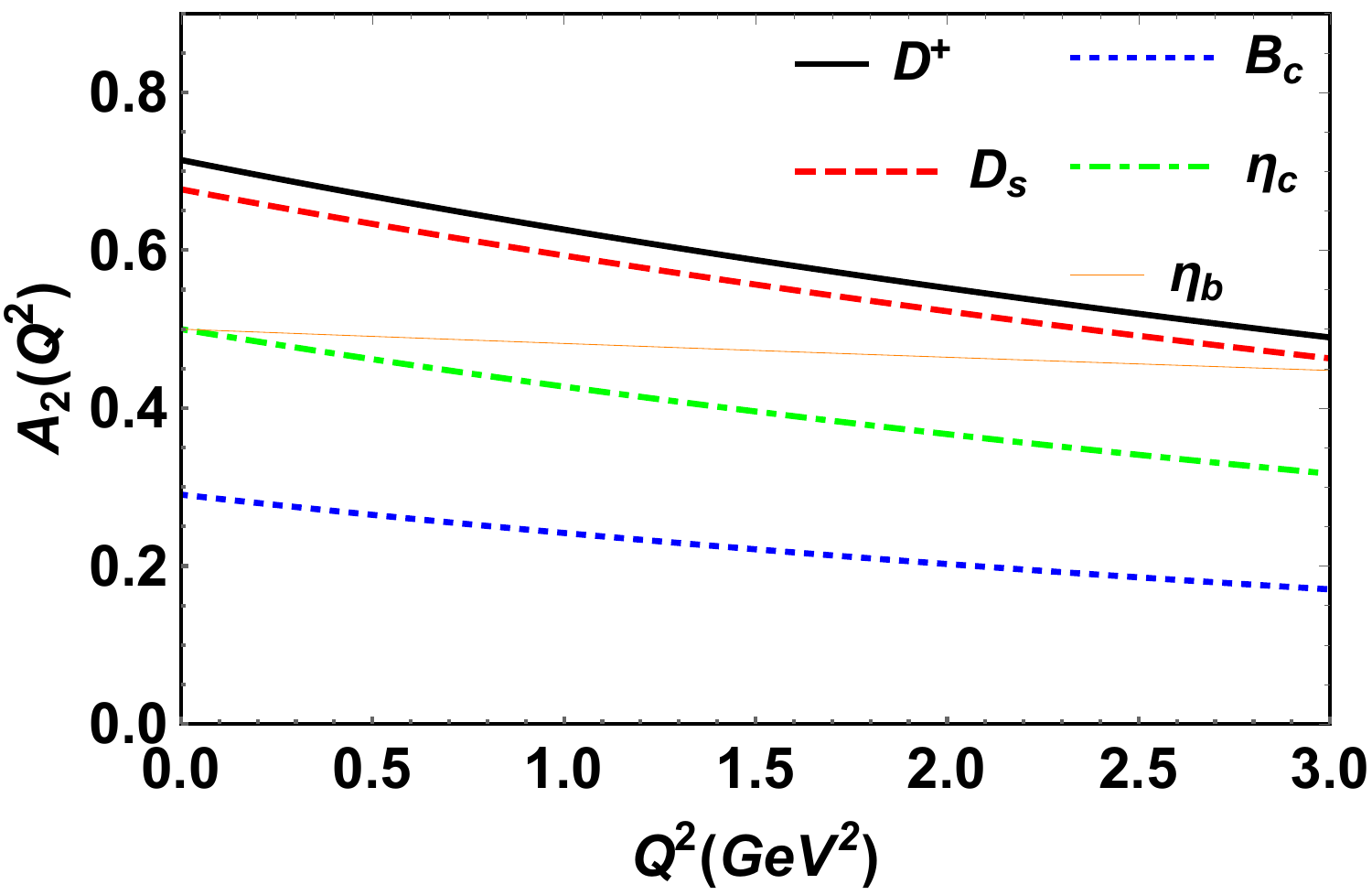}
\hspace{0.03cm}	
\end{minipage}
\caption{\label{GFFs} (Color online) Gravitational form factors as a function of $-t$ for quarks of (a) light-light and light-heavy pair of mesons and (b) heavy-light and heavy-heavy pair of mesons.}
\end{figure*}
%-------------------------------

The first moment of the unpolarized GPD $H_{M}(x,0,-t)$ provides an insight to the contribution of the $q$-quark flavor to the total GFF of meson and can be described as
\be
A_{2(M_q)}(-t)=\int dx ~x~ H_M(x,0,-t).
\ee
Analogous to the EMFF of mesons, one can also obtain the GFF of mesons as expressed in Eq. (\ref{FFs}).
Both EMFFs and GFF obey the FF sum rules
\be
F_{M}(Q^2=0)= 1,\\
A_{2(M_q)}(Q^2=0)+A_{2(M_{\bar q})}(Q^2=0)=1.
\ee
GFFs of light-light and light-heavy pairs of mesons have been presented in Fig. \ref{GFFs} (a). They reflect that both light-light pseudoscalars follow same pattern of fall off with different amplitudes. This is a consequence of massive mass of $\bar{s}$ in $K^+$. Hence, more massive the quark, more is its contribution towards meson's total EMFF. Similar kind of trend is only seen between contribution of valence quarks of $B^+$ and $B_s$ mesons.  Fig. \ref{GFFs} (b) represents the GFFs of heavy-light and heavy-heavy pairs of mesons. $\eta_b$ and $\eta_c$ mesons have same $A_Q(0)$ values  but $\eta_c$ shows a significant fall off with $-t$ as compared to that in than $\eta_b$. $D^+$ and $D_s$ represent a steeper trend with $-t$ when compared to $\eta_c$ mesons but with higher amplitudes. As we move from $D^+$ to $D_s$ meson, the mass difference between the quark-antiquark of a meson decreases, thus the amplitude of $A_c(-t)$. Further,  $B_c$ meson has the least amplitude of $A_c(-t)$ as it contains smallest mass difference between the quark-antiquark pair than $D^+$ and $D_s$. In short, the amplitude of the GFFs of mesons decreases as the mass of the quark of a meson carrying same quark-antiquark flavor decreases or the mass difference between quark-antiquark pairs decreases. 
%----------------------------
\section{Summary}\label{con}
We have considered the light-cone quark model to study the transverse and spatial structure of spin-$0$ light as well as heavy pseudoscalar mesons. Quark quark correlator has been solved in the light-cone framework to study the transverse momentum dependent distributions (TMDs) and generalized parton distributions (GPDs). Three-dimensional valence quark distributions of unpolarized TMD $f_1$($x,\bfk$) as a function of longitudinal momentum fraction $x$ and transverse momentum $\bfk$ implies a dependence of light pseudoscalar mesons on entire region of $x$ with abrupt fall off with increase in $\bfk$. However, heavy pseudoscalar mesons are found to show $x$ dependence on comparatively smaller region with a very smooth fall off with $\bfk$. From unpolarized TMD, an average momenta $\langle \bfk \rangle$ and $\langle \bfk^2 \rangle$ carried by the active quark inside mesons are also computed. These values are found to increase gradually as mesons becomes heavy. Except $\eta_b$ and $B_c$ mesons, the value of $\langle \bfk^2 \rangle$ for all other considered light and heavy mesons comes out to be less than $\langle \bfk \rangle$. Parton distribution functions are also evaluated from their unpolarized TMDs for all the pseudoscalar mesons. These distributions reveal that as the quark becomes heavy, the probability of finding the constituent quark with comparatively higher $x$ is more. Evolved distributions are found to be compatible with available experimental data for pion and BLFQ predictions for kaon valence partons. Average longitudinal momentum $\langle x \rangle$ carried by the active quark from its parent meson is inversely proportional to the mass difference of quark-antiquark pair.

Three-dimensional valence quark distributions of chiral-even unpolarized generalized parton distribution as a function of longitudinal momentum fraction $x$ and invariant momentum transfer $-t$ are also studied. The general trend of intense distributions at zero transverse momentum transfer is same for all the light as well as heavy pseudoscalar mesons. However, the peak value of distributions for heavy mesons is shifted to comparatively larger values of $x$ with tapering down of distributions over $x$-dependence. An attenuated fall off for the distributions with $-t$ is also observed. Narrowed distributions over $x$ reflect minimal impact of dynamical chiral symmetry breaking in heavy mesons. Furthermore, electromagnetic form factors (EMFFs) and gravitational form factors (GFFs) are also investigated. For pions, sufficient experimental and lattice simulated data is available and our results come out to be compatible with them. Kaon and ratio of kaon to pion EMFFs are also compared with  the available data for small values of $-t$ which are also in good agreement. For the case of $D^+$, $D_s$ and $\eta_c$ mesons, only lattice simulated data is available and we found compatible results of EMFFs for them. Comparison of EMFFs among all light and heavy pseudoscalar follows: lighter the active quark, more steeper is the EMFF distribution for it. Similar behavior is also followed by GFFs of active quarks of mesons. However, amplitude of GFFs depends merely on the mass of the constituent quark of a meson.

\section{Acknowledgement}
H.D. would like to thank  the Science and Engineering Research Board, Anusandhan-National Research Foundation, Government of India under the scheme SERB-POWER Fellowship (Ref No. SPF/2023/000116) for financial support.

\section{Reference}

\bibliography{ref}

\begin{thebibliography}{92}%
\makeatletter
\providecommand \@ifxundefined [1]{%
 \@ifx{#1\undefined}
}%
\providecommand \@ifnum [1]{%
 \ifnum #1\expandafter \@firstoftwo
 \else \expandafter \@secondoftwo
 \fi
}%
\providecommand \@ifx [1]{%
 \ifx #1\expandafter \@firstoftwo
 \else \expandafter \@secondoftwo
 \fi
}%
\providecommand \natexlab [1]{#1}%
\providecommand \enquote  [1]{``#1''}%
\providecommand \bibnamefont  [1]{#1}%
\providecommand \bibfnamefont [1]{#1}%
\providecommand \citenamefont [1]{#1}%
\providecommand \href@noop [0]{\@secondoftwo}%
\providecommand \href [0]{\begingroup \@sanitize@url \@href}%
\providecommand \@href[1]{\@@startlink{#1}\@@href}%
\providecommand \@@href[1]{\endgroup#1\@@endlink}%
\providecommand \@sanitize@url [0]{\catcode `\\12\catcode `\$12\catcode
  `\&12\catcode `\#12\catcode `\^12\catcode `\_12\catcode `\%12\relax}%
\providecommand \@@startlink[1]{}%
\providecommand \@@endlink[0]{}%
\providecommand \url  [0]{\begingroup\@sanitize@url \@url }%
\providecommand \@url [1]{\endgroup\@href {#1}{\urlprefix }}%
\providecommand \urlprefix  [0]{URL }%
\providecommand \Eprint [0]{\href }%
\providecommand \doibase [0]{https://doi.org/}%
\providecommand \selectlanguage [0]{\@gobble}%
\providecommand \bibinfo  [0]{\@secondoftwo}%
\providecommand \bibfield  [0]{\@secondoftwo}%
\providecommand \translation [1]{[#1]}%
\providecommand \BibitemOpen [0]{}%
\providecommand \bibitemStop [0]{}%
\providecommand \bibitemNoStop [0]{.\EOS\space}%
\providecommand \EOS [0]{\spacefactor3000\relax}%
\providecommand \BibitemShut  [1]{\csname bibitem#1\endcsname}%
\let\auto@bib@innerbib\@empty
%</preamble>
\bibitem [{\citenamefont {Brodsky}\ \emph {et~al.}(1998)\citenamefont
  {Brodsky}, \citenamefont {Pauli},\ and\ \citenamefont
  {Pinsky}}]{Brodsky:1997de}%
  \BibitemOpen
  \bibfield  {author} {\bibinfo {author} {\bibfnamefont {S.~J.}\ \bibnamefont
  {Brodsky}}, \bibinfo {author} {\bibfnamefont {H.-C.}\ \bibnamefont {Pauli}},\
  and\ \bibinfo {author} {\bibfnamefont {S.~S.}\ \bibnamefont {Pinsky}},\
  }\bibfield  {title} {\bibinfo {title} {{Quantum chromodynamics and other
  field theories on the light cone}},\ }\href
  {https://doi.org/10.1016/S0370-1573(97)00089-6} {\bibfield  {journal}
  {\bibinfo  {journal} {Phys. Rept.}\ }\textbf {\bibinfo {volume} {301}},\
  \bibinfo {pages} {299} (\bibinfo {year} {1998})},\ \Eprint
  {https://arxiv.org/abs/hep-ph/9705477} {arXiv:hep-ph/9705477} \BibitemShut
  {NoStop}%
\bibitem [{\citenamefont {Zhang}(1997)}]{Zhang:1997dd}%
  \BibitemOpen
  \bibfield  {author} {\bibinfo {author} {\bibfnamefont {W.-M.}\ \bibnamefont
  {Zhang}},\ }\bibfield  {title} {\bibinfo {title} {{A Weak coupling treatment
  of nonperturbative QCD dynamics to heavy hadrons}},\ }\href
  {https://doi.org/10.1103/PhysRevD.56.1528} {\bibfield  {journal} {\bibinfo
  {journal} {Phys. Rev. D}\ }\textbf {\bibinfo {volume} {56}},\ \bibinfo
  {pages} {1528} (\bibinfo {year} {1997})},\ \Eprint
  {https://arxiv.org/abs/hep-ph/9705226} {arXiv:hep-ph/9705226} \BibitemShut
  {NoStop}%
\bibitem [{\citenamefont {Collins}\ and\ \citenamefont
  {Soper}(1982)}]{Collins:1981uw}%
  \BibitemOpen
  \bibfield  {author} {\bibinfo {author} {\bibfnamefont {J.~C.}\ \bibnamefont
  {Collins}}\ and\ \bibinfo {author} {\bibfnamefont {D.~E.}\ \bibnamefont
  {Soper}},\ }\bibfield  {title} {\bibinfo {title} {{Parton Distribution and
  Decay Functions}},\ }\href {https://doi.org/10.1016/0550-3213(82)90021-9}
  {\bibfield  {journal} {\bibinfo  {journal} {Nucl. Phys. B}\ }\textbf
  {\bibinfo {volume} {194}},\ \bibinfo {pages} {445} (\bibinfo {year}
  {1982})}\BibitemShut {NoStop}%
\bibitem [{\citenamefont {Martin}\ \emph {et~al.}(1998)\citenamefont {Martin},
  \citenamefont {Roberts}, \citenamefont {Stirling},\ and\ \citenamefont
  {Thorne}}]{Martin:1998sq}%
  \BibitemOpen
  \bibfield  {author} {\bibinfo {author} {\bibfnamefont {A.~D.}\ \bibnamefont
  {Martin}}, \bibinfo {author} {\bibfnamefont {R.~G.}\ \bibnamefont {Roberts}},
  \bibinfo {author} {\bibfnamefont {W.~J.}\ \bibnamefont {Stirling}},\ and\
  \bibinfo {author} {\bibfnamefont {R.~S.}\ \bibnamefont {Thorne}},\ }\bibfield
   {title} {\bibinfo {title} {{Parton distributions: A New global analysis}},\
  }\href {https://doi.org/10.1007/s100520050220} {\bibfield  {journal}
  {\bibinfo  {journal} {Eur. Phys. J. C}\ }\textbf {\bibinfo {volume} {4}},\
  \bibinfo {pages} {463} (\bibinfo {year} {1998})},\ \Eprint
  {https://arxiv.org/abs/hep-ph/9803445} {arXiv:hep-ph/9803445} \BibitemShut
  {NoStop}%
\bibitem [{\citenamefont {Gluck}\ \emph {et~al.}(1995)\citenamefont {Gluck},
  \citenamefont {Reya},\ and\ \citenamefont {Vogt}}]{Gluck:1994uf}%
  \BibitemOpen
  \bibfield  {author} {\bibinfo {author} {\bibfnamefont {M.}~\bibnamefont
  {Gluck}}, \bibinfo {author} {\bibfnamefont {E.}~\bibnamefont {Reya}},\ and\
  \bibinfo {author} {\bibfnamefont {A.}~\bibnamefont {Vogt}},\ }\bibfield
  {title} {\bibinfo {title} {{Dynamical parton distributions of the proton and
  small x physics}},\ }\href {https://doi.org/10.1007/BF01624586} {\bibfield
  {journal} {\bibinfo  {journal} {Z. Phys. C}\ }\textbf {\bibinfo {volume}
  {67}},\ \bibinfo {pages} {433} (\bibinfo {year} {1995})}\BibitemShut
  {NoStop}%
\bibitem [{\citenamefont {Alekhin}(2003)}]{Alekhin:2002fv}%
  \BibitemOpen
  \bibfield  {author} {\bibinfo {author} {\bibfnamefont {S.}~\bibnamefont
  {Alekhin}},\ }\bibfield  {title} {\bibinfo {title} {{Parton distributions
  from deep inelastic scattering data}},\ }\href
  {https://doi.org/10.1103/PhysRevD.68.014002} {\bibfield  {journal} {\bibinfo
  {journal} {Phys. Rev. D}\ }\textbf {\bibinfo {volume} {68}},\ \bibinfo
  {pages} {014002} (\bibinfo {year} {2003})},\ \Eprint
  {https://arxiv.org/abs/hep-ph/0211096} {arXiv:hep-ph/0211096} \BibitemShut
  {NoStop}%
\bibitem [{\citenamefont {Polchinski}\ and\ \citenamefont
  {Strassler}(2003)}]{Polchinski:2002jw}%
  \BibitemOpen
  \bibfield  {author} {\bibinfo {author} {\bibfnamefont {J.}~\bibnamefont
  {Polchinski}}\ and\ \bibinfo {author} {\bibfnamefont {M.~J.}\ \bibnamefont
  {Strassler}},\ }\bibfield  {title} {\bibinfo {title} {{Deep inelastic
  scattering and gauge / string duality}},\ }\href
  {https://doi.org/10.1088/1126-6708/2003/05/012} {\bibfield  {journal}
  {\bibinfo  {journal} {JHEP}\ }\textbf {\bibinfo {volume} {05}},\ \bibinfo
  {pages} {012}},\ \Eprint {https://arxiv.org/abs/hep-th/0209211}
  {arXiv:hep-th/0209211} \BibitemShut {NoStop}%
\bibitem [{\citenamefont {Diehl}(2016)}]{Diehl:2015uka}%
  \BibitemOpen
  \bibfield  {author} {\bibinfo {author} {\bibfnamefont {M.}~\bibnamefont
  {Diehl}},\ }\bibfield  {title} {\bibinfo {title} {{Introduction to GPDs and
  TMDs}},\ }\href {https://doi.org/10.1140/epja/i2016-16149-3} {\bibfield
  {journal} {\bibinfo  {journal} {Eur. Phys. J. A}\ }\textbf {\bibinfo {volume}
  {52}},\ \bibinfo {pages} {149} (\bibinfo {year} {2016})},\ \Eprint
  {https://arxiv.org/abs/1512.01328} {arXiv:1512.01328 [hep-ph]} \BibitemShut
  {NoStop}%
\bibitem [{\citenamefont {Angeles-Martinez}\ \emph {et~al.}(2015)\citenamefont
  {Angeles-Martinez} \emph {et~al.}}]{Angeles-Martinez:2015sea}%
  \BibitemOpen
  \bibfield  {author} {\bibinfo {author} {\bibfnamefont {R.}~\bibnamefont
  {Angeles-Martinez}} \emph {et~al.},\ }\bibfield  {title} {\bibinfo {title}
  {{Transverse Momentum Dependent (TMD) parton distribution functions: status
  and prospects}},\ }\href {https://doi.org/10.5506/APhysPolB.46.2501}
  {\bibfield  {journal} {\bibinfo  {journal} {Acta Phys. Polon. B}\ }\textbf
  {\bibinfo {volume} {46}},\ \bibinfo {pages} {2501} (\bibinfo {year}
  {2015})},\ \Eprint {https://arxiv.org/abs/1507.05267} {arXiv:1507.05267
  [hep-ph]} \BibitemShut {NoStop}%
\bibitem [{\citenamefont {Pasquini}\ \emph {et~al.}(2008)\citenamefont
  {Pasquini}, \citenamefont {Cazzaniga},\ and\ \citenamefont
  {Boffi}}]{Pasquini:2008ax}%
  \BibitemOpen
  \bibfield  {author} {\bibinfo {author} {\bibfnamefont {B.}~\bibnamefont
  {Pasquini}}, \bibinfo {author} {\bibfnamefont {S.}~\bibnamefont
  {Cazzaniga}},\ and\ \bibinfo {author} {\bibfnamefont {S.}~\bibnamefont
  {Boffi}},\ }\bibfield  {title} {\bibinfo {title} {{Transverse momentum
  dependent parton distributions in a light-cone quark model}},\ }\href
  {https://doi.org/10.1103/PhysRevD.78.034025} {\bibfield  {journal} {\bibinfo
  {journal} {Phys. Rev. D}\ }\textbf {\bibinfo {volume} {78}},\ \bibinfo
  {pages} {034025} (\bibinfo {year} {2008})},\ \Eprint
  {https://arxiv.org/abs/0806.2298} {arXiv:0806.2298 [hep-ph]} \BibitemShut
  {NoStop}%
\bibitem [{\citenamefont {Kaur}\ \emph {et~al.}(2024)\citenamefont {Kaur},
  \citenamefont {Puhan}, \citenamefont {Pandey}, \citenamefont {Kumar},
  \citenamefont {Dutt},\ and\ \citenamefont {Dahiya}}]{Kaur:2024wze}%
  \BibitemOpen
  \bibfield  {author} {\bibinfo {author} {\bibfnamefont {N.}~\bibnamefont
  {Kaur}}, \bibinfo {author} {\bibfnamefont {S.}~\bibnamefont {Puhan}},
  \bibinfo {author} {\bibfnamefont {R.}~\bibnamefont {Pandey}}, \bibinfo
  {author} {\bibfnamefont {A.}~\bibnamefont {Kumar}}, \bibinfo {author}
  {\bibfnamefont {S.}~\bibnamefont {Dutt}},\ and\ \bibinfo {author}
  {\bibfnamefont {H.}~\bibnamefont {Dahiya}},\ }\bibfield  {title} {\bibinfo
  {title} {{Does nuclear medium affect the transverse momentum-dependent parton
  distributions of valence quark of pions?}},\ }\href@noop {} {\  (\bibinfo
  {year} {2024})},\ \Eprint {https://arxiv.org/abs/2409.05394}
  {arXiv:2409.05394 [hep-ph]} \BibitemShut {NoStop}%
\bibitem [{\citenamefont {Diehl}(2003)}]{Diehl:2003ny}%
  \BibitemOpen
  \bibfield  {author} {\bibinfo {author} {\bibfnamefont {M.}~\bibnamefont
  {Diehl}},\ }\bibfield  {title} {\bibinfo {title} {{Generalized parton
  distributions}},\ }\href {https://doi.org/10.1016/j.physrep.2003.08.002}
  {\bibfield  {journal} {\bibinfo  {journal} {Phys. Rept.}\ }\textbf {\bibinfo
  {volume} {388}},\ \bibinfo {pages} {41} (\bibinfo {year} {2003})},\ \Eprint
  {https://arxiv.org/abs/hep-ph/0307382} {arXiv:hep-ph/0307382} \BibitemShut
  {NoStop}%
\bibitem [{\citenamefont {Chavez}\ \emph {et~al.}(2022)\citenamefont {Chavez},
  \citenamefont {Bertone}, \citenamefont {De~Soto~Borrero}, \citenamefont
  {Defurne}, \citenamefont {Mezrag}, \citenamefont {Moutarde}, \citenamefont
  {Rodr\'\i{}guez-Quintero},\ and\ \citenamefont {Segovia}}]{Chavez:2021llq}%
  \BibitemOpen
  \bibfield  {author} {\bibinfo {author} {\bibfnamefont {J.~M.~M.}\
  \bibnamefont {Chavez}}, \bibinfo {author} {\bibfnamefont {V.}~\bibnamefont
  {Bertone}}, \bibinfo {author} {\bibfnamefont {F.}~\bibnamefont
  {De~Soto~Borrero}}, \bibinfo {author} {\bibfnamefont {M.}~\bibnamefont
  {Defurne}}, \bibinfo {author} {\bibfnamefont {C.}~\bibnamefont {Mezrag}},
  \bibinfo {author} {\bibfnamefont {H.}~\bibnamefont {Moutarde}}, \bibinfo
  {author} {\bibfnamefont {J.}~\bibnamefont {Rodr\'\i{}guez-Quintero}},\ and\
  \bibinfo {author} {\bibfnamefont {J.}~\bibnamefont {Segovia}},\ }\bibfield
  {title} {\bibinfo {title} {{Pion generalized parton distributions: A path
  toward phenomenology}},\ }\href {https://doi.org/10.1103/PhysRevD.105.094012}
  {\bibfield  {journal} {\bibinfo  {journal} {Phys. Rev. D}\ }\textbf {\bibinfo
  {volume} {105}},\ \bibinfo {pages} {094012} (\bibinfo {year} {2022})},\
  \Eprint {https://arxiv.org/abs/2110.06052} {arXiv:2110.06052 [hep-ph]}
  \BibitemShut {NoStop}%
\bibitem [{\citenamefont {Zhang}\ and\ \citenamefont
  {Ping}(2021)}]{Zhang:2021tnr}%
  \BibitemOpen
  \bibfield  {author} {\bibinfo {author} {\bibfnamefont {J.-L.}\ \bibnamefont
  {Zhang}}\ and\ \bibinfo {author} {\bibfnamefont {J.-L.}\ \bibnamefont
  {Ping}},\ }\bibfield  {title} {\bibinfo {title} {{Kaon generalized parton
  distributions and light-front wave functions in the
  Nambu\textendash{}Jona-Lasinio model}},\ }\href
  {https://doi.org/10.1140/epjc/s10052-021-09600-z} {\bibfield  {journal}
  {\bibinfo  {journal} {Eur. Phys. J. C}\ }\textbf {\bibinfo {volume} {81}},\
  \bibinfo {pages} {814} (\bibinfo {year} {2021})}\BibitemShut {NoStop}%
\bibitem [{\citenamefont {Broniowski}\ \emph {et~al.}(2023)\citenamefont
  {Broniowski}, \citenamefont {Shastry},\ and\ \citenamefont
  {Ruiz~Arriola}}]{Broniowski:2022iip}%
  \BibitemOpen
  \bibfield  {author} {\bibinfo {author} {\bibfnamefont {W.}~\bibnamefont
  {Broniowski}}, \bibinfo {author} {\bibfnamefont {V.}~\bibnamefont
  {Shastry}},\ and\ \bibinfo {author} {\bibfnamefont {E.}~\bibnamefont
  {Ruiz~Arriola}},\ }\bibfield  {title} {\bibinfo {title} {{Off-shell
  generalized parton distributions and form factors of the pion}},\ }\href
  {https://doi.org/10.1016/j.physletb.2023.137872} {\bibfield  {journal}
  {\bibinfo  {journal} {Phys. Lett. B}\ }\textbf {\bibinfo {volume} {840}},\
  \bibinfo {pages} {137872} (\bibinfo {year} {2023})},\ \Eprint
  {https://arxiv.org/abs/2211.11067} {arXiv:2211.11067 [hep-ph]} \BibitemShut
  {NoStop}%
\bibitem [{\citenamefont {Kaur}\ and\ \citenamefont
  {Dahiya}(2024{\natexlab{a}})}]{Kaur:2023zhn}%
  \BibitemOpen
  \bibfield  {author} {\bibinfo {author} {\bibfnamefont {N.}~\bibnamefont
  {Kaur}}\ and\ \bibinfo {author} {\bibfnamefont {H.}~\bibnamefont {Dahiya}},\
  }\bibfield  {title} {\bibinfo {title} {{Generalized parton distributions for
  the lowest-lying octet baryons}},\ }\href
  {https://doi.org/10.1140/epja/s10050-024-01265-y} {\bibfield  {journal}
  {\bibinfo  {journal} {Eur. Phys. J. A}\ }\textbf {\bibinfo {volume} {60}},\
  \bibinfo {pages} {42} (\bibinfo {year} {2024}{\natexlab{a}})},\ \Eprint
  {https://arxiv.org/abs/2310.03462} {arXiv:2310.03462 [hep-ph]} \BibitemShut
  {NoStop}%
\bibitem [{\citenamefont {Guidal}\ \emph {et~al.}(2005)\citenamefont {Guidal},
  \citenamefont {Polyakov}, \citenamefont {Radyushkin},\ and\ \citenamefont
  {Vanderhaeghen}}]{Guidal:2004nd}%
  \BibitemOpen
  \bibfield  {author} {\bibinfo {author} {\bibfnamefont {M.}~\bibnamefont
  {Guidal}}, \bibinfo {author} {\bibfnamefont {M.~V.}\ \bibnamefont
  {Polyakov}}, \bibinfo {author} {\bibfnamefont {A.~V.}\ \bibnamefont
  {Radyushkin}},\ and\ \bibinfo {author} {\bibfnamefont {M.}~\bibnamefont
  {Vanderhaeghen}},\ }\bibfield  {title} {\bibinfo {title} {{Nucleon
  form-factors from generalized parton distributions}},\ }\href
  {https://doi.org/10.1103/PhysRevD.72.054013} {\bibfield  {journal} {\bibinfo
  {journal} {Phys. Rev. D}\ }\textbf {\bibinfo {volume} {72}},\ \bibinfo
  {pages} {054013} (\bibinfo {year} {2005})},\ \Eprint
  {https://arxiv.org/abs/hep-ph/0410251} {arXiv:hep-ph/0410251} \BibitemShut
  {NoStop}%
\bibitem [{\citenamefont {Echevarria}\ \emph {et~al.}(2023)\citenamefont
  {Echevarria}, \citenamefont {Gutierrez~Garcia},\ and\ \citenamefont
  {Scimemi}}]{Echevarria:2022ztg}%
  \BibitemOpen
  \bibfield  {author} {\bibinfo {author} {\bibfnamefont {M.~G.}\ \bibnamefont
  {Echevarria}}, \bibinfo {author} {\bibfnamefont {P.~A.}\ \bibnamefont
  {Gutierrez~Garcia}},\ and\ \bibinfo {author} {\bibfnamefont {I.}~\bibnamefont
  {Scimemi}},\ }\bibfield  {title} {\bibinfo {title} {{GTMDs and the
  factorization of exclusive double Drell-Yan}},\ }\href
  {https://doi.org/10.1016/j.physletb.2023.137881} {\bibfield  {journal}
  {\bibinfo  {journal} {Phys. Lett. B}\ }\textbf {\bibinfo {volume} {840}},\
  \bibinfo {pages} {137881} (\bibinfo {year} {2023})},\ \Eprint
  {https://arxiv.org/abs/2208.00021} {arXiv:2208.00021 [hep-ph]} \BibitemShut
  {NoStop}%
\bibitem [{\citenamefont {Meissner}\ \emph {et~al.}(2009)\citenamefont
  {Meissner}, \citenamefont {Metz},\ and\ \citenamefont
  {Schlegel}}]{Meissner:2009ww}%
  \BibitemOpen
  \bibfield  {author} {\bibinfo {author} {\bibfnamefont {S.}~\bibnamefont
  {Meissner}}, \bibinfo {author} {\bibfnamefont {A.}~\bibnamefont {Metz}},\
  and\ \bibinfo {author} {\bibfnamefont {M.}~\bibnamefont {Schlegel}},\
  }\bibfield  {title} {\bibinfo {title} {{Generalized parton correlation
  functions for a spin-1/2 hadron}},\ }\href
  {https://doi.org/10.1088/1126-6708/2009/08/056} {\bibfield  {journal}
  {\bibinfo  {journal} {JHEP}\ }\textbf {\bibinfo {volume} {08}},\ \bibinfo
  {pages} {056}},\ \Eprint {https://arxiv.org/abs/0906.5323} {arXiv:0906.5323
  [hep-ph]} \BibitemShut {NoStop}%
\bibitem [{\citenamefont {Goeke}\ \emph {et~al.}(2005)\citenamefont {Goeke},
  \citenamefont {Metz},\ and\ \citenamefont {Schlegel}}]{Goeke:2005hb}%
  \BibitemOpen
  \bibfield  {author} {\bibinfo {author} {\bibfnamefont {K.}~\bibnamefont
  {Goeke}}, \bibinfo {author} {\bibfnamefont {A.}~\bibnamefont {Metz}},\ and\
  \bibinfo {author} {\bibfnamefont {M.}~\bibnamefont {Schlegel}},\ }\bibfield
  {title} {\bibinfo {title} {{Parameterization of the quark-quark correlator of
  a spin-1/2 hadron}},\ }\href {https://doi.org/10.1016/j.physletb.2005.05.037}
  {\bibfield  {journal} {\bibinfo  {journal} {Phys. Lett. B}\ }\textbf
  {\bibinfo {volume} {618}},\ \bibinfo {pages} {90} (\bibinfo {year} {2005})},\
  \Eprint {https://arxiv.org/abs/hep-ph/0504130} {arXiv:hep-ph/0504130}
  \BibitemShut {NoStop}%
\bibitem [{\citenamefont {Bacchetta}\ \emph {et~al.}(2017)\citenamefont
  {Bacchetta}, \citenamefont {Delcarro}, \citenamefont {Pisano}, \citenamefont
  {Radici},\ and\ \citenamefont {Signori}}]{Bacchetta:2017gcc}%
  \BibitemOpen
  \bibfield  {author} {\bibinfo {author} {\bibfnamefont {A.}~\bibnamefont
  {Bacchetta}}, \bibinfo {author} {\bibfnamefont {F.}~\bibnamefont {Delcarro}},
  \bibinfo {author} {\bibfnamefont {C.}~\bibnamefont {Pisano}}, \bibinfo
  {author} {\bibfnamefont {M.}~\bibnamefont {Radici}},\ and\ \bibinfo {author}
  {\bibfnamefont {A.}~\bibnamefont {Signori}},\ }\bibfield  {title} {\bibinfo
  {title} {{Extraction of partonic transverse momentum distributions from
  semi-inclusive deep-inelastic scattering, Drell-Yan and Z-boson
  production}},\ }\href {https://doi.org/10.1007/JHEP06(2017)081} {\bibfield
  {journal} {\bibinfo  {journal} {JHEP}\ }\textbf {\bibinfo {volume} {06}},\
  \bibinfo {pages} {081}},\ \bibinfo {note} {[Erratum: JHEP 06, 051 (2019)]},\
  \Eprint {https://arxiv.org/abs/1703.10157} {arXiv:1703.10157 [hep-ph]}
  \BibitemShut {NoStop}%
\bibitem [{\citenamefont {Makris}\ \emph {et~al.}(2021)\citenamefont {Makris},
  \citenamefont {Ringer},\ and\ \citenamefont {Waalewijn}}]{Makris:2020ltr}%
  \BibitemOpen
  \bibfield  {author} {\bibinfo {author} {\bibfnamefont {Y.}~\bibnamefont
  {Makris}}, \bibinfo {author} {\bibfnamefont {F.}~\bibnamefont {Ringer}},\
  and\ \bibinfo {author} {\bibfnamefont {W.~J.}\ \bibnamefont {Waalewijn}},\
  }\bibfield  {title} {\bibinfo {title} {{Joint thrust and TMD resummation in
  electron-positron and electron-proton collisions}},\ }\href
  {https://doi.org/10.1007/JHEP02(2021)070} {\bibfield  {journal} {\bibinfo
  {journal} {JHEP}\ }\textbf {\bibinfo {volume} {02}},\ \bibinfo {pages}
  {070}},\ \Eprint {https://arxiv.org/abs/2009.11871} {arXiv:2009.11871
  [hep-ph]} \BibitemShut {NoStop}%
\bibitem [{\citenamefont {Boer}\ \emph {et~al.}(1997)\citenamefont {Boer},
  \citenamefont {Jakob},\ and\ \citenamefont {Mulders}}]{Boer:1997mf}%
  \BibitemOpen
  \bibfield  {author} {\bibinfo {author} {\bibfnamefont {D.}~\bibnamefont
  {Boer}}, \bibinfo {author} {\bibfnamefont {R.}~\bibnamefont {Jakob}},\ and\
  \bibinfo {author} {\bibfnamefont {P.~J.}\ \bibnamefont {Mulders}},\
  }\bibfield  {title} {\bibinfo {title} {{Asymmetries in polarized hadron
  production in e+ e- annihilation up to order 1/Q}},\ }\href
  {https://doi.org/10.1016/S0550-3213(97)00456-2} {\bibfield  {journal}
  {\bibinfo  {journal} {Nucl. Phys. B}\ }\textbf {\bibinfo {volume} {504}},\
  \bibinfo {pages} {345} (\bibinfo {year} {1997})},\ \Eprint
  {https://arxiv.org/abs/hep-ph/9702281} {arXiv:hep-ph/9702281} \BibitemShut
  {NoStop}%
\bibitem [{\citenamefont {Catani}\ \emph {et~al.}(2015)\citenamefont {Catani},
  \citenamefont {de~Florian}, \citenamefont {Ferrera},\ and\ \citenamefont
  {Grazzini}}]{Catani:2015vma}%
  \BibitemOpen
  \bibfield  {author} {\bibinfo {author} {\bibfnamefont {S.}~\bibnamefont
  {Catani}}, \bibinfo {author} {\bibfnamefont {D.}~\bibnamefont {de~Florian}},
  \bibinfo {author} {\bibfnamefont {G.}~\bibnamefont {Ferrera}},\ and\ \bibinfo
  {author} {\bibfnamefont {M.}~\bibnamefont {Grazzini}},\ }\bibfield  {title}
  {\bibinfo {title} {{Vector boson production at hadron colliders:
  transverse-momentum resummation and leptonic decay}},\ }\href
  {https://doi.org/10.1007/JHEP12(2015)047} {\bibfield  {journal} {\bibinfo
  {journal} {JHEP}\ }\textbf {\bibinfo {volume} {12}},\ \bibinfo {pages}
  {047}},\ \Eprint {https://arxiv.org/abs/1507.06937} {arXiv:1507.06937
  [hep-ph]} \BibitemShut {NoStop}%
\bibitem [{\citenamefont {Kaur}\ and\ \citenamefont
  {Dahiya}(2024{\natexlab{b}})}]{Kaur:2024bgo}%
  \BibitemOpen
  \bibfield  {author} {\bibinfo {author} {\bibfnamefont {N.}~\bibnamefont
  {Kaur}}\ and\ \bibinfo {author} {\bibfnamefont {H.}~\bibnamefont {Dahiya}},\
  }\bibfield  {title} {\bibinfo {title} {{Transverse distortion and single-spin
  asymmetries for low-lying octet baryons}},\ }\href
  {https://doi.org/10.1142/S0217751X24500763} {\bibfield  {journal} {\bibinfo
  {journal} {Int. J. Mod. Phys. A}\ }\textbf {\bibinfo {volume} {39}},\
  \bibinfo {pages} {2450076} (\bibinfo {year} {2024}{\natexlab{b}})},\ \Eprint
  {https://arxiv.org/abs/2405.00445} {arXiv:2405.00445 [hep-ph]} \BibitemShut
  {NoStop}%
\bibitem [{\citenamefont {Freese}\ and\ \citenamefont
  {Clo\"et}(2021)}]{Freese:2020mcx}%
  \BibitemOpen
  \bibfield  {author} {\bibinfo {author} {\bibfnamefont {A.}~\bibnamefont
  {Freese}}\ and\ \bibinfo {author} {\bibfnamefont {I.~C.}\ \bibnamefont
  {Clo\"et}},\ }\bibfield  {title} {\bibinfo {title} {{Quark spin and orbital
  angular momentum from proton generalized parton distributions}},\ }\href
  {https://doi.org/10.1103/PhysRevC.103.045204} {\bibfield  {journal} {\bibinfo
   {journal} {Phys. Rev. C}\ }\textbf {\bibinfo {volume} {103}},\ \bibinfo
  {pages} {045204} (\bibinfo {year} {2021})},\ \Eprint
  {https://arxiv.org/abs/2005.10286} {arXiv:2005.10286 [nucl-th]} \BibitemShut
  {NoStop}%
\bibitem [{\citenamefont {Luan}\ and\ \citenamefont {Lu}(2023)}]{Luan:2023lmt}%
  \BibitemOpen
  \bibfield  {author} {\bibinfo {author} {\bibfnamefont {X.}~\bibnamefont
  {Luan}}\ and\ \bibinfo {author} {\bibfnamefont {Z.}~\bibnamefont {Lu}},\
  }\bibfield  {title} {\bibinfo {title} {{Generalized parton distributions of
  sea quark at zero skewness in the light-cone model}},\ }\href
  {https://doi.org/10.1140/epjc/s10052-023-11637-1} {\bibfield  {journal}
  {\bibinfo  {journal} {Eur. Phys. J. C}\ }\textbf {\bibinfo {volume} {83}},\
  \bibinfo {pages} {504} (\bibinfo {year} {2023})},\ \Eprint
  {https://arxiv.org/abs/2302.11278} {arXiv:2302.11278 [hep-ph]} \BibitemShut
  {NoStop}%
\bibitem [{\citenamefont {Ji}(1997)}]{Ji:1996nm}%
  \BibitemOpen
  \bibfield  {author} {\bibinfo {author} {\bibfnamefont {X.-D.}\ \bibnamefont
  {Ji}},\ }\bibfield  {title} {\bibinfo {title} {{Deeply virtual Compton
  scattering}},\ }\href {https://doi.org/10.1103/PhysRevD.55.7114} {\bibfield
  {journal} {\bibinfo  {journal} {Phys. Rev. D}\ }\textbf {\bibinfo {volume}
  {55}},\ \bibinfo {pages} {7114} (\bibinfo {year} {1997})},\ \Eprint
  {https://arxiv.org/abs/hep-ph/9609381} {arXiv:hep-ph/9609381} \BibitemShut
  {NoStop}%
\bibitem [{\citenamefont {Xie}\ \emph {et~al.}(2023)\citenamefont {Xie},
  \citenamefont {Kou}, \citenamefont {Fu}, \citenamefont {Ye},\ and\
  \citenamefont {Chen}}]{Xie:2023xkz}%
  \BibitemOpen
  \bibfield  {author} {\bibinfo {author} {\bibfnamefont {G.}~\bibnamefont
  {Xie}}, \bibinfo {author} {\bibfnamefont {W.}~\bibnamefont {Kou}}, \bibinfo
  {author} {\bibfnamefont {Q.}~\bibnamefont {Fu}}, \bibinfo {author}
  {\bibfnamefont {Z.}~\bibnamefont {Ye}},\ and\ \bibinfo {author}
  {\bibfnamefont {X.}~\bibnamefont {Chen}},\ }\bibfield  {title} {\bibinfo
  {title} {{Deeply virtual compton scattering at future electron-ion
  colliders}},\ }\href {https://doi.org/10.1140/epjc/s10052-023-12065-x}
  {\bibfield  {journal} {\bibinfo  {journal} {Eur. Phys. J. C}\ }\textbf
  {\bibinfo {volume} {83}},\ \bibinfo {pages} {900} (\bibinfo {year} {2023})},\
  \Eprint {https://arxiv.org/abs/2306.02357} {arXiv:2306.02357 [hep-ph]}
  \BibitemShut {NoStop}%
\bibitem [{\citenamefont {Favart}\ \emph {et~al.}(2016)\citenamefont {Favart},
  \citenamefont {Guidal}, \citenamefont {Horn},\ and\ \citenamefont
  {Kroll}}]{Favart:2015umi}%
  \BibitemOpen
  \bibfield  {author} {\bibinfo {author} {\bibfnamefont {L.}~\bibnamefont
  {Favart}}, \bibinfo {author} {\bibfnamefont {M.}~\bibnamefont {Guidal}},
  \bibinfo {author} {\bibfnamefont {T.}~\bibnamefont {Horn}},\ and\ \bibinfo
  {author} {\bibfnamefont {P.}~\bibnamefont {Kroll}},\ }\bibfield  {title}
  {\bibinfo {title} {{Deeply Virtual Meson Production on the nucleon}},\ }\href
  {https://doi.org/10.1140/epja/i2016-16158-2} {\bibfield  {journal} {\bibinfo
  {journal} {Eur. Phys. J. A}\ }\textbf {\bibinfo {volume} {52}},\ \bibinfo
  {pages} {158} (\bibinfo {year} {2016})},\ \Eprint
  {https://arxiv.org/abs/1511.04535} {arXiv:1511.04535 [hep-ph]} \BibitemShut
  {NoStop}%
\bibitem [{\citenamefont {Brooks}\ \emph {et~al.}(2018)\citenamefont {Brooks},
  \citenamefont {Schmidt},\ and\ \citenamefont {Siddikov}}]{Brooks:2018uqk}%
  \BibitemOpen
  \bibfield  {author} {\bibinfo {author} {\bibfnamefont {W.}~\bibnamefont
  {Brooks}}, \bibinfo {author} {\bibfnamefont {I.}~\bibnamefont {Schmidt}},\
  and\ \bibinfo {author} {\bibfnamefont {M.}~\bibnamefont {Siddikov}},\
  }\bibfield  {title} {\bibinfo {title} {{Deeply virtual meson production on
  neutrons}},\ }\href {https://doi.org/10.1103/PhysRevD.98.116006} {\bibfield
  {journal} {\bibinfo  {journal} {Phys. Rev. D}\ }\textbf {\bibinfo {volume}
  {98}},\ \bibinfo {pages} {116006} (\bibinfo {year} {2018})},\ \Eprint
  {https://arxiv.org/abs/1810.08077} {arXiv:1810.08077 [hep-ph]} \BibitemShut
  {NoStop}%
\bibitem [{\citenamefont {Meissner}\ \emph {et~al.}(2008)\citenamefont
  {Meissner}, \citenamefont {Metz}, \citenamefont {Schlegel},\ and\
  \citenamefont {Goeke}}]{Meissner:2008ay}%
  \BibitemOpen
  \bibfield  {author} {\bibinfo {author} {\bibfnamefont {S.}~\bibnamefont
  {Meissner}}, \bibinfo {author} {\bibfnamefont {A.}~\bibnamefont {Metz}},
  \bibinfo {author} {\bibfnamefont {M.}~\bibnamefont {Schlegel}},\ and\
  \bibinfo {author} {\bibfnamefont {K.}~\bibnamefont {Goeke}},\ }\bibfield
  {title} {\bibinfo {title} {{Generalized parton correlation functions for a
  spin-0 hadron}},\ }\href {https://doi.org/10.1088/1126-6708/2008/08/038}
  {\bibfield  {journal} {\bibinfo  {journal} {JHEP}\ }\textbf {\bibinfo
  {volume} {08}},\ \bibinfo {pages} {038}},\ \Eprint
  {https://arxiv.org/abs/0805.3165} {arXiv:0805.3165 [hep-ph]} \BibitemShut
  {NoStop}%
\bibitem [{\citenamefont {Kaur}\ and\ \citenamefont
  {Dahiya}(2021)}]{Kaur:2019jfa}%
  \BibitemOpen
  \bibfield  {author} {\bibinfo {author} {\bibfnamefont {N.}~\bibnamefont
  {Kaur}}\ and\ \bibinfo {author} {\bibfnamefont {H.}~\bibnamefont {Dahiya}},\
  }\bibfield  {title} {\bibinfo {title} {{Transverse momentum dependent parton
  distributions of pion in the light-front holographic model}},\ }\href
  {https://doi.org/10.1142/S0217751X21500524} {\bibfield  {journal} {\bibinfo
  {journal} {Int. J. Mod. Phys. A}\ }\textbf {\bibinfo {volume} {36}},\
  \bibinfo {pages} {2150052} (\bibinfo {year} {2021})},\ \Eprint
  {https://arxiv.org/abs/1908.08657} {arXiv:1908.08657 [hep-ph]} \BibitemShut
  {NoStop}%
\bibitem [{\citenamefont {Cerutti}\ \emph {et~al.}(2023)\citenamefont
  {Cerutti}, \citenamefont {Rossi}, \citenamefont {Venturini}, \citenamefont
  {Bacchetta}, \citenamefont {Bertone}, \citenamefont {Bissolotti},\ and\
  \citenamefont {Radici}}]{Cerutti:2022lmb}%
  \BibitemOpen
  \bibfield  {author} {\bibinfo {author} {\bibfnamefont {M.}~\bibnamefont
  {Cerutti}}, \bibinfo {author} {\bibfnamefont {L.}~\bibnamefont {Rossi}},
  \bibinfo {author} {\bibfnamefont {S.}~\bibnamefont {Venturini}}, \bibinfo
  {author} {\bibfnamefont {A.}~\bibnamefont {Bacchetta}}, \bibinfo {author}
  {\bibfnamefont {V.}~\bibnamefont {Bertone}}, \bibinfo {author} {\bibfnamefont
  {C.}~\bibnamefont {Bissolotti}},\ and\ \bibinfo {author} {\bibfnamefont
  {M.}~\bibnamefont {Radici}} (\bibinfo {collaboration} {MAP (Multi-dimensional
  Analyses of Partonic distributions)}),\ }\bibfield  {title} {\bibinfo {title}
  {{Extraction of pion transverse momentum distributions from Drell-Yan
  data}},\ }\href {https://doi.org/10.1103/PhysRevD.107.014014} {\bibfield
  {journal} {\bibinfo  {journal} {Phys. Rev. D}\ }\textbf {\bibinfo {volume}
  {107}},\ \bibinfo {pages} {014014} (\bibinfo {year} {2023})},\ \Eprint
  {https://arxiv.org/abs/2210.01733} {arXiv:2210.01733 [hep-ph]} \BibitemShut
  {NoStop}%
\bibitem [{\citenamefont {Chu}\ \emph {et~al.}(2024)\citenamefont {Chu} \emph
  {et~al.}}]{LatticeParton:2023xdl}%
  \BibitemOpen
  \bibfield  {author} {\bibinfo {author} {\bibfnamefont {M.-H.}\ \bibnamefont
  {Chu}} \emph {et~al.} (\bibinfo {collaboration} {Lattice Parton}),\
  }\bibfield  {title} {\bibinfo {title} {{Transverse-momentum-dependent wave
  functions of the pion from lattice QCD}},\ }\href
  {https://doi.org/10.1103/PhysRevD.109.L091503} {\bibfield  {journal}
  {\bibinfo  {journal} {Phys. Rev. D}\ }\textbf {\bibinfo {volume} {109}},\
  \bibinfo {pages} {L091503} (\bibinfo {year} {2024})},\ \Eprint
  {https://arxiv.org/abs/2302.09961} {arXiv:2302.09961 [hep-lat]} \BibitemShut
  {NoStop}%
\bibitem [{\citenamefont {Engelhardt}\ \emph {et~al.}(2016)\citenamefont
  {Engelhardt}, \citenamefont {H\"agler}, \citenamefont {Musch}, \citenamefont
  {Negele},\ and\ \citenamefont {Sch\"afer}}]{Engelhardt:2015xja}%
  \BibitemOpen
  \bibfield  {author} {\bibinfo {author} {\bibfnamefont {M.}~\bibnamefont
  {Engelhardt}}, \bibinfo {author} {\bibfnamefont {P.}~\bibnamefont
  {H\"agler}}, \bibinfo {author} {\bibfnamefont {B.}~\bibnamefont {Musch}},
  \bibinfo {author} {\bibfnamefont {J.}~\bibnamefont {Negele}},\ and\ \bibinfo
  {author} {\bibfnamefont {A.}~\bibnamefont {Sch\"afer}},\ }\bibfield  {title}
  {\bibinfo {title} {{Lattice QCD study of the Boer-Mulders effect in a
  pion}},\ }\href {https://doi.org/10.1103/PhysRevD.93.054501} {\bibfield
  {journal} {\bibinfo  {journal} {Phys. Rev. D}\ }\textbf {\bibinfo {volume}
  {93}},\ \bibinfo {pages} {054501} (\bibinfo {year} {2016})},\ \Eprint
  {https://arxiv.org/abs/1506.07826} {arXiv:1506.07826 [hep-lat]} \BibitemShut
  {NoStop}%
\bibitem [{\citenamefont {Pasquini}\ and\ \citenamefont
  {Schweitzer}(2014)}]{Pasquini:2014ppa}%
  \BibitemOpen
  \bibfield  {author} {\bibinfo {author} {\bibfnamefont {B.}~\bibnamefont
  {Pasquini}}\ and\ \bibinfo {author} {\bibfnamefont {P.}~\bibnamefont
  {Schweitzer}},\ }\bibfield  {title} {\bibinfo {title} {{Pion transverse
  momentum dependent parton distributions in a light-front constituent
  approach, and the Boer-Mulders effect in the pion-induced Drell-Yan
  process}},\ }\href {https://doi.org/10.1103/PhysRevD.90.014050} {\bibfield
  {journal} {\bibinfo  {journal} {Phys. Rev. D}\ }\textbf {\bibinfo {volume}
  {90}},\ \bibinfo {pages} {014050} (\bibinfo {year} {2014})},\ \Eprint
  {https://arxiv.org/abs/1406.2056} {arXiv:1406.2056 [hep-ph]} \BibitemShut
  {NoStop}%
\bibitem [{\citenamefont {Han}\ \emph {et~al.}(2021)\citenamefont {Han},
  \citenamefont {Xie}, \citenamefont {Wang},\ and\ \citenamefont
  {Chen}}]{Han:2020vjp}%
  \BibitemOpen
  \bibfield  {author} {\bibinfo {author} {\bibfnamefont {C.}~\bibnamefont
  {Han}}, \bibinfo {author} {\bibfnamefont {G.}~\bibnamefont {Xie}}, \bibinfo
  {author} {\bibfnamefont {R.}~\bibnamefont {Wang}},\ and\ \bibinfo {author}
  {\bibfnamefont {X.}~\bibnamefont {Chen}},\ }\bibfield  {title} {\bibinfo
  {title} {{An Analysis of Parton Distribution Functions of the Pion and the
  Kaon with the Maximum Entropy Input}},\ }\href
  {https://doi.org/10.1140/epjc/s10052-021-09087-8} {\bibfield  {journal}
  {\bibinfo  {journal} {Eur. Phys. J. C}\ }\textbf {\bibinfo {volume} {81}},\
  \bibinfo {pages} {302} (\bibinfo {year} {2021})},\ \Eprint
  {https://arxiv.org/abs/2010.14284} {arXiv:2010.14284 [hep-ph]} \BibitemShut
  {NoStop}%
\bibitem [{\citenamefont {Kou}\ \emph {et~al.}(2023)\citenamefont {Kou},
  \citenamefont {Shi}, \citenamefont {Chen},\ and\ \citenamefont
  {Jia}}]{Kou:2023ady}%
  \BibitemOpen
  \bibfield  {author} {\bibinfo {author} {\bibfnamefont {W.}~\bibnamefont
  {Kou}}, \bibinfo {author} {\bibfnamefont {C.}~\bibnamefont {Shi}}, \bibinfo
  {author} {\bibfnamefont {X.}~\bibnamefont {Chen}},\ and\ \bibinfo {author}
  {\bibfnamefont {W.}~\bibnamefont {Jia}},\ }\bibfield  {title} {\bibinfo
  {title} {{Transverse momentum dependent parton distributions of pion at
  leading twist}},\ }\href {https://doi.org/10.1103/PhysRevD.108.036021}
  {\bibfield  {journal} {\bibinfo  {journal} {Phys. Rev. D}\ }\textbf {\bibinfo
  {volume} {108}},\ \bibinfo {pages} {036021} (\bibinfo {year} {2023})},\
  \Eprint {https://arxiv.org/abs/2304.09814} {arXiv:2304.09814 [hep-ph]}
  \BibitemShut {NoStop}%
\bibitem [{\citenamefont {Puhan}\ \emph {et~al.}(2024)\citenamefont {Puhan},
  \citenamefont {Sharma}, \citenamefont {Kaur}, \citenamefont {Kumar},\ and\
  \citenamefont {Dahiya}}]{Puhan2023}%
  \BibitemOpen
  \bibfield  {author} {\bibinfo {author} {\bibfnamefont {S.}~\bibnamefont
  {Puhan}}, \bibinfo {author} {\bibfnamefont {S.}~\bibnamefont {Sharma}},
  \bibinfo {author} {\bibfnamefont {N.}~\bibnamefont {Kaur}}, \bibinfo {author}
  {\bibfnamefont {N.}~\bibnamefont {Kumar}},\ and\ \bibinfo {author}
  {\bibfnamefont {H.}~\bibnamefont {Dahiya}},\ }\bibfield  {title} {\bibinfo
  {title} {{T-even TMDs for the spin-0 pseudo-scalar mesons upto twist-4 using
  light-front formalism}},\ }\href {https://doi.org/10.1007/JHEP02(2024)075}
  {\bibfield  {journal} {\bibinfo  {journal} {JHEP}\ }\textbf {\bibinfo
  {volume} {02}},\ \bibinfo {pages} {075}},\ \Eprint
  {https://arxiv.org/abs/2310.03464} {arXiv:2310.03464 [hep-ph]} \BibitemShut
  {NoStop}%
\bibitem [{\citenamefont {Puhan}\ and\ \citenamefont
  {Dahiya}(2024{\natexlab{a}})}]{Puhan:2023hio}%
  \BibitemOpen
  \bibfield  {author} {\bibinfo {author} {\bibfnamefont {S.}~\bibnamefont
  {Puhan}}\ and\ \bibinfo {author} {\bibfnamefont {H.}~\bibnamefont {Dahiya}},\
  }\bibfield  {title} {\bibinfo {title} {{Leading twist T-even TMDs for the
  spin-1 heavy vector mesons}},\ }\href
  {https://doi.org/10.1103/PhysRevD.109.034005} {\bibfield  {journal} {\bibinfo
   {journal} {Phys. Rev. D}\ }\textbf {\bibinfo {volume} {109}},\ \bibinfo
  {pages} {034005} (\bibinfo {year} {2024}{\natexlab{a}})},\ \Eprint
  {https://arxiv.org/abs/2310.03465} {arXiv:2310.03465 [hep-ph]} \BibitemShut
  {NoStop}%
\bibitem [{\citenamefont {Zhang}\ \emph {et~al.}(2019)\citenamefont {Zhang},
  \citenamefont {Chen}, \citenamefont {Jin}, \citenamefont {Lin}, \citenamefont
  {Sch\"afer},\ and\ \citenamefont {Zhao}}]{Zhang:2018nsy}%
  \BibitemOpen
  \bibfield  {author} {\bibinfo {author} {\bibfnamefont {J.-H.}\ \bibnamefont
  {Zhang}}, \bibinfo {author} {\bibfnamefont {J.-W.}\ \bibnamefont {Chen}},
  \bibinfo {author} {\bibfnamefont {L.}~\bibnamefont {Jin}}, \bibinfo {author}
  {\bibfnamefont {H.-W.}\ \bibnamefont {Lin}}, \bibinfo {author} {\bibfnamefont
  {A.}~\bibnamefont {Sch\"afer}},\ and\ \bibinfo {author} {\bibfnamefont
  {Y.}~\bibnamefont {Zhao}},\ }\bibfield  {title} {\bibinfo {title} {{First
  direct lattice-QCD calculation of the $x$-dependence of the pion parton
  distribution function}},\ }\href
  {https://doi.org/10.1103/PhysRevD.100.034505} {\bibfield  {journal} {\bibinfo
   {journal} {Phys. Rev. D}\ }\textbf {\bibinfo {volume} {100}},\ \bibinfo
  {pages} {034505} (\bibinfo {year} {2019})},\ \Eprint
  {https://arxiv.org/abs/1804.01483} {arXiv:1804.01483 [hep-lat]} \BibitemShut
  {NoStop}%
\bibitem [{\citenamefont {Lan}\ \emph {et~al.}(2020)\citenamefont {Lan},
  \citenamefont {Mondal}, \citenamefont {Jia}, \citenamefont {Zhao},\ and\
  \citenamefont {Vary}}]{Lan:2019rba}%
  \BibitemOpen
  \bibfield  {author} {\bibinfo {author} {\bibfnamefont {J.}~\bibnamefont
  {Lan}}, \bibinfo {author} {\bibfnamefont {C.}~\bibnamefont {Mondal}},
  \bibinfo {author} {\bibfnamefont {S.}~\bibnamefont {Jia}}, \bibinfo {author}
  {\bibfnamefont {X.}~\bibnamefont {Zhao}},\ and\ \bibinfo {author}
  {\bibfnamefont {J.~P.}\ \bibnamefont {Vary}},\ }\bibfield  {title} {\bibinfo
  {title} {{Pion and kaon parton distribution functions from basis light front
  quantization and QCD evolution}},\ }\href
  {https://doi.org/10.1103/PhysRevD.101.034024} {\bibfield  {journal} {\bibinfo
   {journal} {Phys. Rev. D}\ }\textbf {\bibinfo {volume} {101}},\ \bibinfo
  {pages} {034024} (\bibinfo {year} {2020})},\ \Eprint
  {https://arxiv.org/abs/1907.01509} {arXiv:1907.01509 [nucl-th]} \BibitemShut
  {NoStop}%
\bibitem [{\citenamefont {Cui}\ \emph {et~al.}(2020)\citenamefont {Cui},
  \citenamefont {Ding}, \citenamefont {Gao}, \citenamefont {Raya},
  \citenamefont {Binosi}, \citenamefont {Chang}, \citenamefont {Roberts},
  \citenamefont {Rodr\'\i{}guez-Quintero},\ and\ \citenamefont
  {Schmidt}}]{Cui:2020tdf}%
  \BibitemOpen
  \bibfield  {author} {\bibinfo {author} {\bibfnamefont {Z.-F.}\ \bibnamefont
  {Cui}}, \bibinfo {author} {\bibfnamefont {M.}~\bibnamefont {Ding}}, \bibinfo
  {author} {\bibfnamefont {F.}~\bibnamefont {Gao}}, \bibinfo {author}
  {\bibfnamefont {K.}~\bibnamefont {Raya}}, \bibinfo {author} {\bibfnamefont
  {D.}~\bibnamefont {Binosi}}, \bibinfo {author} {\bibfnamefont
  {L.}~\bibnamefont {Chang}}, \bibinfo {author} {\bibfnamefont {C.~D.}\
  \bibnamefont {Roberts}}, \bibinfo {author} {\bibfnamefont {J.}~\bibnamefont
  {Rodr\'\i{}guez-Quintero}},\ and\ \bibinfo {author} {\bibfnamefont {S.~M.}\
  \bibnamefont {Schmidt}},\ }\bibfield  {title} {\bibinfo {title} {{Kaon and
  pion parton distributions}},\ }\href
  {https://doi.org/10.1140/epjc/s10052-020-08578-4} {\bibfield  {journal}
  {\bibinfo  {journal} {Eur. Phys. J. C}\ }\textbf {\bibinfo {volume} {80}},\
  \bibinfo {pages} {1064} (\bibinfo {year} {2020})}\BibitemShut {NoStop}%
\bibitem [{\citenamefont {Serna}\ \emph {et~al.}(2024)\citenamefont {Serna},
  \citenamefont {El-Bennich},\ and\ \citenamefont {Krein}}]{Serna:2024vpn}%
  \BibitemOpen
  \bibfield  {author} {\bibinfo {author} {\bibfnamefont {F.~E.}\ \bibnamefont
  {Serna}}, \bibinfo {author} {\bibfnamefont {B.}~\bibnamefont {El-Bennich}},\
  and\ \bibinfo {author} {\bibfnamefont {G.~a.}\ \bibnamefont {Krein}},\
  }\bibfield  {title} {\bibinfo {title} {{Parton distribution functions and
  transverse momentum dependence of heavy mesons}},\ }\href@noop {} {\
  (\bibinfo {year} {2024})},\ \Eprint {https://arxiv.org/abs/2409.01441}
  {arXiv:2409.01441 [hep-ph]} \BibitemShut {NoStop}%
\bibitem [{\citenamefont {Almeida-Zamora}\ \emph {et~al.}(2024)\citenamefont
  {Almeida-Zamora}, \citenamefont {Cobos-Mart\'\i{}nez}, \citenamefont
  {Bashir}, \citenamefont {Raya}, \citenamefont {Rodr\'\i{}guez-Quintero},\
  and\ \citenamefont {Segovia}}]{Almeida-Zamora:2023bqb}%
  \BibitemOpen
  \bibfield  {author} {\bibinfo {author} {\bibfnamefont {B.}~\bibnamefont
  {Almeida-Zamora}}, \bibinfo {author} {\bibfnamefont {J.~J.}\ \bibnamefont
  {Cobos-Mart\'\i{}nez}}, \bibinfo {author} {\bibfnamefont {A.}~\bibnamefont
  {Bashir}}, \bibinfo {author} {\bibfnamefont {K.}~\bibnamefont {Raya}},
  \bibinfo {author} {\bibfnamefont {J.}~\bibnamefont
  {Rodr\'\i{}guez-Quintero}},\ and\ \bibinfo {author} {\bibfnamefont
  {J.}~\bibnamefont {Segovia}},\ }\bibfield  {title} {\bibinfo {title}
  {{Algebraic model to study the internal structure of pseudoscalar mesons with
  heavy-light quark content}},\ }\href
  {https://doi.org/10.1103/PhysRevD.109.014016} {\bibfield  {journal} {\bibinfo
   {journal} {Phys. Rev. D}\ }\textbf {\bibinfo {volume} {109}},\ \bibinfo
  {pages} {014016} (\bibinfo {year} {2024})},\ \Eprint
  {https://arxiv.org/abs/2309.17282} {arXiv:2309.17282 [hep-ph]} \BibitemShut
  {NoStop}%
\bibitem [{\citenamefont {Puhan}\ and\ \citenamefont
  {Dahiya}(2024{\natexlab{b}})}]{Puhan:2024ckp}%
  \BibitemOpen
  \bibfield  {author} {\bibinfo {author} {\bibfnamefont {S.}~\bibnamefont
  {Puhan}}\ and\ \bibinfo {author} {\bibfnamefont {H.}~\bibnamefont {Dahiya}},\
  }\bibfield  {title} {\bibinfo {title} {{Spatial and Transverse structure of
  Heavy B- and D-mesons}},\ }\href {https://doi.org/10.22323/1.462.0089}
  {\bibfield  {journal} {\bibinfo  {journal} {PoS}\ }\textbf {\bibinfo {volume}
  {HQL2023}},\ \bibinfo {pages} {089} (\bibinfo {year} {2024}{\natexlab{b}})},\
  \Eprint {https://arxiv.org/abs/2408.07717} {arXiv:2408.07717 [hep-ph]}
  \BibitemShut {NoStop}%
\bibitem [{\citenamefont {Acharyya}\ \emph
  {et~al.}(2024{\natexlab{a}})\citenamefont {Acharyya}, \citenamefont {Puhan},
  \citenamefont {Kumar},\ and\ \citenamefont {Dahiya}}]{Acharyya:2024tql}%
  \BibitemOpen
  \bibfield  {author} {\bibinfo {author} {\bibfnamefont {R.}~\bibnamefont
  {Acharyya}}, \bibinfo {author} {\bibfnamefont {S.}~\bibnamefont {Puhan}},
  \bibinfo {author} {\bibfnamefont {N.}~\bibnamefont {Kumar}},\ and\ \bibinfo
  {author} {\bibfnamefont {H.}~\bibnamefont {Dahiya}},\ }\bibfield  {title}
  {\bibinfo {title} {{Spectroscopy of excited quarkonium states in the
  light-front quark model}},\ }\href@noop {} {\  (\bibinfo {year}
  {2024}{\natexlab{a}})},\ \Eprint {https://arxiv.org/abs/2408.07715}
  {arXiv:2408.07715 [hep-ph]} \BibitemShut {NoStop}%
\bibitem [{\citenamefont {Chen}\ \emph {et~al.}(2020)\citenamefont {Chen},
  \citenamefont {Lin},\ and\ \citenamefont {Zhang}}]{Chen:2019lcm}%
  \BibitemOpen
  \bibfield  {author} {\bibinfo {author} {\bibfnamefont {J.-W.}\ \bibnamefont
  {Chen}}, \bibinfo {author} {\bibfnamefont {H.-W.}\ \bibnamefont {Lin}},\ and\
  \bibinfo {author} {\bibfnamefont {J.-H.}\ \bibnamefont {Zhang}},\ }\bibfield
  {title} {\bibinfo {title} {{Pion generalized parton distribution from lattice
  QCD}},\ }\href {https://doi.org/10.1016/j.nuclphysb.2020.114940} {\bibfield
  {journal} {\bibinfo  {journal} {Nucl. Phys. B}\ }\textbf {\bibinfo {volume}
  {952}},\ \bibinfo {pages} {114940} (\bibinfo {year} {2020})},\ \Eprint
  {https://arxiv.org/abs/1904.12376} {arXiv:1904.12376 [hep-lat]} \BibitemShut
  {NoStop}%
\bibitem [{\citenamefont {Kaur}\ \emph {et~al.}(2020)\citenamefont {Kaur},
  \citenamefont {Kumar}, \citenamefont {Lan}, \citenamefont {Mondal},\ and\
  \citenamefont {Dahiya}}]{Kaur:2020vkq}%
  \BibitemOpen
  \bibfield  {author} {\bibinfo {author} {\bibfnamefont {S.}~\bibnamefont
  {Kaur}}, \bibinfo {author} {\bibfnamefont {N.}~\bibnamefont {Kumar}},
  \bibinfo {author} {\bibfnamefont {J.}~\bibnamefont {Lan}}, \bibinfo {author}
  {\bibfnamefont {C.}~\bibnamefont {Mondal}},\ and\ \bibinfo {author}
  {\bibfnamefont {H.}~\bibnamefont {Dahiya}},\ }\bibfield  {title} {\bibinfo
  {title} {{Tomography of light mesons in the light-cone quark model}},\ }\href
  {https://doi.org/10.1103/PhysRevD.102.014021} {\bibfield  {journal} {\bibinfo
   {journal} {Phys. Rev. D}\ }\textbf {\bibinfo {volume} {102}},\ \bibinfo
  {pages} {014021} (\bibinfo {year} {2020})},\ \Eprint
  {https://arxiv.org/abs/2002.01199} {arXiv:2002.01199 [hep-ph]} \BibitemShut
  {NoStop}%
\bibitem [{\citenamefont {Kaur}\ \emph {et~al.}(2018)\citenamefont {Kaur},
  \citenamefont {Kumar}, \citenamefont {Mondal},\ and\ \citenamefont
  {Dahiya}}]{Kaur:2018ewq}%
  \BibitemOpen
  \bibfield  {author} {\bibinfo {author} {\bibfnamefont {N.}~\bibnamefont
  {Kaur}}, \bibinfo {author} {\bibfnamefont {N.}~\bibnamefont {Kumar}},
  \bibinfo {author} {\bibfnamefont {C.}~\bibnamefont {Mondal}},\ and\ \bibinfo
  {author} {\bibfnamefont {H.}~\bibnamefont {Dahiya}},\ }\bibfield  {title}
  {\bibinfo {title} {{Generalized Parton Distributions of Pion for Non-Zero
  Skewness in AdS/QCD}},\ }\href
  {https://doi.org/10.1016/j.nuclphysb.2018.07.003} {\bibfield  {journal}
  {\bibinfo  {journal} {Nucl. Phys. B}\ }\textbf {\bibinfo {volume} {934}},\
  \bibinfo {pages} {80} (\bibinfo {year} {2018})},\ \Eprint
  {https://arxiv.org/abs/1807.01076} {arXiv:1807.01076 [hep-ph]} \BibitemShut
  {NoStop}%
\bibitem [{\citenamefont {Abdul~Khalek}\ \emph {et~al.}(2022)\citenamefont
  {Abdul~Khalek} \emph {et~al.}}]{AbdulKhalek:2021gbh}%
  \BibitemOpen
  \bibfield  {author} {\bibinfo {author} {\bibfnamefont {R.}~\bibnamefont
  {Abdul~Khalek}} \emph {et~al.},\ }\bibfield  {title} {\bibinfo {title}
  {{Science Requirements and Detector Concepts for the Electron-Ion Collider}:
  {EIC Yellow Report}},\ }\href
  {https://doi.org/10.1016/j.nuclphysa.2022.122447} {\bibfield  {journal}
  {\bibinfo  {journal} {Nucl. Phys. A}\ }\textbf {\bibinfo {volume} {1026}},\
  \bibinfo {pages} {122447} (\bibinfo {year} {2022})},\ \Eprint
  {https://arxiv.org/abs/2103.05419} {arXiv:2103.05419 [physics.ins-det]}
  \BibitemShut {NoStop}%
\bibitem [{\citenamefont {Adams}\ \emph {et~al.}(2018)\citenamefont {Adams}
  \emph {et~al.}}]{Adams:2018pwt}%
  \BibitemOpen
  \bibfield  {author} {\bibinfo {author} {\bibfnamefont {B.}~\bibnamefont
  {Adams}} \emph {et~al.},\ }\bibfield  {title} {\bibinfo {title} {{Letter of
  Intent: A New QCD facility at the M2 beam line of the CERN SPS
  (COMPASS++/AMBER)}},\ }\href@noop {} {\  (\bibinfo {year} {2018})},\ \Eprint
  {https://arxiv.org/abs/1808.00848} {arXiv:1808.00848 [hep-ex]} \BibitemShut
  {NoStop}%
\bibitem [{\citenamefont {Arifi}\ \emph {et~al.}(2024)\citenamefont {Arifi},
  \citenamefont {Happ}, \citenamefont {Ohno},\ and\ \citenamefont
  {Oka}}]{Arifi:2024mff}%
  \BibitemOpen
  \bibfield  {author} {\bibinfo {author} {\bibfnamefont {A.~J.}\ \bibnamefont
  {Arifi}}, \bibinfo {author} {\bibfnamefont {L.}~\bibnamefont {Happ}},
  \bibinfo {author} {\bibfnamefont {S.}~\bibnamefont {Ohno}},\ and\ \bibinfo
  {author} {\bibfnamefont {M.}~\bibnamefont {Oka}},\ }\bibfield  {title}
  {\bibinfo {title} {{Structure of heavy mesons in the light-front quark
  model}},\ }\href {https://doi.org/10.1103/PhysRevD.110.014020} {\bibfield
  {journal} {\bibinfo  {journal} {Phys. Rev. D}\ }\textbf {\bibinfo {volume}
  {110}},\ \bibinfo {pages} {014020} (\bibinfo {year} {2024})},\ \Eprint
  {https://arxiv.org/abs/2401.07933} {arXiv:2401.07933 [hep-ph]} \BibitemShut
  {NoStop}%
\bibitem [{\citenamefont {Weber}(1992)}]{Weber:1992ww}%
  \BibitemOpen
  \bibfield  {author} {\bibinfo {author} {\bibfnamefont {H.~J.}\ \bibnamefont
  {Weber}},\ }\bibfield  {title} {\bibinfo {title} {{Light cone quark model
  with spin force for the nucleon and Delta (1232)}},\ }\href
  {https://doi.org/10.1016/0370-2693(92)91868-A} {\bibfield  {journal}
  {\bibinfo  {journal} {Phys. Lett. B}\ }\textbf {\bibinfo {volume} {287}},\
  \bibinfo {pages} {14} (\bibinfo {year} {1992})}\BibitemShut {NoStop}%
\bibitem [{\citenamefont {Xiao}\ \emph {et~al.}(2002)\citenamefont {Xiao},
  \citenamefont {Qian},\ and\ \citenamefont {Ma}}]{Xiao:2002iv}%
  \BibitemOpen
  \bibfield  {author} {\bibinfo {author} {\bibfnamefont {B.-W.}\ \bibnamefont
  {Xiao}}, \bibinfo {author} {\bibfnamefont {X.}~\bibnamefont {Qian}},\ and\
  \bibinfo {author} {\bibfnamefont {B.-Q.}\ \bibnamefont {Ma}},\ }\bibfield
  {title} {\bibinfo {title} {{The Kaon form-factor in the light cone quark
  model}},\ }\href {https://doi.org/10.1140/epja/i2002-10059-y} {\bibfield
  {journal} {\bibinfo  {journal} {Eur. Phys. J. A}\ }\textbf {\bibinfo {volume}
  {15}},\ \bibinfo {pages} {523} (\bibinfo {year} {2002})},\ \Eprint
  {https://arxiv.org/abs/hep-ph/0209138} {arXiv:hep-ph/0209138} \BibitemShut
  {NoStop}%
\bibitem [{\citenamefont {Acharyya}\ \emph
  {et~al.}(2024{\natexlab{b}})\citenamefont {Acharyya}, \citenamefont {Puhan},\
  and\ \citenamefont {Dahiya}}]{Acharyya:2024enp}%
  \BibitemOpen
  \bibfield  {author} {\bibinfo {author} {\bibfnamefont {R.}~\bibnamefont
  {Acharyya}}, \bibinfo {author} {\bibfnamefont {S.}~\bibnamefont {Puhan}},\
  and\ \bibinfo {author} {\bibfnamefont {H.}~\bibnamefont {Dahiya}},\
  }\bibfield  {title} {\bibinfo {title} {{Quark spin-orbit correlations in
  spin-0 and spin-1 mesons using the light-front quark model}},\ }\href
  {https://doi.org/10.1103/PhysRevD.110.034020} {\bibfield  {journal} {\bibinfo
   {journal} {Phys. Rev. D}\ }\textbf {\bibinfo {volume} {110}},\ \bibinfo
  {pages} {034020} (\bibinfo {year} {2024}{\natexlab{b}})},\ \Eprint
  {https://arxiv.org/abs/2405.00446} {arXiv:2405.00446 [hep-ph]} \BibitemShut
  {NoStop}%
\bibitem [{\citenamefont {Xiao}\ and\ \citenamefont {Ma}(2003)}]{Xiao:2003wf}%
  \BibitemOpen
  \bibfield  {author} {\bibinfo {author} {\bibfnamefont {B.-W.}\ \bibnamefont
  {Xiao}}\ and\ \bibinfo {author} {\bibfnamefont {B.-Q.}\ \bibnamefont {Ma}},\
  }\bibfield  {title} {\bibinfo {title} {{Pion photon and photon pion
  transition form-factors in the light cone formalism}},\ }\href
  {https://doi.org/10.1103/PhysRevD.68.034020} {\bibfield  {journal} {\bibinfo
  {journal} {Phys. Rev. D}\ }\textbf {\bibinfo {volume} {68}},\ \bibinfo
  {pages} {034020} (\bibinfo {year} {2003})},\ \Eprint
  {https://arxiv.org/abs/hep-ph/0312162} {arXiv:hep-ph/0312162} \BibitemShut
  {NoStop}%
\bibitem [{\citenamefont {Puhan}\ and\ \citenamefont
  {Dahiya}(2024{\natexlab{c}})}]{Puhan_2024}%
  \BibitemOpen
  \bibfield  {author} {\bibinfo {author} {\bibfnamefont {S.}~\bibnamefont
  {Puhan}}\ and\ \bibinfo {author} {\bibfnamefont {H.}~\bibnamefont {Dahiya}},\
  }\bibfield  {title} {\bibinfo {title} {Spatial and transverse structure of
  heavy b- and d-mesons},\ }in\ \href@noop {} {\emph {\bibinfo {booktitle}
  {Proceedings of 16th International Conference on Heavy Quarks and Leptons —
  PoS(HQL2023)}}},\ \bibinfo {series and number} {HQL2023}\ (\bibinfo
  {publisher} {Sissa Medialab},\ \bibinfo {year} {2024})\ p.\ \bibinfo {pages}
  {089}\BibitemShut {NoStop}%
\bibitem [{\citenamefont {Shi}\ \emph {et~al.}(2022)\citenamefont {Shi},
  \citenamefont {Li}, \citenamefont {Li}, \citenamefont {Chen},\ and\
  \citenamefont {Jia}}]{Shi:2022erw}%
  \BibitemOpen
  \bibfield  {author} {\bibinfo {author} {\bibfnamefont {C.}~\bibnamefont
  {Shi}}, \bibinfo {author} {\bibfnamefont {J.}~\bibnamefont {Li}}, \bibinfo
  {author} {\bibfnamefont {M.}~\bibnamefont {Li}}, \bibinfo {author}
  {\bibfnamefont {X.}~\bibnamefont {Chen}},\ and\ \bibinfo {author}
  {\bibfnamefont {W.}~\bibnamefont {Jia}},\ }\bibfield  {title} {\bibinfo
  {title} {{Transverse momentum distributions of valence quarks in light and
  heavy vector mesons}},\ }\href {https://doi.org/10.1103/PhysRevD.106.014026}
  {\bibfield  {journal} {\bibinfo  {journal} {Phys. Rev. D}\ }\textbf {\bibinfo
  {volume} {106}},\ \bibinfo {pages} {014026} (\bibinfo {year} {2022})},\
  \Eprint {https://arxiv.org/abs/2205.02757} {arXiv:2205.02757 [hep-ph]}
  \BibitemShut {NoStop}%
\bibitem [{\citenamefont {Lepage}\ and\ \citenamefont
  {Brodsky}(1980)}]{Lepage:1980fj}%
  \BibitemOpen
  \bibfield  {author} {\bibinfo {author} {\bibfnamefont {G.~P.}\ \bibnamefont
  {Lepage}}\ and\ \bibinfo {author} {\bibfnamefont {S.~J.}\ \bibnamefont
  {Brodsky}},\ }\bibfield  {title} {\bibinfo {title} {{Exclusive Processes in
  Perturbative Quantum Chromodynamics}},\ }\href
  {https://doi.org/10.1103/PhysRevD.22.2157} {\bibfield  {journal} {\bibinfo
  {journal} {Phys. Rev. D}\ }\textbf {\bibinfo {volume} {22}},\ \bibinfo
  {pages} {2157} (\bibinfo {year} {1980})}\BibitemShut {NoStop}%
\bibitem [{\citenamefont {Pasquini}\ \emph {et~al.}(2023)\citenamefont
  {Pasquini}, \citenamefont {Rodini},\ and\ \citenamefont
  {Venturini}}]{Pasquini:2023aaf}%
  \BibitemOpen
  \bibfield  {author} {\bibinfo {author} {\bibfnamefont {B.}~\bibnamefont
  {Pasquini}}, \bibinfo {author} {\bibfnamefont {S.}~\bibnamefont {Rodini}},\
  and\ \bibinfo {author} {\bibfnamefont {S.}~\bibnamefont {Venturini}}
  (\bibinfo {collaboration} {MAP (Multi-dimensional Analyses of Partonic
  distributions)}),\ }\bibfield  {title} {\bibinfo {title} {{Valence quark,
  sea, and gluon content of the pion from the parton distribution functions and
  the electromagnetic form factor}},\ }\href
  {https://doi.org/10.1103/PhysRevD.107.114023} {\bibfield  {journal} {\bibinfo
   {journal} {Phys. Rev. D}\ }\textbf {\bibinfo {volume} {107}},\ \bibinfo
  {pages} {114023} (\bibinfo {year} {2023})},\ \Eprint
  {https://arxiv.org/abs/2303.01789} {arXiv:2303.01789 [hep-ph]} \BibitemShut
  {NoStop}%
\bibitem [{\citenamefont {Qian}\ and\ \citenamefont {Ma}(2008)}]{Qian:2008px}%
  \BibitemOpen
  \bibfield  {author} {\bibinfo {author} {\bibfnamefont {W.}~\bibnamefont
  {Qian}}\ and\ \bibinfo {author} {\bibfnamefont {B.-Q.}\ \bibnamefont {Ma}},\
  }\bibfield  {title} {\bibinfo {title} {{Vector meson omega-phi mixing and
  their form factors in light-cone quark model}},\ }\href
  {https://doi.org/10.1103/PhysRevD.78.074002} {\bibfield  {journal} {\bibinfo
  {journal} {Phys. Rev. D}\ }\textbf {\bibinfo {volume} {78}},\ \bibinfo
  {pages} {074002} (\bibinfo {year} {2008})},\ \Eprint
  {https://arxiv.org/abs/0809.4411} {arXiv:0809.4411 [hep-ph]} \BibitemShut
  {NoStop}%
\bibitem [{\citenamefont {Brodsky}\ \emph {et~al.}(2001)\citenamefont
  {Brodsky}, \citenamefont {Diehl},\ and\ \citenamefont
  {Hwang}}]{Brodsky:2000xy}%
  \BibitemOpen
  \bibfield  {author} {\bibinfo {author} {\bibfnamefont {S.~J.}\ \bibnamefont
  {Brodsky}}, \bibinfo {author} {\bibfnamefont {M.}~\bibnamefont {Diehl}},\
  and\ \bibinfo {author} {\bibfnamefont {D.~S.}\ \bibnamefont {Hwang}},\
  }\bibfield  {title} {\bibinfo {title} {{Light cone wave function
  representation of deeply virtual Compton scattering}},\ }\href
  {https://doi.org/10.1016/S0550-3213(00)00695-7} {\bibfield  {journal}
  {\bibinfo  {journal} {Nucl. Phys. B}\ }\textbf {\bibinfo {volume} {596}},\
  \bibinfo {pages} {99} (\bibinfo {year} {2001})},\ \Eprint
  {https://arxiv.org/abs/hep-ph/0009254} {arXiv:hep-ph/0009254} \BibitemShut
  {NoStop}%
\bibitem [{\citenamefont {Choi}\ and\ \citenamefont {Ji}(1997)}]{Choi:1996mq}%
  \BibitemOpen
  \bibfield  {author} {\bibinfo {author} {\bibfnamefont {H.~M.}\ \bibnamefont
  {Choi}}\ and\ \bibinfo {author} {\bibfnamefont {C.-R.}\ \bibnamefont {Ji}},\
  }\bibfield  {title} {\bibinfo {title} {{Light cone quark model predictions
  for radiative meson decays}},\ }\href
  {https://doi.org/10.1016/S0375-9474(97)00052-3} {\bibfield  {journal}
  {\bibinfo  {journal} {Nucl. Phys. A}\ }\textbf {\bibinfo {volume} {618}},\
  \bibinfo {pages} {291} (\bibinfo {year} {1997})}\BibitemShut {NoStop}%
\bibitem [{\citenamefont {Arifi}\ \emph {et~al.}(2022)\citenamefont {Arifi},
  \citenamefont {Choi}, \citenamefont {ji},\ and\ \citenamefont
  {Oh}}]{Arifi:2022pal}%
  \BibitemOpen
  \bibfield  {author} {\bibinfo {author} {\bibfnamefont {A.~J.}\ \bibnamefont
  {Arifi}}, \bibinfo {author} {\bibfnamefont {H.-M.}\ \bibnamefont {Choi}},
  \bibinfo {author} {\bibfnamefont {C.-R.}\ \bibnamefont {ji}},\ and\ \bibinfo
  {author} {\bibfnamefont {Y.}~\bibnamefont {Oh}},\ }\bibfield  {title}
  {\bibinfo {title} {{Mixing effects on 1S and 2S state heavy mesons in the
  light-front quark model}},\ }\href
  {https://doi.org/10.1103/PhysRevD.106.014009} {\bibfield  {journal} {\bibinfo
   {journal} {Phys. Rev. D}\ }\textbf {\bibinfo {volume} {106}},\ \bibinfo
  {pages} {014009} (\bibinfo {year} {2022})},\ \Eprint
  {https://arxiv.org/abs/2205.04075} {arXiv:2205.04075 [hep-ph]} \BibitemShut
  {NoStop}%
\bibitem [{\citenamefont {Bacchetta}\ \emph {et~al.}(2020)\citenamefont
  {Bacchetta}, \citenamefont {Celiberto}, \citenamefont {Radici},\ and\
  \citenamefont {Taels}}]{Bacchetta:2020vty}%
  \BibitemOpen
  \bibfield  {author} {\bibinfo {author} {\bibfnamefont {A.}~\bibnamefont
  {Bacchetta}}, \bibinfo {author} {\bibfnamefont {F.~G.}\ \bibnamefont
  {Celiberto}}, \bibinfo {author} {\bibfnamefont {M.}~\bibnamefont {Radici}},\
  and\ \bibinfo {author} {\bibfnamefont {P.}~\bibnamefont {Taels}},\ }\bibfield
   {title} {\bibinfo {title} {{Transverse-momentum-dependent gluon distribution
  functions in a spectator model}},\ }\href
  {https://doi.org/10.1140/epjc/s10052-020-8327-6} {\bibfield  {journal}
  {\bibinfo  {journal} {Eur. Phys. J. C}\ }\textbf {\bibinfo {volume} {80}},\
  \bibinfo {pages} {733} (\bibinfo {year} {2020})},\ \Eprint
  {https://arxiv.org/abs/2005.02288} {arXiv:2005.02288 [hep-ph]} \BibitemShut
  {NoStop}%
\bibitem [{\citenamefont {Lorc\'e}\ \emph {et~al.}(2016)\citenamefont
  {Lorc\'e}, \citenamefont {Pasquini},\ and\ \citenamefont
  {Schweitzer}}]{Lorce:2016ugb}%
  \BibitemOpen
  \bibfield  {author} {\bibinfo {author} {\bibfnamefont {C.}~\bibnamefont
  {Lorc\'e}}, \bibinfo {author} {\bibfnamefont {B.}~\bibnamefont {Pasquini}},\
  and\ \bibinfo {author} {\bibfnamefont {P.}~\bibnamefont {Schweitzer}},\
  }\bibfield  {title} {\bibinfo {title} {{Transverse pion structure beyond
  leading twist in constituent models}},\ }\href
  {https://doi.org/10.1140/epjc/s10052-016-4257-8} {\bibfield  {journal}
  {\bibinfo  {journal} {Eur. Phys. J. C}\ }\textbf {\bibinfo {volume} {76}},\
  \bibinfo {pages} {415} (\bibinfo {year} {2016})},\ \Eprint
  {https://arxiv.org/abs/1605.00815} {arXiv:1605.00815 [hep-ph]} \BibitemShut
  {NoStop}%
\bibitem [{\citenamefont {Lorc\'e}\ \emph {et~al.}(2015)\citenamefont
  {Lorc\'e}, \citenamefont {Pasquini},\ and\ \citenamefont
  {Schweitzer}}]{Lorce:2014hxa}%
  \BibitemOpen
  \bibfield  {author} {\bibinfo {author} {\bibfnamefont {C.}~\bibnamefont
  {Lorc\'e}}, \bibinfo {author} {\bibfnamefont {B.}~\bibnamefont {Pasquini}},\
  and\ \bibinfo {author} {\bibfnamefont {P.}~\bibnamefont {Schweitzer}},\
  }\bibfield  {title} {\bibinfo {title} {{Unpolarized transverse momentum
  dependent parton distribution functions beyond leading twist in quark
  models}},\ }\href {https://doi.org/10.1007/JHEP01(2015)103} {\bibfield
  {journal} {\bibinfo  {journal} {JHEP}\ }\textbf {\bibinfo {volume} {01}},\
  \bibinfo {pages} {103}},\ \Eprint {https://arxiv.org/abs/1411.2550}
  {arXiv:1411.2550 [hep-ph]} \BibitemShut {NoStop}%
\bibitem [{\citenamefont {Shi}\ \emph {et~al.}(2024)\citenamefont {Shi},
  \citenamefont {Liu}, \citenamefont {Du},\ and\ \citenamefont
  {Jia}}]{Shi:2024laj}%
  \BibitemOpen
  \bibfield  {author} {\bibinfo {author} {\bibfnamefont {C.}~\bibnamefont
  {Shi}}, \bibinfo {author} {\bibfnamefont {P.}~\bibnamefont {Liu}}, \bibinfo
  {author} {\bibfnamefont {Y.-L.}\ \bibnamefont {Du}},\ and\ \bibinfo {author}
  {\bibfnamefont {W.}~\bibnamefont {Jia}},\ }\bibfield  {title} {\bibinfo
  {title} {{Heavy flavor-asymmetric pseudoscalar mesons on the light front}},\
  }\href@noop {} {\  (\bibinfo {year} {2024})},\ \Eprint
  {https://arxiv.org/abs/2409.05098} {arXiv:2409.05098 [hep-ph]} \BibitemShut
  {NoStop}%
\bibitem [{\citenamefont {Albino}\ \emph {et~al.}(2022)\citenamefont {Albino},
  \citenamefont {Higuera-Angulo}, \citenamefont {Raya},\ and\ \citenamefont
  {Bashir}}]{Albino:2022gzs}%
  \BibitemOpen
  \bibfield  {author} {\bibinfo {author} {\bibfnamefont {L.}~\bibnamefont
  {Albino}}, \bibinfo {author} {\bibfnamefont {I.~M.}\ \bibnamefont
  {Higuera-Angulo}}, \bibinfo {author} {\bibfnamefont {K.}~\bibnamefont
  {Raya}},\ and\ \bibinfo {author} {\bibfnamefont {A.}~\bibnamefont {Bashir}},\
  }\bibfield  {title} {\bibinfo {title} {{Pseudoscalar mesons: Light front wave
  functions, GPDs, and PDFs}},\ }\href
  {https://doi.org/10.1103/PhysRevD.106.034003} {\bibfield  {journal} {\bibinfo
   {journal} {Phys. Rev. D}\ }\textbf {\bibinfo {volume} {106}},\ \bibinfo
  {pages} {034003} (\bibinfo {year} {2022})},\ \Eprint
  {https://arxiv.org/abs/2207.06550} {arXiv:2207.06550 [hep-ph]} \BibitemShut
  {NoStop}%
\bibitem [{\citenamefont {Zhu}\ \emph {et~al.}(2023)\citenamefont {Zhu},
  \citenamefont {Hu}, \citenamefont {Lan}, \citenamefont {Mondal},
  \citenamefont {Zhao},\ and\ \citenamefont {Vary}}]{Zhu:2023lst}%
  \BibitemOpen
  \bibfield  {author} {\bibinfo {author} {\bibfnamefont {Z.}~\bibnamefont
  {Zhu}}, \bibinfo {author} {\bibfnamefont {Z.}~\bibnamefont {Hu}}, \bibinfo
  {author} {\bibfnamefont {J.}~\bibnamefont {Lan}}, \bibinfo {author}
  {\bibfnamefont {C.}~\bibnamefont {Mondal}}, \bibinfo {author} {\bibfnamefont
  {X.}~\bibnamefont {Zhao}},\ and\ \bibinfo {author} {\bibfnamefont {J.~P.}\
  \bibnamefont {Vary}} (\bibinfo {collaboration} {BLFQ}),\ }\bibfield  {title}
  {\bibinfo {title} {{Transverse structure of the pion beyond leading twist
  with basis light-front quantization}},\ }\href
  {https://doi.org/10.1016/j.physletb.2023.137808} {\bibfield  {journal}
  {\bibinfo  {journal} {Phys. Lett. B}\ }\textbf {\bibinfo {volume} {839}},\
  \bibinfo {pages} {137808} (\bibinfo {year} {2023})},\ \Eprint
  {https://arxiv.org/abs/2301.12994} {arXiv:2301.12994 [hep-ph]} \BibitemShut
  {NoStop}%
\bibitem [{\citenamefont {Miyama}\ and\ \citenamefont
  {Kumano}(1996)}]{Miyama:1995bd}%
  \BibitemOpen
  \bibfield  {author} {\bibinfo {author} {\bibfnamefont {M.}~\bibnamefont
  {Miyama}}\ and\ \bibinfo {author} {\bibfnamefont {S.}~\bibnamefont
  {Kumano}},\ }\bibfield  {title} {\bibinfo {title} {{Numerical solution of
  Q**2 evolution equations in a brute force method}},\ }\href
  {https://doi.org/10.1016/0010-4655(96)00013-6} {\bibfield  {journal}
  {\bibinfo  {journal} {Comput. Phys. Commun.}\ }\textbf {\bibinfo {volume}
  {94}},\ \bibinfo {pages} {185} (\bibinfo {year} {1996})},\ \Eprint
  {https://arxiv.org/abs/hep-ph/9508246} {arXiv:hep-ph/9508246} \BibitemShut
  {NoStop}%
\bibitem [{\citenamefont {Hirai}\ \emph
  {et~al.}(1998{\natexlab{a}})\citenamefont {Hirai}, \citenamefont {Kumano},\
  and\ \citenamefont {Miyama}}]{Hirai:1997gb}%
  \BibitemOpen
  \bibfield  {author} {\bibinfo {author} {\bibfnamefont {M.}~\bibnamefont
  {Hirai}}, \bibinfo {author} {\bibfnamefont {S.}~\bibnamefont {Kumano}},\ and\
  \bibinfo {author} {\bibfnamefont {M.}~\bibnamefont {Miyama}},\ }\bibfield
  {title} {\bibinfo {title} {{Numerical solution of Q**2 evolution equations
  for polarized structure functions}},\ }\href
  {https://doi.org/10.1016/S0010-4655(97)00129-X} {\bibfield  {journal}
  {\bibinfo  {journal} {Comput. Phys. Commun.}\ }\textbf {\bibinfo {volume}
  {108}},\ \bibinfo {pages} {38} (\bibinfo {year} {1998}{\natexlab{a}})},\
  \Eprint {https://arxiv.org/abs/hep-ph/9707220} {arXiv:hep-ph/9707220}
  \BibitemShut {NoStop}%
\bibitem [{\citenamefont {Hirai}\ \emph
  {et~al.}(1998{\natexlab{b}})\citenamefont {Hirai}, \citenamefont {Kumano},\
  and\ \citenamefont {Miyama}}]{Hirai:1997mm}%
  \BibitemOpen
  \bibfield  {author} {\bibinfo {author} {\bibfnamefont {M.}~\bibnamefont
  {Hirai}}, \bibinfo {author} {\bibfnamefont {S.}~\bibnamefont {Kumano}},\ and\
  \bibinfo {author} {\bibfnamefont {M.}~\bibnamefont {Miyama}},\ }\bibfield
  {title} {\bibinfo {title} {{Numerical solution of Q**2 evolution equation for
  the transversity distribution Delta(T)q}},\ }\href
  {https://doi.org/10.1016/S0010-4655(98)00028-9} {\bibfield  {journal}
  {\bibinfo  {journal} {Comput. Phys. Commun.}\ }\textbf {\bibinfo {volume}
  {111}},\ \bibinfo {pages} {150} (\bibinfo {year} {1998}{\natexlab{b}})},\
  \Eprint {https://arxiv.org/abs/hep-ph/9712410} {arXiv:hep-ph/9712410}
  \BibitemShut {NoStop}%
\bibitem [{\citenamefont {Hirai}\ and\ \citenamefont
  {Kumano}(2012)}]{Hirai:2011si}%
  \BibitemOpen
  \bibfield  {author} {\bibinfo {author} {\bibfnamefont {M.}~\bibnamefont
  {Hirai}}\ and\ \bibinfo {author} {\bibfnamefont {S.}~\bibnamefont {Kumano}},\
  }\bibfield  {title} {\bibinfo {title} {{Numerical solution of $Q^2$ evolution
  equations for fragmentation functions}},\ }\href
  {https://doi.org/10.1016/j.cpc.2011.12.022} {\bibfield  {journal} {\bibinfo
  {journal} {Comput. Phys. Commun.}\ }\textbf {\bibinfo {volume} {183}},\
  \bibinfo {pages} {1002} (\bibinfo {year} {2012})},\ \Eprint
  {https://arxiv.org/abs/1106.1553} {arXiv:1106.1553 [hep-ph]} \BibitemShut
  {NoStop}%
\bibitem [{\citenamefont {Aicher}\ \emph {et~al.}(2010)\citenamefont {Aicher},
  \citenamefont {Schafer},\ and\ \citenamefont {Vogelsang}}]{Aicher:2010cb}%
  \BibitemOpen
  \bibfield  {author} {\bibinfo {author} {\bibfnamefont {M.}~\bibnamefont
  {Aicher}}, \bibinfo {author} {\bibfnamefont {A.}~\bibnamefont {Schafer}},\
  and\ \bibinfo {author} {\bibfnamefont {W.}~\bibnamefont {Vogelsang}},\
  }\bibfield  {title} {\bibinfo {title} {{Soft-gluon resummation and the
  valence parton distribution function of the pion}},\ }\href
  {https://doi.org/10.1103/PhysRevLett.105.252003} {\bibfield  {journal}
  {\bibinfo  {journal} {Phys. Rev. Lett.}\ }\textbf {\bibinfo {volume} {105}},\
  \bibinfo {pages} {252003} (\bibinfo {year} {2010})},\ \Eprint
  {https://arxiv.org/abs/1009.2481} {arXiv:1009.2481 [hep-ph]} \BibitemShut
  {NoStop}%
\bibitem [{\citenamefont {Bourrely}\ \emph {et~al.}(2024)\citenamefont
  {Bourrely}, \citenamefont {Buccella}, \citenamefont {Chang},\ and\
  \citenamefont {Peng}}]{Bourrely:2023yzi}%
  \BibitemOpen
  \bibfield  {author} {\bibinfo {author} {\bibfnamefont {C.}~\bibnamefont
  {Bourrely}}, \bibinfo {author} {\bibfnamefont {F.}~\bibnamefont {Buccella}},
  \bibinfo {author} {\bibfnamefont {W.-C.}\ \bibnamefont {Chang}},\ and\
  \bibinfo {author} {\bibfnamefont {J.-C.}\ \bibnamefont {Peng}},\ }\bibfield
  {title} {\bibinfo {title} {{Extraction of kaon partonic distribution
  functions from Drell-Yan and J/\ensuremath{\psi} production data}},\ }\href
  {https://doi.org/10.1016/j.physletb.2023.138395} {\bibfield  {journal}
  {\bibinfo  {journal} {Phys. Lett. B}\ }\textbf {\bibinfo {volume} {848}},\
  \bibinfo {pages} {138395} (\bibinfo {year} {2024})},\ \Eprint
  {https://arxiv.org/abs/2305.18117} {arXiv:2305.18117 [hep-ph]} \BibitemShut
  {NoStop}%
\bibitem [{\citenamefont {Bednar}\ \emph {et~al.}(2020)\citenamefont {Bednar},
  \citenamefont {Clo\"et},\ and\ \citenamefont {Tandy}}]{Bednar:2018mtf}%
  \BibitemOpen
  \bibfield  {author} {\bibinfo {author} {\bibfnamefont {K.~D.}\ \bibnamefont
  {Bednar}}, \bibinfo {author} {\bibfnamefont {I.~C.}\ \bibnamefont
  {Clo\"et}},\ and\ \bibinfo {author} {\bibfnamefont {P.~C.}\ \bibnamefont
  {Tandy}},\ }\bibfield  {title} {\bibinfo {title} {{Distinguishing Quarks and
  Gluons in Pion and Kaon Parton Distribution Functions}},\ }\href
  {https://doi.org/10.1103/PhysRevLett.124.042002} {\bibfield  {journal}
  {\bibinfo  {journal} {Phys. Rev. Lett.}\ }\textbf {\bibinfo {volume} {124}},\
  \bibinfo {pages} {042002} (\bibinfo {year} {2020})},\ \Eprint
  {https://arxiv.org/abs/1811.12310} {arXiv:1811.12310 [nucl-th]} \BibitemShut
  {NoStop}%
\bibitem [{\citenamefont {Brodsky}\ \emph {et~al.}(2007)\citenamefont
  {Brodsky}, \citenamefont {Llanes-Estrada},\ and\ \citenamefont
  {Szczepaniak}}]{Brodsky:2007fr}%
  \BibitemOpen
  \bibfield  {author} {\bibinfo {author} {\bibfnamefont {S.~J.}\ \bibnamefont
  {Brodsky}}, \bibinfo {author} {\bibfnamefont {F.~J.}\ \bibnamefont
  {Llanes-Estrada}},\ and\ \bibinfo {author} {\bibfnamefont {A.~P.}\
  \bibnamefont {Szczepaniak}},\ }\bibfield  {title} {\bibinfo {title}
  {{Illuminating the 1/x moment of parton distribution functions}},\
  }\href@noop {} {\bibfield  {journal} {\bibinfo  {journal} {eConf}\ }\textbf
  {\bibinfo {volume} {C070910}},\ \bibinfo {pages} {149} (\bibinfo {year}
  {2007})},\ \Eprint {https://arxiv.org/abs/0710.0981} {arXiv:0710.0981
  [nucl-th]} \BibitemShut {NoStop}%
\bibitem [{\citenamefont {Horn}\ \emph {et~al.}(2006)\citenamefont {Horn} \emph
  {et~al.}}]{JeffersonLabFpi-2:2006ysh}%
  \BibitemOpen
  \bibfield  {author} {\bibinfo {author} {\bibfnamefont {T.}~\bibnamefont
  {Horn}} \emph {et~al.} (\bibinfo {collaboration} {Jefferson Lab F(pi)-2}),\
  }\bibfield  {title} {\bibinfo {title} {{Determination of the Charged Pion
  Form Factor at Q**2 = 1.60 and 2.45-(GeV/c)**2}},\ }\href
  {https://doi.org/10.1103/PhysRevLett.97.192001} {\bibfield  {journal}
  {\bibinfo  {journal} {Phys. Rev. Lett.}\ }\textbf {\bibinfo {volume} {97}},\
  \bibinfo {pages} {192001} (\bibinfo {year} {2006})},\ \Eprint
  {https://arxiv.org/abs/nucl-ex/0607005} {arXiv:nucl-ex/0607005} \BibitemShut
  {NoStop}%
\bibitem [{\citenamefont {Amendolia}\ \emph
  {et~al.}(1986{\natexlab{a}})\citenamefont {Amendolia} \emph
  {et~al.}}]{NA7:1986vav}%
  \BibitemOpen
  \bibfield  {author} {\bibinfo {author} {\bibfnamefont {S.~R.}\ \bibnamefont
  {Amendolia}} \emph {et~al.} (\bibinfo {collaboration} {NA7}),\ }\bibfield
  {title} {\bibinfo {title} {{A Measurement of the Space - Like Pion
  Electromagnetic Form-Factor}},\ }\href
  {https://doi.org/10.1016/0550-3213(86)90437-2} {\bibfield  {journal}
  {\bibinfo  {journal} {Nucl. Phys. B}\ }\textbf {\bibinfo {volume} {277}},\
  \bibinfo {pages} {168} (\bibinfo {year} {1986}{\natexlab{a}})}\BibitemShut
  {NoStop}%
\bibitem [{\citenamefont {Dally}\ \emph {et~al.}(1982)\citenamefont {Dally}
  \emph {et~al.}}]{Dally:1982zk}%
  \BibitemOpen
  \bibfield  {author} {\bibinfo {author} {\bibfnamefont {E.~B.}\ \bibnamefont
  {Dally}} \emph {et~al.},\ }\bibfield  {title} {\bibinfo {title} {{Elastic
  Scattering Measurement of the Negative Pion Radius}},\ }\href
  {https://doi.org/10.1103/PhysRevLett.48.375} {\bibfield  {journal} {\bibinfo
  {journal} {Phys. Rev. Lett.}\ }\textbf {\bibinfo {volume} {48}},\ \bibinfo
  {pages} {375} (\bibinfo {year} {1982})}\BibitemShut {NoStop}%
\bibitem [{\citenamefont {Br\"ommel}\ \emph {et~al.}(2007)\citenamefont
  {Br\"ommel} \emph {et~al.}}]{QCDSFUKQCD:2006gmg}%
  \BibitemOpen
  \bibfield  {author} {\bibinfo {author} {\bibfnamefont {D.}~\bibnamefont
  {Br\"ommel}} \emph {et~al.} (\bibinfo {collaboration} {QCDSF/UKQCD}),\
  }\bibfield  {title} {\bibinfo {title} {{The Pion form-factor from lattice QCD
  with two dynamical flavours}},\ }\href
  {https://doi.org/10.1140/epjc/s10052-007-0295-6} {\bibfield  {journal}
  {\bibinfo  {journal} {Eur. Phys. J. C}\ }\textbf {\bibinfo {volume} {51}},\
  \bibinfo {pages} {335} (\bibinfo {year} {2007})},\ \Eprint
  {https://arxiv.org/abs/hep-lat/0608021} {arXiv:hep-lat/0608021} \BibitemShut
  {NoStop}%
\bibitem [{\citenamefont {Volmer}\ \emph {et~al.}(2001)\citenamefont {Volmer}
  \emph {et~al.}}]{JeffersonLabFpi:2000nlc}%
  \BibitemOpen
  \bibfield  {author} {\bibinfo {author} {\bibfnamefont {J.}~\bibnamefont
  {Volmer}} \emph {et~al.} (\bibinfo {collaboration} {Jefferson Lab F(pi)}),\
  }\bibfield  {title} {\bibinfo {title} {{Measurement of the Charged Pion
  Electromagnetic Form-Factor}},\ }\href
  {https://doi.org/10.1103/PhysRevLett.86.1713} {\bibfield  {journal} {\bibinfo
   {journal} {Phys. Rev. Lett.}\ }\textbf {\bibinfo {volume} {86}},\ \bibinfo
  {pages} {1713} (\bibinfo {year} {2001})},\ \Eprint
  {https://arxiv.org/abs/nucl-ex/0010009} {arXiv:nucl-ex/0010009} \BibitemShut
  {NoStop}%
\bibitem [{\citenamefont {Tadevosyan}\ \emph {et~al.}(2007)\citenamefont
  {Tadevosyan} \emph {et~al.}}]{JeffersonLabFpi:2007vir}%
  \BibitemOpen
  \bibfield  {author} {\bibinfo {author} {\bibfnamefont {V.}~\bibnamefont
  {Tadevosyan}} \emph {et~al.} (\bibinfo {collaboration} {Jefferson Lab
  F(pi)}),\ }\bibfield  {title} {\bibinfo {title} {{Determination of the pion
  charge form-factor for Q**2 = 0.60-GeV**2 - 1.60-GeV**2}},\ }\href
  {https://doi.org/10.1103/PhysRevC.75.055205} {\bibfield  {journal} {\bibinfo
  {journal} {Phys. Rev. C}\ }\textbf {\bibinfo {volume} {75}},\ \bibinfo
  {pages} {055205} (\bibinfo {year} {2007})},\ \Eprint
  {https://arxiv.org/abs/nucl-ex/0607007} {arXiv:nucl-ex/0607007} \BibitemShut
  {NoStop}%
\bibitem [{\citenamefont {Dally}\ \emph {et~al.}(1980)\citenamefont {Dally}
  \emph {et~al.}}]{Dally:1980dj}%
  \BibitemOpen
  \bibfield  {author} {\bibinfo {author} {\bibfnamefont {E.~B.}\ \bibnamefont
  {Dally}} \emph {et~al.},\ }\bibfield  {title} {\bibinfo {title} {{DIRECT
  MEASUREMENT OF THE NEGATIVE KAON FORM-FACTOR}},\ }\href
  {https://doi.org/10.1103/PhysRevLett.45.232} {\bibfield  {journal} {\bibinfo
  {journal} {Phys. Rev. Lett.}\ }\textbf {\bibinfo {volume} {45}},\ \bibinfo
  {pages} {232} (\bibinfo {year} {1980})}\BibitemShut {NoStop}%
\bibitem [{\citenamefont {Amendolia}\ \emph
  {et~al.}(1986{\natexlab{b}})\citenamefont {Amendolia} \emph
  {et~al.}}]{Amendolia:1986ui}%
  \BibitemOpen
  \bibfield  {author} {\bibinfo {author} {\bibfnamefont {S.~R.}\ \bibnamefont
  {Amendolia}} \emph {et~al.},\ }\bibfield  {title} {\bibinfo {title} {{A
  Measurement of the Kaon Charge Radius}},\ }\href
  {https://doi.org/10.1016/0370-2693(86)91407-3} {\bibfield  {journal}
  {\bibinfo  {journal} {Phys. Lett. B}\ }\textbf {\bibinfo {volume} {178}},\
  \bibinfo {pages} {435} (\bibinfo {year} {1986}{\natexlab{b}})}\BibitemShut
  {NoStop}%
\bibitem [{\citenamefont {Dias}\ \emph {et~al.}(2010)\citenamefont {Dias},
  \citenamefont {Filho},\ and\ \citenamefont {de~Melo}}]{Dias:2010sg}%
  \BibitemOpen
  \bibfield  {author} {\bibinfo {author} {\bibfnamefont {O.~A.~T.}\
  \bibnamefont {Dias}}, \bibinfo {author} {\bibfnamefont {V.~S.}\ \bibnamefont
  {Filho}},\ and\ \bibinfo {author} {\bibfnamefont {J.~P. B.~C.}\ \bibnamefont
  {de~Melo}},\ }\bibfield  {title} {\bibinfo {title} {{Kaon and Pion
  Electromagnetic Form Factor Ratios in the Light-Front}},\ }\href
  {https://doi.org/10.1016/j.nuclphysbps.2010.02.044} {\bibfield  {journal}
  {\bibinfo  {journal} {Nucl. Phys. B Proc. Suppl.}\ }\textbf {\bibinfo
  {volume} {199}},\ \bibinfo {pages} {281} (\bibinfo {year} {2010})},\ \Eprint
  {https://arxiv.org/abs/1001.4039} {arXiv:1001.4039 [hep-ph]} \BibitemShut
  {NoStop}%
\bibitem [{\citenamefont {Can}\ \emph {et~al.}(2013)\citenamefont {Can},
  \citenamefont {Erkol}, \citenamefont {Oka}, \citenamefont {Ozpineci},\ and\
  \citenamefont {Takahashi}}]{Can:2012tx}%
  \BibitemOpen
  \bibfield  {author} {\bibinfo {author} {\bibfnamefont {K.~U.}\ \bibnamefont
  {Can}}, \bibinfo {author} {\bibfnamefont {G.}~\bibnamefont {Erkol}}, \bibinfo
  {author} {\bibfnamefont {M.}~\bibnamefont {Oka}}, \bibinfo {author}
  {\bibfnamefont {A.}~\bibnamefont {Ozpineci}},\ and\ \bibinfo {author}
  {\bibfnamefont {T.~T.}\ \bibnamefont {Takahashi}},\ }\bibfield  {title}
  {\bibinfo {title} {{Vector and axial-vector couplings of D and D* mesons in
  2+1 flavor Lattice QCD}},\ }\href
  {https://doi.org/10.1016/j.physletb.2012.12.050} {\bibfield  {journal}
  {\bibinfo  {journal} {Phys. Lett. B}\ }\textbf {\bibinfo {volume} {719}},\
  \bibinfo {pages} {103} (\bibinfo {year} {2013})},\ \Eprint
  {https://arxiv.org/abs/1210.0869} {arXiv:1210.0869 [hep-lat]} \BibitemShut
  {NoStop}%
\bibitem [{\citenamefont {Li}\ and\ \citenamefont {Wu}(2017)}]{Li:2017eic}%
  \BibitemOpen
  \bibfield  {author} {\bibinfo {author} {\bibfnamefont {N.}~\bibnamefont
  {Li}}\ and\ \bibinfo {author} {\bibfnamefont {Y.-J.}\ \bibnamefont {Wu}},\
  }\bibfield  {title} {\bibinfo {title} {{Lattice study of D and D$_{s}$ meson
  form factors with twisted boundary conditions}},\ }\href
  {https://doi.org/10.1140/epja/i2017-12243-4} {\bibfield  {journal} {\bibinfo
  {journal} {Eur. Phys. J. A}\ }\textbf {\bibinfo {volume} {53}},\ \bibinfo
  {pages} {56} (\bibinfo {year} {2017})}\BibitemShut {NoStop}%
\bibitem [{\citenamefont {Li}\ \emph {et~al.}(2020)\citenamefont {Li},
  \citenamefont {Liu},\ and\ \citenamefont {Wu}}]{Li:2020gau}%
  \BibitemOpen
  \bibfield  {author} {\bibinfo {author} {\bibfnamefont {N.}~\bibnamefont
  {Li}}, \bibinfo {author} {\bibfnamefont {C.-C.}\ \bibnamefont {Liu}},\ and\
  \bibinfo {author} {\bibfnamefont {Y.-J.}\ \bibnamefont {Wu}},\ }\bibfield
  {title} {\bibinfo {title} {{Lattice study of form factors for charmonium}},\
  }\href {https://doi.org/10.1140/epja/s10050-020-00253-2} {\bibfield
  {journal} {\bibinfo  {journal} {Eur. Phys. J. A}\ }\textbf {\bibinfo {volume}
  {56}},\ \bibinfo {pages} {242} (\bibinfo {year} {2020})}\BibitemShut
  {NoStop}%
\end{thebibliography}%


\bibitem{Lepage:1980fj}Lepage, G. \& Brodsky, S. Exclusive Processes in Perturbative Quantum Chromodynamics. {\em Phys. Rev. D}. \textbf{22} pp. 2157 (1980)
\bibitem{Brodsky:1997de}Brodsky, S., Pauli, H. \& Pinsky, S. Quantum chromodynamics and other field theories on the light cone. {\em Phys. Rept.}. \textbf{301} pp. 299-486 (1998)
\bibitem{Pasquini:2023aaf}Pasquini, B., Rodini, S. \& Venturini, S. Valence quark, sea, and gluon content of the pion from the parton distribution functions and the electromagnetic form factor. {\em Phys. Rev. D}. \textbf{107}, 114023 (2023)
\bibitem{Shi:2022erw}Shi, C., Li, J., Li, M., Chen, X. \& Jia, W. Transverse momentum distributions of valence quarks in light and heavy vector mesons. {\em Phys. Rev. D}. \textbf{106}, 014026 (2022)
\bibitem{Puhan_2024}Puhan, S. \& Dahiya, H. Spatial and Transverse structure of Heavy B- and D-mesons. {\em Proceedings Of 16th International Conference On Heavy Quarks And Leptons — PoS(HQL2023)}. pp. 089 (2024,4)
\bibitem{Qian:2008px}Qian, W. \& Ma, B. Vector meson omega-phi mixing and their form factors in light-cone quark model. {\em Phys. Rev. D}. \textbf{78} pp. 074002 (2008)
\bibitem{Xiao:2002iv}Xiao, B., Qian, X. \& Ma, B. The Kaon form-factor in the light cone quark model. {\em Eur. Phys. J. A}. \textbf{15} pp. 523-527 (2002)
\bibitem{Kaur:2020vkq}Kaur, S., Kumar, N., Lan, J., Mondal, C. \& Dahiya, H. Tomography of light mesons in the light-cone quark model. {\em Phys. Rev. D}. \textbf{102}, 014021 (2020)
\bibitem{Ma:1993ht}Ma, B. Spin structure of the pion in a light cone representation. {\em Z. Phys. A}. \textbf{345} pp. 321-325 (1993)
\bibitem{Yu:2007hp}Yu, J., Xiao, B. \& Ma, B. Space-like and time-like pion-rho transition form factors in the light-cone formalism. {\em J. Phys. G}. \textbf{34} pp. 1845-1860 (2007)
\bibitem{Arifi:2022pal}Arifi, A., Choi, H., Ji, C. \& Oh, Y. Mixing effects on 1S and 2S state heavy mesons in the light-front quark model. {\em Phys. Rev. D}. \textbf{106}, 014009 (2022)
\bibitem{Choi:1996mq}Choi, H. \& Ji, C. Light cone quark model predictions for radiative meson decays. {\em Nucl. Phys. A}. \textbf{618} pp. 291-316 (1997)
\bibitem{Diehl:2003ny}Diehl, M. Generalized parton distributions. {\em Phys. Rept.}. \textbf{388} pp. 41-277 (2003)
\bibitem{Bacchetta:2020vty}Bacchetta, A., Celiberto, F., Radici, M. \& Taels, P. Transverse-momentum-dependent gluon distribution functions in a spectator model. {\em Eur. Phys. J. C}. \textbf{80}, 733 (2020)
\bibitem{Kaur:2019jfa}Kaur, N. \& Dahiya, H. Transverse momentum dependent parton distributions of pion in the light-front holographic model. {\em Int. J. Mod. Phys. A}. \textbf{36}, 2150052 (2021)
\bibitem{Serna:2024vpn}Serna, F., El-Bennich, B. \& Krein, G. Parton distribution functions and transverse momentum dependence of heavy mesons.  (2024,9)
\bibitem{Almeida-Zamora:2023bqb}Almeida-Zamora, B., Cobos-Martı́nez, J., Bashir, A., Raya, K., Rodrı́guez-Quintero, J. \& Segovia, J. Algebraic model to study the internal structure of pseudoscalar mesons with heavy-light quark content. {\em Phys. Rev. D}. \textbf{109}, 014016 (2024)
\bibitem{Shi:2024laj}Shi, C., Liu, P., Du, Y. \& Jia, W. Heavy flavor-asymmetric pseudoscalar mesons on the light front.  (2024,9)
\bibitem{Lorce:2016ugb}Lorcé, C., Pasquini, B. \& Schweitzer, P. Transverse pion structure beyond leading twist in constituent models. {\em Eur. Phys. J. C}. \textbf{76}, 415 (2016)
\bibitem{Lorce:2014hxa}Lorcé, C., Pasquini, B. \& Schweitzer, P. Unpolarized transverse momentum dependent parton distribution functions beyond leading twist in quark models. {\em JHEP}. \textbf{1} pp. 103 (2015)
\bibitem{Albino:2022gzs}Albino, L., Higuera-Angulo, I., Raya, K. \& Bashir, A. Pseudoscalar mesons: Light front wave functions, GPDs, and PDFs. {\em Phys. Rev. D}. \textbf{106}, 034003 (2022)
\bibitem{Puhan2023}Puhan, S., Sharma, S., Kaur, N., Kumar, N. \& Dahiya, H. T-even TMDs for the spin-0 pseudo-scalar mesons upto twist-4 using light-front formalism. {\em JHEP}. \textbf{2} pp. 075 (2024)
\bibitem{Zhu:2023lst}Zhu, Z., Hu, Z., Lan, J., Mondal, C., Zhao, X. \& Vary, J. Transverse structure of the pion beyond leading twist with basis light-front quantization. {\em Phys. Lett. B}. \textbf{839} pp. 137808 (2023)
\bibitem{Aicher:2010cb}Aicher, M., Schafer, A. \& Vogelsang, W. Soft-gluon resummation and the valence parton distribution function of the pion. {\em Phys. Rev. Lett.}. \textbf{105} pp. 252003 (2010)
\bibitem{Lan:2019rba}Lan, J., Mondal, C., Jia, S., Zhao, X. \& Vary, J. Pion and kaon parton distribution functions from basis light front quantization and QCD evolution. {\em Phys. Rev. D}. \textbf{101}, 034024 (2020)
\bibitem{Cui:2020tdf}Cui, Z., Ding, M., Gao, F., Raya, K., Binosi, D., Chang, L., Roberts, C., Rodrı́guez-Quintero, J. \& Schmidt, S. Kaon and pion parton distributions. {\em Eur. Phys. J. C}. \textbf{80}, 1064 (2020)
\bibitem{Miyama:1995bd}Miyama, M. \& Kumano, S. Numerical solution of Q**2 evolution equations in a brute force method. {\em Comput. Phys. Commun.}. \textbf{94} pp. 185-215 (1996)
\bibitem{Hirai:1997gb}Hirai, M., Kumano, S. \& Miyama, M. Numerical solution of Q**2 evolution equations for polarized structure functions. {\em Comput. Phys. Commun.}. \textbf{108} pp. 38 (1998)
\bibitem{Hirai:1997mm}Hirai, M., Kumano, S. \& Miyama, M. Numerical solution of Q**2 evolution equation for the transversity distribution Delta(T)q. {\em Comput. Phys. Commun.}. \textbf{111} pp. 150-166 (1998)
\bibitem{Hirai:2011si}Hirai, M. \& Kumano, S. Numerical solution of Q² evolution equations for fragmentation functions. {\em Comput. Phys. Commun.}. \textbf{183} pp. 1002-1013 (2012)
\bibitem{Bourrely:2023yzi}Bourrely, C., Buccella, F., Chang, W. \& Peng, J. Extraction of kaon partonic distribution functions from Drell-Yan and J/\ensuremathψ production data. {\em Phys. Lett. B}. \textbf{848} pp. 138395 (2024)
\bibitem{Han:2020vjp}Han, C., Xie, G., Wang, R. \& Chen, X. An Analysis of Parton Distribution Functions of the Pion and the Kaon with the Maximum Entropy Input. {\em Eur. Phys. J. C}. \textbf{81}, 302 (2021)
\bibitem{Bednar:2018mtf}Bednar, K., Cloët, I. \& Tandy, P. Distinguishing Quarks and Gluons in Pion and Kaon Parton Distribution Functions. {\em Phys. Rev. Lett.}. \textbf{124}, 042002 (2020)
\bibitem{Brodsky:2007fr}Brodsky, S., Llanes-Estrada, F. \& Szczepaniak, A. Illuminating the 1/x moment of parton distribution functions. {\em EConf}. \textbf{C070910} pp. 149 (2007)
\bibitem{Brodsky:2000xy}Brodsky, S., Diehl, M. \& Hwang, D. Light cone wave function representation of deeply virtual Compton scattering. {\em Nucl. Phys. B}. \textbf{596} pp. 99-124 (2001)
\bibitem{Zhang:1997dd}Zhang, W. A Weak coupling treatment of nonperturbative QCD dynamics to heavy hadrons. {\em Phys. Rev. D}. \textbf{56} pp. 1528-1548 (1997)
\bibitem{Collins:1981uw}Collins, J. \& Soper, D. Parton Distribution and Decay Functions. {\em Nucl. Phys. B}. \textbf{194} pp. 445-492 (1982)
\bibitem{Martin:1998sq}Martin, A., Roberts, R., Stirling, W. \& Thorne, R. Parton distributions: A New global analysis. {\em Eur. Phys. J. C}. \textbf{4} pp. 463-496 (1998)
\bibitem{Gluck:1994uf}Gluck, M., Reya, E. \& Vogt, A. Dynamical parton distributions of the proton and small x physics. {\em Z. Phys. C}. \textbf{67} pp. 433-448 (1995)
\bibitem{Alekhin:2002fv}Alekhin, S. Parton distributions from deep inelastic scattering data. {\em Phys. Rev. D}. \textbf{68} pp. 014002 (2003)
\bibitem{Polchinski:2002jw}Polchinski, J. \& Strassler, M. Deep inelastic scattering and gauge / string duality. {\em JHEP}. \textbf{5} pp. 012 (2003)
\bibitem{Can:2012tx}Can, K., Erkol, G., Oka, M., Ozpineci, A. \& Takahashi, T. Vector and axial-vector couplings of D and D* mesons in 2+1 flavor Lattice QCD. {\em Phys. Lett. B}. \textbf{719} pp. 103-109 (2013)
\bibitem{NA7:1986vav}Amendolia, S. \& Others A Measurement of the Space - Like Pion Electromagnetic Form-Factor. {\em Nucl. Phys. B}. \textbf{277} pp. 168 (1986)
\bibitem{Dally:1982zk}Dally, E. \& Others Elastic Scattering Measurement of the Negative Pion Radius. {\em Phys. Rev. Lett.}. \textbf{48} pp. 375-378 (1982)
\bibitem{QCDSFUKQCD:2006gmg}Brömmel, D. \& Others The Pion form-factor from lattice QCD with two dynamical flavours. {\em Eur. Phys. J. C}. \textbf{51} pp. 335-345 (2007)
\bibitem{JeffersonLabFpi:2000nlc}Volmer, J. \& Others Measurement of the Charged Pion Electromagnetic Form-Factor. {\em Phys. Rev. Lett.}. \textbf{86} pp. 1713-1716 (2001)
\bibitem{JeffersonLabFpi-2:2006ysh}Horn, T. \& Others Determination of the Charged Pion Form Factor at Q**2 = 1.60 and 2.45-(GeV/c)**2. {\em Phys. Rev. Lett.}. \textbf{97} pp. 192001 (2006)
\bibitem{JeffersonLabFpi:2007vir}Tadevosyan, V. \& Others Determination of the pion charge form-factor for Q**2 = 0.60-GeV**2 - 1.60-GeV**2. {\em Phys. Rev. C}. \textbf{75} pp. 055205 (2007)
\bibitem{Dally:1980dj}Dally, E. \& Others DIRECT MEASUREMENT OF THE NEGATIVE KAON FORM-FACTOR. {\em Phys. Rev. Lett.}. \textbf{45} pp. 232-235 (1980)
\bibitem{Dias:2010sg}Dias, O., Filho, V. \& Melo, J. Kaon and Pion Electromagnetic Form Factor Ratios in the Light-Front. {\em Nucl. Phys. B Proc. Suppl.}. \textbf{199} pp. 281-284 (2010)
\bibitem{Amendolia:1986ui}Amendolia, S. \& Others A Measurement of the Kaon Charge Radius. {\em Phys. Lett. B}. \textbf{178} pp. 435-440 (1986)
\bibitem{Li:2017eic}Li, N. \& Wu, Y. Lattice study of D and Dₚ meson form factors with twisted boundary conditions. {\em Eur. Phys. J. A}. \textbf{53}, 56 (2017)
\bibitem{Li:2020gau}Li, N., Liu, C. \& Wu, Y. Lattice study of form factors for charmonium. {\em Eur. Phys. J. A}. \textbf{56}, 242 (2020)
\bibitem{Diehl:2015uka}Diehl, M. Introduction to GPDs and TMDs. {\em Eur. Phys. J. A}. \textbf{52}, 149 (2016)
\bibitem{Angeles-Martinez:2015sea}Angeles-Martinez, R. \& Others Transverse Momentum Dependent (TMD) parton distribution functions: status and prospects. {\em Acta Phys. Polon. B}. \textbf{46}, 2501-2534 (2015)
\bibitem{Pasquini:2008ax}Pasquini, B., Cazzaniga, S. \& Boffi, S. Transverse momentum dependent parton distributions in a light-cone quark model. {\em Phys. Rev. D}. \textbf{78} pp. 034025 (2008)
\bibitem{Chavez:2021llq}Chavez, J., Bertone, V., De Soto Borrero, F., Defurne, M., Mezrag, C., Moutarde, H., Rodrı́guez-Quintero, J. \& Segovia, J. Pion generalized parton distributions: A path toward phenomenology. {\em Phys. Rev. D}. \textbf{105}, 094012 (2022)
\bibitem{Zhang:2021tnr}Zhang, J. \& Ping, J. Kaon generalized parton distributions and light-front wave functions in the Nambu–Jona-Lasinio model. {\em Eur. Phys. J. C}. \textbf{81}, 814 (2021)
\bibitem{Broniowski:2022iip}Broniowski, W., Shastry, V. \& Ruiz Arriola, E. Off-shell generalized parton distributions and form factors of the pion. {\em Phys. Lett. B}. \textbf{840} pp. 137872 (2023)
\bibitem{Kaur:2023zhn}Kaur, N. \& Dahiya, H. Generalized parton distributions for the lowest-lying octet baryons. {\em Eur. Phys. J. A}. \textbf{60}, 42 (2024)
\bibitem{Guidal:2004nd}Guidal, M., Polyakov, M., Radyushkin, A. \& Vanderhaeghen, M. Nucleon form-factors from generalized parton distributions. {\em Phys. Rev. D}. \textbf{72} pp. 054013 (2005)
\bibitem{Kaur:2024wze}Kaur, N., Puhan, S., Pandey, R., Kumar, A., Dutt, S. \& Dahiya, H. Does nuclear medium affect the transverse momentum-dependent parton distributions of valence quark of pions?.  (2024,9)
\bibitem{Echevarria:2022ztg}Echevarria, M., Gutierrez Garcia, P. \& Scimemi, I. GTMDs and the factorization of exclusive double Drell-Yan. {\em Phys. Lett. B}. \textbf{840} pp. 137881 (2023)
\bibitem{Meissner:2009ww}Meissner, S., Metz, A. \& Schlegel, M. Generalized parton correlation functions for a spin-1/2 hadron. {\em JHEP}. \textbf{8} pp. 056 (2009)
\bibitem{Goeke:2005hb}Goeke, K., Metz, A. \& Schlegel, M. Parameterization of the quark-quark correlator of a spin-1/2 hadron. {\em Phys. Lett. B}. \textbf{618} pp. 90-96 (2005)
\bibitem{Bacchetta:2017gcc}Bacchetta, A., Delcarro, F., Pisano, C., Radici, M. \& Signori, A. Extraction of partonic transverse momentum distributions from semi-inclusive deep-inelastic scattering, Drell-Yan and Z-boson production. {\em JHEP}. \textbf{6} pp. 081 (2017), [Erratum: JHEP 06, 051 (2019)]
\bibitem{Makris:2020ltr}Makris, Y., Ringer, F. \& Waalewijn, W. Joint thrust and TMD resummation in electron-positron and electron-proton collisions. {\em JHEP}. \textbf{2} pp. 070 (2021)
\bibitem{Boer:1997mf}Boer, D., Jakob, R. \& Mulders, P. Asymmetries in polarized hadron production in e+ e- annihilation up to order 1/Q. {\em Nucl. Phys. B}. \textbf{504} pp. 345-380 (1997)
\bibitem{Catani:2015vma}Catani, S., Florian, D., Ferrera, G. \& Grazzini, M. Vector boson production at hadron colliders: transverse-momentum resummation and leptonic decay. {\em JHEP}. \textbf{12} pp. 047 (2015)
\bibitem{Ji:1996nm}Ji, X. Deeply virtual Compton scattering. {\em Phys. Rev. D}. \textbf{55} pp. 7114-7125 (1997)
\bibitem{Xie:2023xkz}Xie, G., Kou, W., Fu, Q., Ye, Z. \& Chen, X. Deeply virtual compton scattering at future electron-ion colliders. {\em Eur. Phys. J. C}. \textbf{83}, 900 (2023)
\bibitem{Favart:2015umi}Favart, L., Guidal, M., Horn, T. \& Kroll, P. Deeply Virtual Meson Production on the nucleon. {\em Eur. Phys. J. A}. \textbf{52}, 158 (2016)
\bibitem{Brooks:2018uqk}Brooks, W., Schmidt, I. \& Siddikov, M. Deeply virtual meson production on neutrons. {\em Phys. Rev. D}. \textbf{98}, 116006 (2018)
\bibitem{Kaur:2024bgo}Kaur, N. \& Dahiya, H. Transverse distortion and single-spin asymmetries for low-lying octet baryons. {\em Int. J. Mod. Phys. A}. \textbf{39}, 2450076 (2024)
\bibitem{Freese:2020mcx}Freese, A. \& Cloët, I. Quark spin and orbital angular momentum from proton generalized parton distributions. {\em Phys. Rev. C}. \textbf{103}, 045204 (2021)
\bibitem{Luan:2023lmt}Luan, X. \& Lu, Z. Generalized parton distributions of sea quark at zero skewness in the light-cone model. {\em Eur. Phys. J. C}. \textbf{83}, 504 (2023)
\bibitem{Meissner:2008ay}Meissner, S., Metz, A., Schlegel, M. \& Goeke, K. Generalized parton correlation functions for a spin-0 hadron. {\em JHEP}. \textbf{8} pp. 038 (2008)
\bibitem{Cerutti:2022lmb}Cerutti, M., Rossi, L., Venturini, S., Bacchetta, A., Bertone, V., Bissolotti, C. \& Radici, M. Extraction of pion transverse momentum distributions from Drell-Yan data. {\em Phys. Rev. D}. \textbf{107}, 014014 (2023)
\bibitem{LatticeParton:2023xdl}Chu, M. \& Others Transverse-momentum-dependent wave functions of the pion from lattice QCD. {\em Phys. Rev. D}. \textbf{109}, L091503 (2024)
\bibitem{Engelhardt:2015xja}Engelhardt, M., Hägler, P., Musch, B., Negele, J. \& Schäfer, A. Lattice QCD study of the Boer-Mulders effect in a pion. {\em Phys. Rev. D}. \textbf{93}, 054501 (2016)
\bibitem{Kou:2023ady}Kou, W., Shi, C., Chen, X. \& Jia, W. Transverse momentum dependent parton distributions of pion at leading twist. {\em Phys. Rev. D}. \textbf{108}, 036021 (2023)
\bibitem{Pasquini:2014ppa}Pasquini, B. \& Schweitzer, P. Pion transverse momentum dependent parton distributions in a light-front constituent approach, and the Boer-Mulders effect in the pion-induced Drell-Yan process. {\em Phys. Rev. D}. \textbf{90}, 014050 (2014)
\bibitem{Puhan:2023hio}Puhan, S. \& Dahiya, H. Leading twist T-even TMDs for the spin-1 heavy vector mesons. {\em Phys. Rev. D}. \textbf{109}, 034005 (2024)
\bibitem{Chen:2019lcm}Chen, J., Lin, H. \& Zhang, J. Pion generalized parton distribution from lattice QCD. {\em Nucl. Phys. B}. \textbf{952} pp. 114940 (2020)
\bibitem{Zhang:2018nsy}Zhang, J., Chen, J., Jin, L., Lin, H., Schäfer, A. \& Zhao, Y. First direct lattice-QCD calculation of the x-dependence of the pion parton distribution function. {\em Phys. Rev. D}. \textbf{100}, 034505 (2019)
\bibitem{Kaur:2018ewq}Kaur, N., Kumar, N., Mondal, C. \& Dahiya, H. Generalized Parton Distributions of Pion for Non-Zero Skewness in AdS/QCD. {\em Nucl. Phys. B}. \textbf{934} pp. 80-95 (2018)
\bibitem{AbdulKhalek:2021gbh}Abdul Khalek, R. \& Others Science Requirements and Detector Concepts for the Electron-Ion Collider: EIC Yellow Report. {\em Nucl. Phys. A}. \textbf{1026} pp. 122447 (2022)
\bibitem{Adams:2018pwt}Adams, B. \& Others Letter of Intent: A New QCD facility at the M2 beam line of the CERN SPS (COMPASS++/AMBER).  (2018,8)
\bibitem{Arifi:2024mff}Arifi, A., Happ, L., Ohno, S. \& Oka, M. Structure of heavy mesons in the light-front quark model. {\em Phys. Rev. D}. \textbf{110}, 014020 (2024)
\bibitem{Weber:1992ww}Weber, H. Light cone quark model with spin force for the nucleon and Delta (1232). {\em Phys. Lett. B}. \textbf{287} pp. 14-17 (1992)
\bibitem{Acharyya:2024enp}Acharyya, R., Puhan, S. \& Dahiya, H. Quark spin-orbit correlations in spin-0 and spin-1 mesons using the light-front quark model. {\em Phys. Rev. D}. \textbf{110}, 034020 (2024)
\bibitem{Xiao:2003wf}Xiao, B. \& Ma, B. Pion photon and photon pion transition form-factors in the light cone formalism. {\em Phys. Rev. D}. \textbf{68} pp. 034020 (2003)
\bibitem{Puhan:2024ckp}Puhan, S. \& Dahiya, H. Spatial and Transverse structure of Heavy B- and D-mesons. {\em PoS}. \textbf{HQL2023} pp. 089 (2024)
\bibitem{Acharyya:2024tql}Acharyya, R., Puhan, S., Kumar, N. \& Dahiya, H. Spectroscopy of excited quarkonium states in the light-front quark model.  (2024,8)

%---------------------------------------------------
\end{document}